\documentclass[a4paper]{article}

\usepackage{graphicx}    
\usepackage{amsmath,amssymb}
\usepackage[numbers]{natbib}
\usepackage{verbatim}
\usepackage{bbm}
\usepackage{mdwlist}
\usepackage{enumerate}
\usepackage[a4paper,bindingoffset=0.2in,%
            left=1in,right=1in,top=1in,bottom=1in,%
            footskip=.25in]{geometry}
            
\def\b0{\mbox{\bf{0}}}

\def\bx{\mbox{\boldmath $x$}}

\newcommand{\X}{\mathcal{X}}
\newcommand{\R}{\mathbb{R}}
\newcommand{\Expec}{\mathbb{E}}
\newcommand{\T}{\mathrm{T}}

\newcommand{\1}{\mathbbm{1}}

\newcommand{\mat}[1]{\mbox{\textbf{#1}}}
\newcommand{\vect}[1]{\mbox{\textbf{#1}}}
\newcommand{\vectg}[1]{\boldsymbol{#1}}

\newcommand*{\RDVSscale}{\@empty}
\newcommand*{\EEscale}{\@empty}
\newcommand*{\DSscale}{\@empty} 
\newcommand*{\SFRDscale}{\@empty}

\begin{document}

\title{Design of Experiments for Screening}
\author{David C. Woods and Susan M. Lewis \\
Southampton Statistical Sciences Research Institute\\ 
University of Southampton \\
Southampton SO17 1BJ UK \\
\texttt{\{D.Woods,S.M.Lewis\}@southampton.ac.uk}}
\date{}
\maketitle


\textbf{Abstract} 
The aim of this paper is to review methods of designing screening experiments, ranging from designs originally developed for physical experiments to those especially tailored to experiments on numerical models. The strengths and weaknesses of the various designs for screening variables in numerical models are discussed. First, classes of factorial designs for experiments to estimate main effects and interactions through a linear statistical model are described, specifically regular and non-regular fractional factorial designs, supersaturated designs and systematic fractional replicate designs. Generic issues of aliasing, bias and cancellation of factorial effects are discussed. Second, group screening experiments are considered including factorial group screening and sequential bifurcation. Third, random sampling plans are discussed including Latin hypercube sampling and sampling plans to estimate elementary effects. Fourth, a variety of modelling methods commonly employed with screening designs are briefly described. Finally, a novel study demonstrates six screening methods on two frequently-used exemplars, and their performances are compared.

\section{Introduction}\label{introduction}

Screening \cite{DeanLewis2006} is the process of discovering, through statistical design of experiments and modelling, those controllable factors or input variables that have a substantive impact on the response or output which is either calculated from a numerical model or observed from a physical process. 

Knowledge of these \textit{active} input variables is key to optimisation and control of the numerical model or process. In many areas of science and industry, there is often a large number of potentially important variables. Effective screening experiments are then needed to identify the active variables as economically as possible. This may be achieved through careful choice of experiment size and the set of combinations of input variable values (the design) to be run in the experiment. Each run determines an evaluation of the numerical model or an observation to be made on the physical process. The variables found to be active from the experiment are further investigated in one or more follow-up experiments that enable estimation of a detailed predictive statistical model of the output variable. 

The need to screen a large number of input variables in a relatively small experiment presents challenges for both design and modelling. Crucial to success is the principle of \textit{factor sparsity} \cite{BoxMeyer86} which states that only a small proportion of the input variables have a substantive influence on the output. If this widely-observed principle does not hold, then a small screening experiment may fail to reliably detect the active variables and a much larger investigation will be required.   

Whilst most literature has focussed on designs for physical experiments, screening is also important in the study of numerical models via computer experiments \cite{SantnerWilliamsNotz2003}. Such models often describe complex input-output relationships and have numerous input variables. A primary reason for building a numerical model is to gain better understanding of the nature of these relationships, especially the identification of the active input variables. If a small set of active variables can be identified, then the computational costs of subsequent exploration and exploitation of the numerical model are reduced. Construction of a surrogate model from the active variables requires less experimentation, and smaller Monte Carlo samples may suffice for uncertainty analysis and uncertainty propagation.

The effectiveness of screening can be evaluated in a variety of ways. Suppose there are $d$ input variables held in vector $\vect{x} = (x_1,\ldots,x_d)^{\T}$ and that $\X\subset\R^d$ contains all possible values of $\vect{x}$, i.e. all possible combinations of input variable values. Let $A_T\subseteq\{1,\ldots,d\}$ be the set of indices of the truly active variables and $A_S\subseteq\{1,\ldots,d\}$ consist of the indices of those variables selected as active through screening. Then the following measures may be defined: (i) sensitivity, $\phi_{\mbox{s}} = |A_S\cap A_T|/|A_T|$, the proportion of active variables that are successfully detected, where $\phi_{\mbox{s}}$ is defined as $1$ when $A_T=\emptyset$; (ii) false discovery rate \cite{BenjaminiHochberg95}, $\phi_{\mbox{fdr}} = |A_S\cap\bar{A}_T|/|A_S|$, where $\bar{A}_T$ is the complement of $A_T$,
the proportion of variables selected as active that are actually inactive and $\phi_{\mbox{fdr}}$ is defined as $0$ when $A_S=\emptyset$; (iii) type I error rate, $\phi_{\mbox{I}} = |A_S\cap \bar{A}_T|/|\bar{A}_T|$, the proportion of inactive variables that are selected as active. In practice, high sensitivity is often considered more important than a low type I error rate or false discovery rate \cite{Draguljicetal2014} because failure to detect an active input variable results in no further investigation of the variable and no exploitation of its effect on the output for purposes of optimisation and control.    

The majority of designs for screening experiments are tailored to the identification and estimation of a surrogate model that approximates an output variable $Y(\vect{x})$. A class of surrogate models which has been successfully applied in a variety of fields \cite{OverstallWoods2015} has the form 
 \begin{align}
Y(\vect{x}) & =  \vect{h}^\T(\vect{x})\vectg{\beta} + \varepsilon(\vect{x})\,,\label{eq:lm}
 \end{align}
where $\vect{h}$ is a $p\times 1$ vector of known functions of $\vect{x}$, $\vectg{\beta} = (\beta_0,\ldots,\beta_{p-1})^\T$ are unknown parameters, and $\varepsilon(\vect{x})$ is a random variable with a $N(0,\sigma^2)$ distribution for constant $\sigma^2$. Note that if multiple responses are obtained from each run of the experiment, then the simplest and most common approach is separate screening of the variables for each response using individual models of the form~\eqref{eq:lm}. 

An important decision in planning a screening experiment is the level of fidelity, or accuracy, required of a surrogate model for effective screening including the choice of the elements of $\vect{h}$ in~\eqref{eq:lm}. Two forms of~\eqref{eq:lm} are commonly used for screening variables in numerical models: linear regression models and Gaussian process models.

\subsection{Linear regression models}\label{sec:linreg}
Linear regression models assume that $\varepsilon(\vect{x}_0)$ and $\varepsilon(\vect{x}_0^\prime)$, $\vect{x}_0\ne\vect{x}_0^\prime\in\X$, are independent random variables. Estimation of detailed mean functions $\vect{h}^\T(\vect{x})\vectg{\beta}$ with a large number of terms requires large experiments which can be prohibitively expensive. Hence, many popular screening strategies investigate each input variable $x_i$ at two levels, often coded $+1$, $-1$ and referred to as ``high'' and ``low'', respectively \cite[][chs. 6 and 7]{BoxHunterHunter}. Interest is then in identifying those variables that have a large \textit{main effect}, defined for variable $x_i$ as the difference between the average expected responses for the $2^{d-1}$ combinations of variable values with $x_i=+1$ and the average for the $2^{d-1}$ combinations with $x_i=-1$. Main effects may be estimated via a \textit{first-order surrogate model}
\begin{equation}\label{eq:firstorder}
\vect{h}^\T(\vect{x})\vectg{\beta} = \beta_0 + \beta_1x_1 + \ldots + \beta_dx_d\,,
\end{equation}
where $p=d+1$. Such a ``main-effects screening'' strategy relies on a firm belief in \textit{strong effect heredity} \cite{HamadaWu92}, that is, important interactions or other nonlinearities involve only those input variables that have large main effects. Without this property, active variables may be overlooked.
 
There is evidence, particularly from industrial experiments \cite{Brenneman2014,Scintoetal2014}, that strong effect heredity may fail to hold in practice. This has led to the recent development and assessment of design and data analysis methodology that also allows screening of interactions between pairs of variables \cite{LewisDean2001,Draguljicetal2014}. For two-level variables, the interaction between $x_i$ and $x_j$ ($i,j=1,\ldots,d;\,i\ne j$) is defined as one-half of the difference between the conditional main effect for $x_i$ given $x_j=+1$ and the conditional main effect of $x_i$ given $x_j=-1$. Main effects and two-variable interactions can be estimated via a \textit{first-order surrogate model with two-variable product terms}  
\begin{equation}\label{eq:interactions}
\vect{h}^\T(\vect{x})\vectg{\beta} = \beta_0 + \beta_1x_1 + \ldots + \beta_dx_d + \beta_{12}x_1x_2 + \ldots + \beta_{(d-1)d}x_{d-1}x_d\,,
\end{equation}
where $p=1+d(d+1)/2$ and $\beta_{d+1},\ldots,\beta_{p-1}$ in~\eqref{eq:lm} are relabelled $\beta_{12},\ldots,\beta_{(d-1)d}$ for notational clarity. 

The main effects and interactions are collectively known as the factorial effects and can be shown to be the elements of $2\vectg{\beta}$. The screening problem may be cast as variable or model selection, that is choosing a statistical model composed of a subset of the terms in~\eqref{eq:interactions}.

The parameters in $\vectg{\beta}$ can be estimated by least squares. Let $x_i^{(j)}$ be the value taken by the $i$th variable in the $j$th run $(i=1,\ldots,d;\,j=1,\ldots,n)$. Then the rows of the $n\times d$ \textit{design matrix} 
$$\mat{X}^n = \left(x_1^{(j)},\ldots,x^{(j)}_d\right)_{j=1,\ldots,n}$$ 
each hold one run of the design. Let $\vect{Y}^n = (Y^{(1)},\ldots,Y^{(n)})$ be the output vector. Then the least squares estimator of $\vectg{\beta}$ is
\begin{equation}\label{eq:ls}
\hat{\vectg{\beta}} = \left(\mat{H}^\T\mat{H}\right)^{-1}\mat{H}^\T\vect{Y}^n\,,
\end{equation}
where $\mat{H} = (\vect{h}(\vect{x}^n_1),\ldots,\vect{h}(\vect{x}^n_n))^\T$ is the \textit{model matrix}, and $(\vect{x}^n_j)^\T$ is the $j$th row of $\mat{X}^n$. For the least squares estimators to be uniquely defined, $\mat{H}$ must be of full column rank. 

In physical screening experiments, often no attempt is made to estimate nonlinear effects other than two-variable interactions. This practice is underpinned by the principle of \textit{effect hierarchy} \cite[][]{WuHamada} which states that low-order factorial effects, such as main effects and two-variable interactions, are more likely to be important than higher-order effects. This principle is supported by substantial empirical evidence from physical experiments. 

However, excluding higher-order terms from surrogate model~\eqref{eq:lm} can result in biased estimators~\eqref{eq:ls}. Understanding, and minimising, this bias is key to linear model screening. Suppose that a more appropriate surrogate model is 
$$Y(\vect{x}) = \beta_0 + \vect{h}^\T(\bx)\vectg{\beta} + \tilde{\vect{h}}^\T(\bx)\tilde{\vectg{\beta}} + \varepsilon\,,$$
where $\tilde{\vect{h}}(\vect{x})$ is a $\tilde{p}$-vector of model terms, additional to those held in $\vect{h}(\vect{x})$, and $\tilde{\vectg{\beta}}$ is a $\tilde{p}$-vector of constants. Then the expected value of $\hat{\vectg{\beta}}$ is given by
\begin{equation}\label{eq:lmbias}
\Expec(\hat{\vectg{\beta}}) = \vectg{\beta} + \mat{A}\tilde{\vectg{\beta}}\,,
\end{equation}
where 
\begin{equation}\label{eq:A}
\mat{A}=(\mat{H}^\T\mat{H})^{-1}\mat{H}^\T\tilde{\mat{H}}\,,
\end{equation}
and $\tilde{\mat{H}} = (\tilde{\vect{h}}(\vect{x}^n_j))_j$. The \textit{alias matrix} $\mat{A}$ determines the pattern of bias in $\hat{\vectg{\beta}}$ due to omitting the terms $\tilde{\vect{h}}^\T(\bx)\tilde{\vectg{\beta}}$ from the surrogate model, and can be controlled through the choice of design. The size of the bias is determined by $\tilde{\vectg{\beta}}$ which is outside the experimenter's control.

\subsection{Gaussian process models}\label{sec:GPmodels}
Gaussian process (GP) models are used when it is anticipated that understanding more complex relationships between the input and output variables is necessary for screening. Under a GP model, it is assumed that $\varepsilon(\vect{x}_0),\varepsilon(\vect{x}_0^\prime)$ follow a bivariate normal distribution with correlation dependent on a distance metric applied to $\vect{x}_0,\vect{x}_0^\prime$, see \cite{RasmussenWilliams2006}.

Screening with a GP model requires interrogation of the parameters that control this correlation. A common correlation function employed for GP screening has the form:
\begin{equation}\label{eq:corrfun}
\mbox{cor}(\vect{x},\vect{x}^\prime) = \prod_{i=1}^d \exp\left(-\theta_i|x_i-x_i^\prime|^{\alpha_i}\right)\,,\qquad \theta_i\ge 0,\,0<\alpha_i\le 2\,.
\end{equation}
Conditional on $\theta_1,\ldots,\theta_d$, closed-form maximum likelihood or generalised least squares estimators for $\vectg{\beta}$ and $\sigma^2$ are available. However, $\theta_i$ requires numerical estimation. Reliable estimation of these more sophisticated and flexible surrogate models for a large number of variables requires larger experiments and may incur an impractically large number of evaluations of the numerical model. 

\subsection{Screening without a surrogate model}\label{sec:model-free}
The selection of active variables using a surrogate model relies on the model assumptions and their validation. An alternative model-free approach is the estimation of elementary effects \cite{morris91}. The elementary effect for the $i$th input variable for a combination of input values $\vect{x}_0\in\X$ is an approximation to the derivative of $Y(\vect{x}_0)$ in the direction of the $i$th variable. More formally,
\begin{equation}\label{eq:elem}
 \mbox{EE}_i(\vect{x}_0) = \frac{Y(\vect{x}_0 + \Delta\vect{e}_{id}) - Y(\vect{x}_0)}{\Delta}\,,\qquad i=1,\ldots,d\,,
 \end{equation}
where $\vect{e}_{id}$ is the $i$th unit vector of length $d$ (the $i$th column of the $d\times d$ identity matrix) and $\Delta>0$ is a given constant such that $\vect{x}+\Delta\vect{e}_{id}\in\X$. Repeated random draws of $\vect{x}_0$ from $\X$ according to a chosen distribution enable an empirical, model-free distribution for the elementary effect of the $i$th variable to be estimated. The moments (for example, mean and variance) of this distribution may be used to identify active effects, as discussed later.

In the remainder of the paper, a variety of screening methods are reviewed and discussed, starting with (regular and nonregular) factorial and fractional factorial designs in the next section. Later sections cover methods of screening groups of variables, such as factorial group screening and sequential bifurcation; random sampling plans and space-filling designs, including sampling plans for estimating elementary effects; and model selection methods. The paper finishes by comparing and contrasting the performance of six screening methods on two examples from the literature.  

\section{Factorial screening designs}\label{factorial}

In a full factorial design, each of the $d$ input variables is assigned a fixed number of values or levels and the design consists of one run of each of the distinct combinations of these values. Designs in which each variable has two values are mainly considered here, giving $n=2^d$ runs in the full factorial design. For even moderate values of $d$, experiments using such designs may be infeasibly large due to the costs or computing resources required. Further, such designs can be wasteful as they allow estimation of all interactions amongst the $d$ variables, whereas effect hierarchy suggests that low-order factorial effects (main effects and two-variable interactions) will be the most important. These problems may be overcome by using a carefully chosen subset, or \textit{fraction}, of the combinations of variable values in the full factorial design. Such fractional factorial designs have a long history of use in physical experiments \cite{Finney1943} and, more recently, have also been used in the study of numerical models \cite{Dupuyetal2014}. However, they bring the complication that the individual main effects and interactions cannot be estimated independently. Two classes of designs are discussed here.

\subsection{Regular fractional factorial designs}\label{regular}

The most widely used two-level fractional factorial designs are $1/2^{q}$ fractions of the $2^d$ full factorial design, known as $2^{d-q}$ designs \cite[][ch. 5]{WuHamada} $(1\le q<d$ is integer). As the outputs from all the combinations of variable values are not available from the experiment, the individual main effects and interactions cannot be estimated. However, in a \textit{regular} fractional factorial design, $2^{d-q}$ linear combinations of the factorial effects can be estimated. Two factorial effects that occur in the same linear combination cannot be independently estimated and are said to be \textit{aliased}. The designs are constructed by choosing which factorial effects should be aliased together.

\begin{table}
\begin{center}
\caption{A $2^{4-1}$ fractional factorial design constructed from the $2^3$ full factorial design showing the aliased effects.}
\label{tab:regfracex}       
\begin{tabular}{c@{\hskip 10pt}c@{\hskip 10pt}c@{\hskip 10pt}c@{\hskip 10pt}c@{\hskip 10pt}c@{\hskip 10pt}c@{\hskip 10pt}c@{\hskip 10pt}c@{\hskip 10pt}c}
\hline\noalign{\smallskip}
Run & $I$ & $x_1$ & $x_2$ & $x_3$ & $x_1x_2$ & $x_1x_3$ & $x_2x_3$ & $x_1x_2x_3$ \\
 & $=x_1x_2x_3x_4$ & $=x_2x_3x_4$ & $=x_1x_3x_4$ & $=x_1x_2x_4$  & $=x_3x_4$ & $=x_2x_4$ & $=x_1x_4$ & $=x_4$ \\ 
\noalign{\smallskip}\hline\noalign{\smallskip}
1 & +1 & -1 & -1 & -1 & +1 & +1 & +1 & -1 & \\
2 & +1 & -1 & -1 & +1 & +1 & -1 & -1 & +1 & \\
3 & +1 & -1 & +1 & -1 & -1 & +1 & -1 & +1 & \\
4 & +1 & -1 & +1 & +1 & -1 & -1 & +1 & -1 & \\
5 & +1 & +1 & -1 & -1 & -1 & -1 & +1 & +1 & \\
6 & +1 & +1 & -1 & +1 & -1 & +1 & -1 & -1 & \\
7 & +1 & +1 & +1 & -1 & +1 & -1 & -1 & -1 & \\
8 & +1 & +1 & +1 & +1 & +1 & +1 & +1 & +1 & \\
\noalign{\smallskip}\hline
\end{tabular}
\end{center}
\end{table}

The following example illustrates a full factorial design, the construction of a regular fractional factorial design and the resulting aliasing amongst the factorial effects. Consider first a $2^3$ fractional factorial design in variables $x_1$, $x_2$, $x_3$. Each run of this design is shown as a row across the columns 3 to 5 in Table~\ref{tab:regfracex}. Thus these three columns form the design matrix. The entries in these columns are the coefficients of the expected responses in the linear combinations that constitute the main effects, ignoring constants. Where interactions are involved, as in model~\eqref{eq:interactions}, their corresponding coefficients are obtained as elementwise products of columns 3 to 5. Thus columns 2--8 of Table~\ref{tab:regfracex} give the model matrix for model~\eqref{eq:interactions}. 

A $2^{4-1}$ regular fractional factorial design in $n=8$ runs may be constructed from the $2^3$ design by assigning the fourth variable, $x_4$, to one of the interaction columns. In Table~\ref{tab:regfracex}, $x_4$ is assigned to the column corresponding to the highest order interaction, $x_1x_2x_3$. Each of the eight runs now has the property that $x_{1}x_{2}x_{3}x_4=+1$ and hence, as each variable can only take values $\pm 1$, it follows that $x_1=x_{2}x_{3}x_4$, $x_2=x_{1}x_{3}x_4$ and $x_3=x_{1}x_{2}x_4$. Similarly, $x_1x_2=x_3x_4$, $x_1x_3=x_2x_4$ and $x_1x_4=x_2x_3$. Two consequences are: (i) each main effect is aliased with a three-variable interaction; (ii) each two-variable interaction is aliased with another two-variable interaction. However, for each variable, the \textit{sum} of the main effect and the three-variable interaction not involving that variable can be estimated. These two effects are said to be aliased. The other pairs of aliased effects are shown in Table~\ref{tab:regfracex}. The four-variable interaction cannot be estimated and is said to be aliased with the mean, denoted by $I=x_1x_2x_3x_4$ (column 2 of Table~\ref{tab:regfracex}). 

An estimable model for this $2^{4-1}$ design is 
\begin{equation*}\label{eq:exmodel}
\vect{h}^\T(\vect{x})\vectg{\beta} = \beta_0 + \beta_1x_1 + \ldots + \beta_4x_4 + \beta_{12}x_1x_2 + \beta_{13}x_{1}x_3 + \beta_{23}x_{2}x_3\,,
\end{equation*}
with model matrix $\mat{H}$ given by columns 2--9 of Table~\ref{tab:regfracex}. The columns of $\mat{H}$ are mutually orthogonal, $\vect{h}(\vect{x}_j^n)^\T\vect{h}(\vect{x}_j^n)=0$ for $j\ne k;\,j,k=1,\ldots,8$. The aliasing in the design will result in a biased estimator of $\vectg{\beta}$. This can be seen by setting
\begin{equation*}\label{eq:exmodel2}
\tilde{\vect{h}}^\T(\vect{x})\tilde{\vectg{\beta}} =  \beta_{14}x_1x_4 + \beta_{24}x_{2}x_4 + \beta_{34}x_{3}x_4 + \sum_{1\le j<k<l}\beta_{jkl}x_jx_kx_l + \beta_{1234}x_1x_2x_3x_4\,,
\end{equation*}
which leads to the alias matrix $\mat{A} = \sum_{j=1}^8\vect{e}_{j8}\vect{e}_{(8-j+1)8}$, which is an anti-diagonal identity matrix, and
\begin{align*}
\Expec(\hat{\beta}_0) = & \beta_0 + \beta_{1234}\,, & & \\
\Expec(\hat{\beta}_1) = & \beta_1 + \beta_{234}\,, & \Expec(\hat{\beta}_2) = & \beta_2 + \beta_{134}\, & \Expec(\hat{\beta}_3) = & \beta_3 + \beta_{124}\,, & \Expec(\hat{\beta}_4) = & \beta_4 + \beta_{123} \\
\Expec(\hat{\beta}_{12}) = & \beta_{12} + \beta_{34}\,, & \Expec(\hat{\beta}_{13}) = & \beta_{13} + \beta_{24}\,, & \Expec(\hat{\beta}_{23}) = & \beta_{23} + \beta_{14}\,. 
\end{align*}
 
More generally, to construct a $2^{d-q}$ fractional factorial design, a set $\{v_1,\ldots,v_q\}$ of \textit{defining words}, such as $x_1x_2x_3x_4$, must be chosen and the corresponding factorial effects aliased with the mean. That is, the product of variable values defined by each of these words is constant in the design (and equal to either -1 or +1). As the product of any two columns of constants in the design must also be constant, there is a total of $2^q-1$ effects aliased with the mean. The list of all effects aliased with the mean is called the defining relation, and is written as $I=v_1=\ldots=v_q=v_1v_2=\ldots=v_1\cdots v_q$. Products of the defining words are straightforward to calculate as $x_i^2=1$, so that $v_j^2=1$ ($i=1,\ldots,d;\,j=1,\ldots,2^d$).

The aliasing scheme for a design is easily obtained from the defining relation. A factorial effect with corresponding word $v_j$ is aliased with each factorial effect corresponding to the words $v_jv_1,v_jv_2,\ldots,v_jv_1\cdots v_q$ formed by the product of $v_j$ with every word in the defining relation. Hence, the defining relation $I=x_1x_2x_3x_4$ results in $x_1=x_2x_3x_4$, $x_2=x_1x_3x_4$ and so on, see Table~\ref{tab:regfracex}. 

As demonstrated above, the impact of aliasing is bias in the estimators of the regression coefficients in~\eqref{eq:lm} which can be formulated through the alias matrix. For a regular fractional factorial design, the columns of $\mat{H}$ are mutually orthogonal and hence $\mat{A} = \frac{1}{n}\mat{H}^\T\tilde{\mat{H}}$. If the functions in $\tilde{\vect{h}}$ correspond to those high order interactions not included in $\vect{h}$, then the elements of $A$ are all either 0 or $\pm 1$. This is because the aliasing of factorial effects ensures that each column of $\tilde{\mat{H}}$ is either orthogonal to all columns in $\mat{H}$ or identical to a column of $\mat{H}$ up to a change of sign. Thus $\mat{A}$ identifies aliasing amongst the factorial effects.

Crucial to fractional factorial design is the choice of a defining relation to ensure that effects of interest are not aliased together. Typically, this involves choosing defining words to ensure that only words corresponding to higher-order factorial effects are included in the defining relation. 

Regular fractional factorial designs are classed according to their \textit{resolution}. A \textit{resolution III} design has at least one main effect aliased with a two-variable interaction. A \textit{resolution IV} design has no main effects aliased with interactions but at least one pair of two-variable interactions aliased together. A \textit{resolution V} design has no main effects or two-variable interactions aliased with any other main effects or two-variable interactions. A more detailed and informative classification of regular fractional factorial designs is obtained via the aberration criterion \cite{ChengTang2005}.

Although resolution V designs allow for the estimation of higher fidelity surrogate models, they typically require too many runs for screening studies. The most common regular fractional factorial designs used in screening are resolution III designs as part of a main-effects screening strategy. The design in Table~\ref{tab:regfracex} has resolution IV.

\subsection{Nonregular fractional factorial designs}\label{non-regular}

The regular designs discussed above require $n$ to be a power of two, which limits their application to some experiments. Further, even resolution III regular fractional factorials may require too many runs to be feasible for large numbers of variables. For example, with $11$ variables, a resolution III regular fractional design requires $n=16$ runs. Smaller experiments with $n$ not equal to a power of two can often be performed by using the wider class of \textit{non-regular} fractional factorial designs \cite{XuPhoaWong2009} that cannot be constructed via a set of defining words. For $11$ variables, a design with $n=12$ runs can be constructed that can estimate all 11 main effects independently of each other. Whilst these designs are more flexible in their run size, the cost is a more complex aliasing scheme that makes interpretation of experimental results more challenging and requires the use of more sophisticated modelling methods.

Many non-regular designs are constructed via \textit{orthogonal arrays} \cite{Rao47}. A symmetric orthogonal array of strength $t$, denoted OA($n$, $s^d$, $t$), is an $n\times d$ matrix of $s$ different symbols such that all ordered $t$-tuples of the symbols occur equally often as rows of any $n\times t$ submatrix of the array. Each such array defines an $n$-run factorial design in $d$ variables, each having $s$ levels. Here, only arrays with $s=2$ symbols, $\pm 1$, will be discussed. The strength of the array is closely related to the resolution of the design. An array of strength $t=2$ allows estimation of all main effects independently of each other but not of the two-variable interactions (cf resolution III); a strength 3 array allows estimation of main effects independently of two-variable interactions (cf resolution IV). Clearly, the two-level regular fractional factorial designs are all orthogonal arrays. However, the class of orthogonal arrays is wider and includes many other designs that cannot be obtained via a defining relation.    

An important class of orthogonal arrays are constructed from Hadamard matrices \cite{Hall67}. A Hadamard matrix $\mat{C}$ of order $n$ is an $n\times n$ matrix with entries $\pm 1$ such that $\mat{C}^\T\mat{C} = n\mat{I}_n$, where $\mat{I}_n$ is the $n\times n$ identity matrix. An OA($n$, $2^{n-1}$, 2) can be obtained by multiplying rows of $\mat{C}$ by -1 as necessary to make all entries in the first column equal +1, and then removing the first column. Such a design can estimate the main effects of all $d=n-1$ variables independently, assuming negligible interactions. This class of designs includes the regular fractional factorials (e.g. for $n=4,8,16,\ldots$) but also other designs with $n$ a multiple of four but not a power of two ($n=12, 20, 24,\ldots$). These designs were first proposed by Plackett and Burman \cite{PlackettBurman}. Table~\ref{tab:PB12} gives the $n=12$ run Plackett-Burman (PB) design, one of the most frequently used for screening. 

\begin{table}
\begin{center}
\caption{The $n=12$ run non-regular Plackett-Burman design.}
\label{tab:PB12}       
\begin{tabular}{c@{\hskip 10pt}r@{\hskip 10pt}r@{\hskip 10pt}r@{\hskip 10pt}r@{\hskip 10pt}r@{\hskip 10pt}r@{\hskip 10pt}r@{\hskip 10pt}r@{\hskip 10pt}r@{\hskip 10pt}r@{\hskip 10pt}r}
\hline\noalign{\smallskip}
Run & $x_1$ & $x_2$ & $x_3$ & $x_4$ & $x_5$ & $x_6$ & $x_7$ & $x_8$ & $x_9$ & $x_{10}$ & $x_{11}$ \\
\noalign{\smallskip}\hline\noalign{\smallskip}
1 & -1 & -1 & -1 & -1 & -1 & -1 & -1 & -1 & -1 & -1 & -1 \\ 
  2 & -1 & -1 & -1 & -1 & -1 & 1 & 1 & 1 & 1 & 1 & 1 \\ 
  3 & -1 & -1 & 1 & 1 & 1 & -1 & -1 & -1 & 1 & 1 & 1 \\ 
  4 & -1 & 1 & -1 & 1 & 1 & -1 & 1 & 1 & -1 & -1 & 1 \\ 
  5 & -1 & 1 & 1 & -1 & 1 & 1 & -1 & 1 & -1 & 1 & -1 \\ 
  6 & -1 & 1 & 1 & 1 & -1 & 1 & 1 & -1 & 1 & -1 & -1 \\ 
  7 & 1 & -1 & 1 & 1 & -1 & -1 & 1 & 1 & -1 & 1 & -1 \\ 
  8 & 1 & -1 & 1 & -1 & 1 & 1 & 1 & -1 & -1 & -1 & 1 \\ 
  9 & 1 & -1 & -1 & 1 & 1 & 1 & -1 & 1 & 1 & -1 & -1 \\ 
  10 & 1 & 1 & 1 & -1 & -1 & -1 & -1 & 1 & 1 & -1 & 1 \\ 
  11 & 1 & 1 & -1 & 1 & -1 & 1 & -1 & -1 & -1 & 1 & 1 \\ 
  12 & 1 & 1 & -1 & -1 & 1 & -1 & 1 & -1 & 1 & 1 & -1 \\
\noalign{\smallskip}\hline
\end{tabular}
\end{center}
\end{table}

The price paid for the greater economy of run size offered by non-regular designs is more complex aliasing. Although designs formed from orthogonal arrays, including PB designs, allow estimation of each main effect independently of all other main effects, these estimators will usually be \textit{partially aliased} with many two-variable interactions. That is, the alias matrix $\mat{A}$ will contain many entries with $0<|a_{ij}|<1$. For example, consider the aliasing between main effects and two-variable interactions for the 12-run PB design in Table~\ref{tab:PB12}, as summarised in the $11\times 55$ alias matrix. The main effect of each variable is partially aliased with all 45 interactions that do not include that variable. That is, for the $i$th variable
$$
\Expec(\hat{\beta}_i) = \beta_i + \frac{1}{3}\sum_{j=1}^{11}\sum_{k>j}^{11}(-1)^{b_{ijk}}(1-\1_{i=j\cap i=k})\beta_{jk}\,,
$$
where $\1_A$ is the indicator function for the set $A$ and $b_{ijk}=0$ or $1$ ($i,j,k=1,\ldots,11$). For this design, each interaction is partially aliased with 9 main effects. The competing 16-run resolution III $2^{11-7}$ regular fraction has each main effect aliased with at most four two-variable interactions and each interaction aliased only with at most one main effect. Hence while an active interaction would bias only one main effect for the regular design, it would bias 9 main effects for the PB design, albeit to a lesser extent.

However, an important advantage of partial aliasing is that it allows interactions to be considered through the use of variable selection methods (discussed later) without requiring a large increase in the number of runs. For example, the 12-run PB design has been used to identify important interactions \cite{HamadaWu92,Chipmanetal97}.

A wide range of non-regular designs can be constructed. An algorithm has been developed for constructing designs which allow orthogonal estimation of all main effects together with catalogues of designs for $n=12,16,20$ \cite{SunLiYe2002} . Other authors have used computer search and criteria based on model selection properties to find non-regular designs \cite{Li2006}. A common approach is to use criteria derived from $D$-optimality \cite[][ch. 11]{ADT2007} to find fractional factorial designs for differing numbers of variables and runs \cite{DuMouchelJones94}. Designs from these methods may or may not allow independent estimation of the variable main effects dependent on the models under investigated and the number of runs available.   

Most screening experiments use designs at two-levels, possibly with the addition of one or more centre points to provide a portmanteau test for curvature. Recently, an economic class of three-level screening designs have been proposed, called ``Definitive Screening Designs'' (DSDs) \cite{JonesNachtsheim2011}, to investigate $d$ variables, generally in as few as $n=2d+1$ runs. The structure of the designs is illustrated in Table~\ref{tab:dsdex} for $d=6$. The design has a single centre point and $2d$ runs formed from $d$ mirrored pairs. The $j$th pair has the $j$th variable set to zero and the other $d-1$ variables set to $\pm 1$. The second run in the pair is formed by multiplying all the elements in the first run by -1. That is, the $2d$ runs form a \textit{foldover} design \cite{BoxWilson51}. This foldover property ensures that main effects and two-variable interactions are orthogonal and hence main effects are estimated independently from these interactions, unlike for resolution III or PB designs. Further, all quadratic effects are estimated independently of the main effects but not independently of the two-variable interactions. Finally, the two-variable interactions will be partially aliased with each other. These designs are growing in popularity, with a sizeable literature available on their construction \cite{NguyenStyllianou2012,XiaoLinBai2012,PhoaLin2015}.

\begin{table}
\begin{center}
\caption{The definitive screening design for $d=6$ variables.}
\label{tab:dsdex}       
\begin{tabular}{c@{\hskip 10pt}r@{\hskip 10pt}r@{\hskip 10pt}r@{\hskip 10pt}r@{\hskip 10pt}r@{\hskip 10pt}r@{\hskip 10pt}r}
\hline\noalign{\smallskip}
Run & $x_1$ & $x_2$ & $x_3$ & $x_4$ & $x_5$ & $x_6$ \\
\noalign{\smallskip}\hline\noalign{\smallskip}
1 & 0 & 1 & -1 & -1 & -1 & -1 \\ 
  2 & 0 & -1 & 1 & 1 & 1 & 1 \\ 
  3 & 1 & 0 & -1 & 1 & 1 & -1\\ 
  4 &  -1 & 0 & 1 & -1 & -1 & 1\\ 
  5 & -1 & -1 & 0 & 1 & -1 & -1 \\ 
  6 &  1 & 1 & 0 & -1 & 1 & 1\\ 
  7 &  -1 & 1 & 1 & 0 & 1 & -1 \\ 
  8 &  1 & -1 & -1 & 0 & -1 & 1 \\ 
  9 &  1 & -1 & 1 & -1 & 0 & -1 \\ 
  10 & -1 & 1 & -1 & 1 & 0 & 1 \\ 
  11 & 1 & 1 & 1 & 1 & -1 & 0 \\ 
  12 & -1 & -1 & -1 & -1 & 1 & 0 \\
  13 & 0 & 0 & 0 & 0 & 0 & 0 \\
\noalign{\smallskip}\hline
\end{tabular}
\end{center}
\end{table}

\subsection{Supersaturated designs for main effects screening}\label{SSD}

For experiments with a large number of variables or runs that are very expensive or time consuming, \textit{supersaturated} designs have been proposed as a low-resource (small $n$) solution to the screening problem \cite{Gilmour2006}. Originally, supersaturated designs were defined as having too few runs to estimate the intercept and the $d$ main effects in model~\eqref{eq:firstorder}, that is, $n<d+1$. The resulting partial aliasing is more complicated than for the designs discussed so far, in that at least one main effect estimator is biased by one or more other main effects. Consequently, there has been some controversy about the use of these designs \cite{Abrahametal99}. Recently, evidence has been provided for the effectiveness of the designs when factor sparsity holds and the active main effects are large \cite{MarleyWoods2010,Draguljicetal2014}. Applications of supersaturated designs include screening variables in numerical models for circuit design \cite{LiuFang2006}, extraterrestrial atmospheric science \cite{Claeys-Brunoetal2011} and simulation models for maritime terrorism \cite{Xingetal2013}. 

Supersaturated designs were first proposed in the discussion \cite{Box59} of random balance designs \cite{Satterthwaite59}. The first systematic construction method \cite{BoothCox62} found designs via computer search that have pairs of columns of the design matrix $\mat{X}^n$ as nearly orthogonal as possible through use of the $\Expec (s^2)$ design selection criterion (defined below). There was no further research in the area for 30 years until Lin \cite{Lin93} and Wu \cite{Wu93} independently revived interest in the construction of these designs. Both their methods are based on Hadamard matrices, and can be understood, respectively, as (i) selecting a half-fraction from a Hadamard matrix (Lin), and (ii) appending one or more interaction columns to a Hadamard matrix and assigning a new variable to each of these columns (Wu). 

Both methods can be illustrated using the $n=12$ run PB design in Table~\ref{tab:PB12}. To construct a supersaturated design for $d=10$ variables in $n=6$ runs by method (i), all 6 runs of the PB design with $x_{11}=-1$ are removed, followed by deletion of the $x_{11}$ column. The resulting design is shown in Figure~\ref{tab:LinSSD}. To obtain a design by method (ii) for $d=21$ variables in $n=12$ runs, 10 columns are appended that correspond to the interactions of $x_1$ with variables $x_2$ to $x_{11}$, and variables $x_{12}$ to $x_{21}$ are assigned to these columns, see Table~\ref{tab:WuSSD}. 

\begin{table}
\begin{center}
\caption{An $n=6$ run supersaturated design for $d=10$ variables obtained by the method of Lin \cite{Lin93}.}
\label{tab:LinSSD}       
\begin{tabular}{c@{\hskip 10pt}r@{\hskip 10pt}r@{\hskip 10pt}r@{\hskip 10pt}r@{\hskip 10pt}r@{\hskip 10pt}r@{\hskip 10pt}r@{\hskip 10pt}r@{\hskip 10pt}r@{\hskip 10pt}r}
\hline\noalign{\smallskip}
Run & $x_1$ & $x_2$ & $x_3$ & $x_4$ & $x_5$ & $x_6$ & $x_7$ & $x_8$ & $x_9$ & $x_{10}$ \\
\noalign{\smallskip}\hline\noalign{\smallskip}
  1 & -1 & -1 & -1 & -1 & -1 & 1 & 1 & 1 & 1 & 1 \\ 
  2 & -1 & -1 & 1 & 1 & 1 & -1 & -1 & -1 & 1 & 1 \\ 
  3 & -1 & 1 & -1 & 1 & 1 & -1 & 1 & 1 & -1 & -1 \\ 
  4 & 1 & -1 & 1 & -1 & 1 & 1 & 1 & -1 & -1 & -1 \\ 
  5 & 1 & 1 & 1 & -1 & -1 & -1 & -1 & 1 & 1 & -1 \\ 
  6 & 1 & 1 & -1 & 1 & -1 & 1 & -1 & -1 & -1 & 1 \\ 
\noalign{\smallskip}\hline
\end{tabular}
\end{center}
\end{table}

\begin{table}
\begin{center}
\caption{An $n=12$ run supersaturated design for $d=21$ obtained by the method of Wu \cite{Wu93}.}
\label{tab:WuSSD}       
\begin{tabular}{c@{\hskip 7pt}r@{\hskip 7pt}r@{\hskip 7pt}r@{\hskip 7pt}r@{\hskip 7pt}r@{\hskip 7pt}r@{\hskip 7pt}r@{\hskip 7pt}r@{\hskip 7pt}r@{\hskip 7pt}r@{\hskip 7pt}r@{\hskip 7pt}r@{\hskip 7pt}r@{\hskip 7pt}r@{\hskip 7pt}r@{\hskip 7pt}r@{\hskip 7pt}r@{\hskip 7pt}r@{\hskip 7pt}r@{\hskip 7pt}r@{\hskip 7pt}r}
\hline\noalign{\smallskip}
Run & $x_1$ & $x_2$ & $x_3$ & $x_4$ & $x_5$ & $x_6$ & $x_7$ & $x_8$ & $x_9$ & $x_{10}$ & $x_{11}$ & $x_{12}$ & $x_{13}$ & $x_{14}$ & $x_{15}$ & $x_{16}$ & $x_{17}$ & $x_{18}$ & $x_{19}$ & $x_{20}$ & $x_{21}$   \\
\noalign{\smallskip}\hline\noalign{\smallskip}
1 & -1 & -1 & -1 & -1 & -1 & -1 & -1 & -1 & -1 & -1 & -1 & 1 & 1 & 1 & 1 & 1 & 1 & 1 & 1 & 1 & 1 \\ 
  2 & -1 & -1 & -1 & -1 & -1 & 1 & 1 & 1 & 1 & 1 & 1 & 1 & 1 & 1 & 1 & -1 & -1 & -1 & -1 & -1 & -1 \\ 
  3 & -1 & -1 & 1 & 1 & 1 & -1 & -1 & -1 & 1 & 1 & 1 & 1 & -1 & -1 & -1 & 1 & 1 & 1 & -1 & -1 & -1 \\ 
  4 & -1 & 1 & -1 & 1 & 1 & -1 & 1 & 1 & -1 & -1 & 1 & -1 & 1 & -1 & -1 & 1 & -1 & -1 & 1 & 1 & -1 \\ 
  5 & -1 & 1 & 1 & -1 & 1 & 1 & -1 & 1 & -1 & 1 & -1 & -1 & -1 & 1 & -1 & -1 & 1 & -1 & 1 & -1 & 1 \\ 
  6 & -1 & 1 & 1 & 1 & -1 & 1 & 1 & -1 & 1 & -1 & -1 & -1 & -1 & -1 & 1 & -1 & -1 & 1 & -1 & 1 & 1 \\ 
  7 & 1 & -1 & 1 & 1 & -1 & -1 & 1 & 1 & -1 & 1 & -1 & -1 & 1 & 1 & -1 & -1 & 1 & 1 & -1 & 1 & -1 \\ 
  8 & 1 & -1 & 1 & -1 & 1 & 1 & 1 & -1 & -1 & -1 & 1 & -1 & 1 & -1 & 1 & 1 & 1 & -1 & -1 & -1 & 1 \\ 
  9 & 1 & -1 & -1 & 1 & 1 & 1 & -1 & 1 & 1 & -1 & -1 & -1 & -1 & 1 & 1 & 1 & -1 & 1 & 1 & -1 & -1 \\ 
  10 & 1 & 1 & 1 & -1 & -1 & -1 & -1 & 1 & 1 & -1 & 1 & 1 & 1 & -1 & -1 & -1 & -1 & 1 & 1 & -1 & 1 \\ 
  11 & 1 & 1 & -1 & 1 & -1 & 1 & -1 & -1 & -1 & 1 & 1 & 1 & -1 & 1 & -1 & 1 & -1 & -1 & -1 & 1 & 1 \\ 
  12 & 1 & 1 & -1 & -1 & 1 & -1 & 1 & -1 & 1 & 1 & -1 & 1 & -1 & -1 & 1 & -1 & 1 & -1 & 1 & 1 & -1 \\ 
\noalign{\smallskip}\hline
\end{tabular}
\end{center}
\end{table}

Since 1993, there has been a substantial research effort on construction methods for supersaturated designs, see for example \cite{Lin95,Nguyen96,NguyenCheng2008}. The most commonly used criterion for design selection in the literature is $\Expec (s^2)$-optimality \cite{BoothCox62}. More recently, the Bayesian $D$-optimality criterion \cite{DuMouchelJones94,JonesLinNachtsheim2008} has become popular.

\paragraph{$\Expec (s^2)$-optimality} This criterion selects a \textit{balanced} design, that is a design with $n=2m$ for some integer $m>0$ where each column of $\mat{X}^n$ contains $m$ entries equal to -1 and $m$ entries equal to +1. The $\Expec (s^2)$-optimal design minimises the average of the squared inner-products between columns $i$ and $j$ of $\mat{X}^n$ $(i,j=1,\ldots,d;\,i\ne j)$, 
\begin{equation}\label{eq:es2}
\Expec (s^2)=\frac{2}{d(d-1)} \sum_{i<j} s_{ij}^2\,,
\end{equation}
where $s_{ij}$ is the $ij$th element of $\left(\mat{X}^n\right)^\T\mat{X}^n$ $(i,j=1, \ldots, d)$. A lower bound on $\Expec (s^2)$ is available \cite{BulutogluCheng2004,RyanBulutoglu2007}. The designs in Tables~\ref{tab:LinSSD} and~\ref{tab:WuSSD} achieve the lower bound and hence are $\Expec (s^2)$-optimal. For the design in Table~\ref{tab:LinSSD}, each $s_{ij}^2 = 4$. For the design in Table~\ref{tab:WuSSD}, $\Expec (s^2) = 6.85714$, with 120 pairs of columns being orthogonal ($s_{ij}^2 = 0$) and the remaining 90 pairs of columns having $s_{ij}^2=16$. Recently, the definition of $\Expec (s^2)$ has been extended to unbalanced designs \cite{MarleyWoods2010,JonesMajumdar2014} by including the inner-product between each column of $\mat{X}^n$ and the vector $\vect{1}_n$, the $n\times 1$ vector with every entry 1, which corresponds to the intercept term in model~\eqref{eq:lm}. This extension widens the class of available designs. 

\paragraph{Bayesian $D$-optimality} This criterion selects a design that maximises the determinant of the posterior variance-covariance matrix for $(\beta_0,\vectg{\beta}^\T)^\T$,
\begin{equation}\label{eq:BayesD}
\Psi_{D}=\left|(\mat{H}^\star)^\T\mat{H}^\star+\mat{K}/\tau^2\right|^{1/(d+1)}\,,
\end{equation}
where $\mat{H}^\star = [\vect{1}_n|\vect{X}^n]$, $\mat{K} = \mat{I}_{d+1} - \vect{e}_{1(d+1)}\vect{e}_{1(d+1)}^\T$, $\tau^2>0$ and $\tau^2\mat{K}^{-1}$ is the prior variance-covariance matrix for $\vectg{\beta}$. Equation~\eqref{eq:BayesD} results from assuming an informative prior distribution for each $\beta_i$ ($i=1,\ldots,d$) with mean zero and small prior variance, to reflect factor sparsity, and a non-informative prior distribution for $\beta_0$. The prior information can be regarded as equivalent to having sufficient additional runs to allow estimation of all parameters $\beta_0,\ldots,\beta_d$, with the value of $\tau^2$ reflecting the quantity of available prior information. However, the optimal designs obtained tend to be insensitive to the choice of $\tau^2$ \cite{MarleyWoods2010}.

Both $\Expec (s^2)$- and $D$-optimal designs can be found numerically, using algorithms such as columnwise-pairwise \cite{LiWu97} or coordinate exchange \cite{MeyerNachtsheim95}. From simulation studies, it has been shown that there is little difference in the performance of $\Expec (s^2)$- and Bayesian $D$-optimal designs assessed by, for example, sensitivity and type I error rate \cite{MarleyWoods2010}.

Supersaturated designs have also been constructed that allow the detection of two-variable interactions \cite{LiuDean2004}. Here the definition of supersaturated has been widened to include designs that have fewer runs than the total number of factorial effects to be investigated. In particular, Bayesian $D$-optimal designs have been shown to be effective in identifying active interactions \cite{Draguljicetal2014}. Note that under this expanded definition of supersaturated designs, all fractional factorial designs are supersaturated under model~\eqref{eq:lm} when $p<n$. 

\subsection{Common issues with factorial screening designs}\label{factorialissues}

The analysis of unreplicated factorial designs commonly used for screening experiments has been a topic of much research \cite{lenth89,HamadaBalakrishnan98,VossWang2006}. In a physical experiment, the lack of replication to provide a model-free estimate of $\sigma^2$ can make it difficult to assess the importance of individual factorial effects. The most commonly applied method for orthogonal designs treats this problem as analogous to the identification of outliers, and makes use of (half-) normal plots of the factorial effects. For many non-regular and supersaturated designs, more advanced analysis methods are necessary; see later. For studies on numerical models, provided all the input variables are controlled, the problem of assessing statistical significance does not occur as no unusually large observations can have arisen due to ``chance''. Here, factorial effects can be ranked by size and those variables whose effects lead to a substantive change in the response declared active.

Biased estimators of factorial effects, however, are an issue for experiments on both numerical models and physical processes. Complex (partial) aliasing can produce two types of bias in the estimated parameters in model~\eqref{eq:lm}: upwards bias so that a type I error may occur (amalgamation), or downward bias leading to missing active variables (cancellation). Simulation studies have been used to assess these risks \cite{DeanLewis2002,MarleyWoods2010,Draguljicetal2014}. 

Bias may also, of course, be induced by assuming a form of the surrogate model that is too simple, for example, through the surrogate having too few turning points (e.g. being a polynomial of too low order) or lacking the detail to explain the local behaviour of the numerical model. This kind of bias is potentially the primary source of mistakes in screening variables in numerical models. When prior scientific knowledge suggests that the numerical model is highly nonlinear, screening methods should be employed that have fewer restrictions on the surrogate model or are model-free. Such methods, including designs for the estimation of elementary effects~\eqref{eq:elem}, are described later in this paper. Typically, they require larger experiments than the designs in the present section.

\subsection{Systematic fractional replicate designs}\label{cotter}

Systematic fractional replicate designs \cite{Cotter79} enable expressions to be estimated that indicate the influence of each variable on the output, through main effects and interactions, without assumptions in model~\eqref{eq:lm} on the order of interactions that may be important. These designs have had considerable use for screening inputs to numerical models, especially in the medical and biological sciences \cite{ScottDrechseletal2012,Yangetal2014}. In these designs, each variable takes two levels and there are $n=2d+2$ runs. 

The designs are simple to construct as (i) one run with all variables set to -1; (ii) $d$ runs with each variable in turn set to +1 and the other variables set to -1; (iii) $d$ runs with each variable in turn set to -1 and the other variables set to +1; and (iv) one run with all variables set to +1. Let the elements of vector $\vect{Y}^n$ be such that $Y^{(1)}$ is the output from the run in (i), $Y^{(2)},\ldots,Y^{(d+1)}$ are the outputs from the runs in (ii),  $Y^{(d+2)},\ldots,Y^{(2d+1)}$ are from the runs in (iii), and $Y^{2d+2}$ is from the run in (iv). In such a design, each main effect can be estimated independently of all two-variable interactions. This can easily be seen from the alternative construction as a foldover from a \textit{one-factor-at-a-time} (OFAAT) design with $n=d+1$. That is, a design having one run with each variable set to -1, and $d$ runs with each variable in turn set to +1 with all other variables set to -1. 

For each variable $x_i$ ($i=1,\ldots,d$) two linear combinations, $S_o(i)$ and $S_e(i)$, of ``odd order'' and ``even order'' model parameters, respectively, can be estimated:
\begin{equation}\label{So}
S_o(i) = \beta_i + \underset{i\ne j\ne k}{\sum_{j=1}^d\sum_{k=1}^d}\beta_{ijk} + \ldots\,,
\end{equation}
and
\begin{equation}\label{Se}
S_e(i) = \underset{i\ne j}{\sum_{j=1}^d}\beta_{ij} +  \underset{i\ne j\ne k\ne l}{\sum_{j=1}^d\sum_{k=1}^d\sum_{l=1}^d}\beta_{ijkl} +\ldots\,,
\end{equation}
with respective unbiased estimators
\begin{equation*}\label{Co}
C_0(i) = \frac{1}{4}\left\{\left(Y^{(2d+2)}-Y^{(d+i+1)}\right) + \left(Y^{(i+1)}-Y^{(1)}\right)\right\}\,,
\end{equation*} 
and
\begin{equation*}\label{Ce}
C_e(i) = \frac{1}{4}\left\{\left(Y^{(2d+2)}-Y^{(d+i+1)}\right) - \left(Y^{(i+1)}-Y^{(1)}\right)\right\}\,.
\end{equation*} 
Under effect hierarchy, it may be anticipated that a large absolute value of $C_o(i)$ is due to a large main effect for the $i$th variable, and a large absolute value of $C_e(i)$ is due to large two-variable interactions. A design that also enables estimation of two-variable interactions independently of each other is obtained by appending $(d-1)(d-2)/2$ runs, each having two variables set to +1 and $d-2$ variables set to -1 \cite{QuWu2005}.

For numerical models, where observations are not subject to random error, active variables can be selected by ranking the sensitivity indices defined by 
\begin{equation}\label{SI}
S(i) = \frac{M(i)}{\sum_{j=1}^dM(j)}\,,\qquad i=1,\ldots,d\,,
\end{equation}
where $M(i)=|C_o(i)|+|C_e(i)|$. This methodology is potentially sensitive to the cancellation or amalgamation of factorial effects, discussed in the previous section.  
 
From~\eqref{eq:elem}, it can also be seen that use of a systematic fractional replicate design is equivalent to calculating two elementary effects (with $\Delta = 2$) for each variable at the extremes of the design region. Let $\mbox{EE}_{1i} = \left(Y^{(2d+2)}-Y^{(d+i+1)}\right)/2$ and $\mbox{EE}_{2i} = \left(Y^{(i+1)}-Y^{(1)}\right)/2$ be these elementary effects for the $i$th variable. Then it follows directly that $S(i) \propto \max \left( |EE_{1i}|, |EE_{2i}|\right)$, and the above method selects as active those variables with elementary effects that are large in absolute value.
 
\section{Screening groups of variables}\label{groupscreening}
	 
Early work on group screening used pooled blood samples to detect individuals with a disease as economically as possible \cite{Dorfman43}. The technique was extended, almost 20 years later, to screening large numbers of two-level variables in factorial experiments where a main effects only model is assumed for the output \cite{Watson61}. For an overview of this work and several other strategies, see \cite{Morris2006}.

In group screening, the set of variables is partitioned into groups and the values of the variables within each group are varied together. Smaller designs can then be used to experiment on these groups. This strategy deliberately aliases the main effects of the individual variables. Hence, follow-up experimentation is needed on those variables in the groups found to be important in order to detect the individual active variables. The main screening techniques that employ grouping of variables are described below. 
	 
\subsection{Factorial group screening}\label{facGS}
The majority of factorial group screening methods apply to variables with two levels and use two stages of experimentation. At the first stage, the $d$ variables are partitioned into $g$ groups, where the $j$th group contains $g_j\ge 1$ variables $(j=1,\ldots,g)$. High and low levels for each of the $g$ grouped variables are defined by setting all the individual variables in a group to either their high level or to their low level simultaneously. The first experiment with $n_1$ runs, is performed on the relatively small number of grouped variables. \textit{Classical group screening} then estimates the main effects for each of the grouped variables, and takes those variables involved in groups that have large estimated main effects through to a second stage experiment. Individual variables are investigated at this stage and their main effects, and possibly interactions, are estimated.

For sufficiently large groups of variables, highly resource-efficient designs can be employed at stage 1 of classical group screening for even very large numbers of factors. Under the assumption of negligible interactions, orthogonal non-regular designs, such as PB designs, can be used. For screening variables from a deterministic numerical model, designs in which the columns corresponding to the grouped main effects are not orthogonal can be effective \cite{Bettonvil95} provided $n_1>g+1$, as the precision of factorial effect estimators is not a concern.   

Effective classical group screening depends on strong effect heredity, namely, that important two-variable interactions occur only between variables both having important main effects. More recently, strategies for group screening that also investigate interactions at stage 1 have been developed \cite{LewisDean2001}. In \textit{interaction group screening} both main effects and two-variable interactions between the grouped variables are estimated at stage 1. The interaction between two grouped variables is the summation of the interactions between all pairs of variables where one variable comes from each group; interactions between two variables in the same group are aliased with the mean. Variables in groups found to have large estimated main effects or to be involved in large interactions are carried forward to the second stage. From the second stage experiment, main effects and interactions are examined between the individual variables within each group declared active. Where the first stage has identified a large interaction between two grouped variables,  the interactions between pairs of individual variables, one from each group, are also investigated. For this strategy, larger resolution V designs, capable of independently estimating all grouped main effects and two-variable interactions, have so far been used at stage 1, when decisions to drop groups of variables are made.

Group screening experiments can be viewed as supersaturated experiments in the individual variables. However,  when orthogonal designs are used for the stage 1 experiment, decisions on which groups of variables to take forward can be made using $t$-tests on the grouped main effects and interactions. When smaller designs are used, particularly if $n_1$ is less than the number of grouped effects of interest, more advanced modelling methods are required, in common with other supersaturated designs (see later). Incorrectly discarding active variables at stage 1 may result in missed opportunities to improve process control or product quality. Hence it is common to be conservative in the choice of design at stage 1, for example, in the number of runs, and also to allow a higher type I error rate.   

In the two-stage process, the design for the second experiment cannot be decided until the stage 1 data have been collected and the groups of factors deemed active have been identified. In fact, the size, $N_2$, of the second stage experiment required by the group screening strategy is a random variable. The distribution of $N_2$ is determined by features under the experimenter's control, such as $d$, $g$, $g_1,\ldots,g_g$, $n_1$, the first stage design and decision rules for declaring a grouped variable active at stage 1. It also depends on features outside the experimenter's control, such as the number of active individual variables and the size and nature of their effects, and the signal-to-noise ratio if the process is noisy. Given prior knowledge of these uncontrollable features, the grouping strategy, designs and analysis methods can be tailored, for example, to produce a smaller expected experiment size, $n_1+\Expec(N_2)$, or to minimise the probability of missing active variables \cite{LewisDean2001,VineLewisDean2005}. Of course, these two goals are usually in conflict and hence a trade-off has to be made.  In practice, the design used at stage 2 depends on the number of variables brought forward and the particular effects requiring estimation; options include regular or non-regular fractional factorial designs and $D$-optimal designs. 

Original descriptions of classical group screening made the assumption that all the active variable main effects have the same sign to avoid the possibility of cancellation of the main effects of two or more active variables in the same group. As discussed previously, cancellation can affect any fractional factorial experiment. Group screening is often viewed as particularly susceptible due to the complete aliasing of main effects of individual variables and the screening out of whole groups of variables at stage 1. Often, particularly for numerical models, prior knowledge makes reasonable the assumption of active main effects having the same sign. Otherwise, the risks of missing active variables should be assessed by simulation \cite{MauroSmith82} and, in fact, the risk can be modest under factor sparsity \cite{Draguljicetal2014}.

\subsection{Sequential bifurcation}
Screening groups of variables is also used in sequential bifurcation, proposed originally for deterministic simulation experiments \cite{BettonvilKleijnen96}. The technique can investigate a very large number of variables, each having two levels, when a sufficiently accurate surrogate for the output is a first-order model~\eqref{eq:firstorder}. It is assumed that each parameter $\beta_i$ ($i=1,\ldots,d$) is positive (or can be made positive by interchanging the variable levels) to avoid cancellation of effects.

The procedure starts with a single group composed of all the variables which is split into two new groups (bifurcation). For a deterministic numerical model, the initial experiment has just two runs: all variables set to the low levels $(\vect{x}^{(1)})$ and all variables set to the high levels $(\vect{x}^{(2)})$. If the output $Y^{(2)}>Y^{(1)}$, then the group is split, with variables $x_1,\ldots,x_{d_1}$ placed in group 1 and $x_{d_1+1},\ldots,x_{d}$ placed in group 2. At the next stage, a single further run $\vect{x}^{(3)}$ is made which has all group 1 variables set to their high levels and all group 2 variables set low. If $Y^{(3)}>Y^{(1)}$, then group 1 is split further, and group 2 is split if $Y^{(2)}>Y^{(3)}$. These comparisons can be replaced by $Y^{(3)} - Y^{(1)} > \delta$ and $Y^{(2)} - Y^{(3)} > \delta$, where $\delta$ is an elicited threshold. This procedure of performing one new run and assessing the split of each subsequent group continues until singleton groups, containing variables deemed to be active, have been identified. Finally, these individual variables are investigated. If the output variable is stochastic, the replications of each run are made and a two-sample $t$-tests can be used to decide whether or not to split a group. 

Typically, if $d=2^k$ for some integer $k>0$, then at each split, half the variables are assigned to one of the groups, and the other half are assigned to the second group. Otherwise, use of unequal group sizes can increase the efficiency (in terms of overall experiment size) of sequential bifurcation when there is prior knowledge of effect sizes. Then, at each split, the first new group should have size equal to the largest possible power of $2$. For example, if the group to be split contains $m$ variables, then the first new group should contain $2^l$ variables such that $2^l<m$. The remaining $m-2^l$ variables are assigned to the second group. If variables have been ordered by an a priori assessment of increasing importance, the most important variables will be in the second, smaller group and hence more variables can be ruled out as unimportant more quickly.

The importance of two-variable interactions may be investigated by using the output from the following runs to assess each split. The first is run $\vect{x}$ used in the standard sequential bifurcation method; the second is the mirror image of $\vect{x}$ in which each variable is set low that is set high in $\vect{x}$, and vice versa. This foldover structure ensures that any two-variable interactions will not bias estimators of grouped main effects at each stage. This alternative design also permits standard sequential bifurcation to be performed and, if the variables deemed active differ from those found via the foldover, then the presence of active interactions is indicated. Again, successful identification of the active variables relies on the principle of strong effect heredity.   

A variety of adaptations of sequential bifurcation have been proposed, including methods of controlling type I and type II error rates \cite{Wanetal2006,Wanetal2010} and a procedure to identify dispersion effects in robust parameter design \cite{AnkenmanChengLewis2014}. For further details, see \cite[][ch. 4]{Kleijnen2015}.

\subsection{Iterated fractional factorial designs}
These designs \cite{AndresHajas93} also group variables in a sequence of applications of the same fractional factorial design. Unlike factorial group screening and sequential bifurcation, the variables are assigned at random to the groups at each stage. Individual variables are identified as active if they are in the intersection of those groups having important main effects at each stage.

Suppose there are $g=2^l$ groups, for integer $l>0$. The initial design has $2g$ runs obtained as a foldover of a $g\times g$ Hadamard matrix; for details see \cite{Campolongoetal2000}. This construction gives a design in which main effects are not aliased with two-variable interactions (a Resolution IV design). The $d\ge g$ variables are assigned at random to the groups, and each grouped variable is then assigned at random to a column of the design. The experiment is performed and analysed as a stage 1 group screening design. Subsequent stages repeat this procedure, using the same design but with different, independent assignments of variables to groups and groups to columns. Individual variables which are common to groups of variables found to be active across several stages of experimentation are deemed to be active. Estimates of the main effects using data from all the stages can also be constructed.

There are two other differences from the other grouping methods discussed in this section. First, for a proportion of stages, the variables are set to a mid-level value (0), rather than high (+1) or low (-1). These runs allow an estimate of curvature to be made, and some screening of quadratic effects to be undertaken. Second, to mitigate cancellation of main effects, the coding of the high and low levels may be swapped at random, that is, the sign of the main effect reversed.

The use of iterated fractional factorial designs requires a larger total number of runs than other group screening methods, as a sequence of factorial screening designs is implemented. However, the method has been suggested for use when there are many variables (thousands) arranged in a few large groups. Simulation studies \cite{SaltelliAndresHomma93,SaltelliAndresHomma95} have shown it can be effective here, provided there are very few active variables.

\subsection{Two-stage group screening for Gaussian process models}
More recently, methodology for group screening in two stages of experimentation using Gaussian process modelling to identify the active variables has been developed for numerical models \cite{MoonDeanSantner2012}. At the first stage, an initial experiment that employs an orthogonal space-filling design (see the next section) is used to identify variables to be grouped together. Examples are variables that are inert or those having a similar effect on the output, such as having a common sign and a similarly sized linear or quadratic effect. A sensitivity analysis on the grouped variables is then performed using a Gaussian process model, built from the first-stage data. Groups of variables identified as active in this analysis are investigated in a second-stage experiment in which the variables found to be unimportant are kept constant. The second-stage data are then combined with the first-stage data and a further sensitivity analysis performed to make a final selection of the active variables. An important advantage of this method is the reduced computational cost of performing a sensitivity study on the grouped variables at the first stage.
	 
\section{Random sampling plans and space-filling}\label{random sampling}

\subsection{Latin hypercube sampling}
The most common experimental design used to study deterministic numerical models is the Latin Hypercube Sample (LHS) \cite{McKayBeckmanConover79}. These designs address the difficult problem of space-filling in high dimensions, that is, when there are many controllable variables. Even when adequate space-filling in $d$ dimensions with $n$ points may be impossible, a LHS design offers $n$ points that have good one-dimensional space-filling properties for a chosen distribution, usually a uniform distribution. Thus use of a LHS at least implicitly invokes the principle of factor sparsity, and hence is potentially suited for use in screening experiments.

\begin{figure}
\centering
\begin{tabular}{ccc}
(a) & (b) & (c) \\[-3ex]
\includegraphics[scale=0.25]{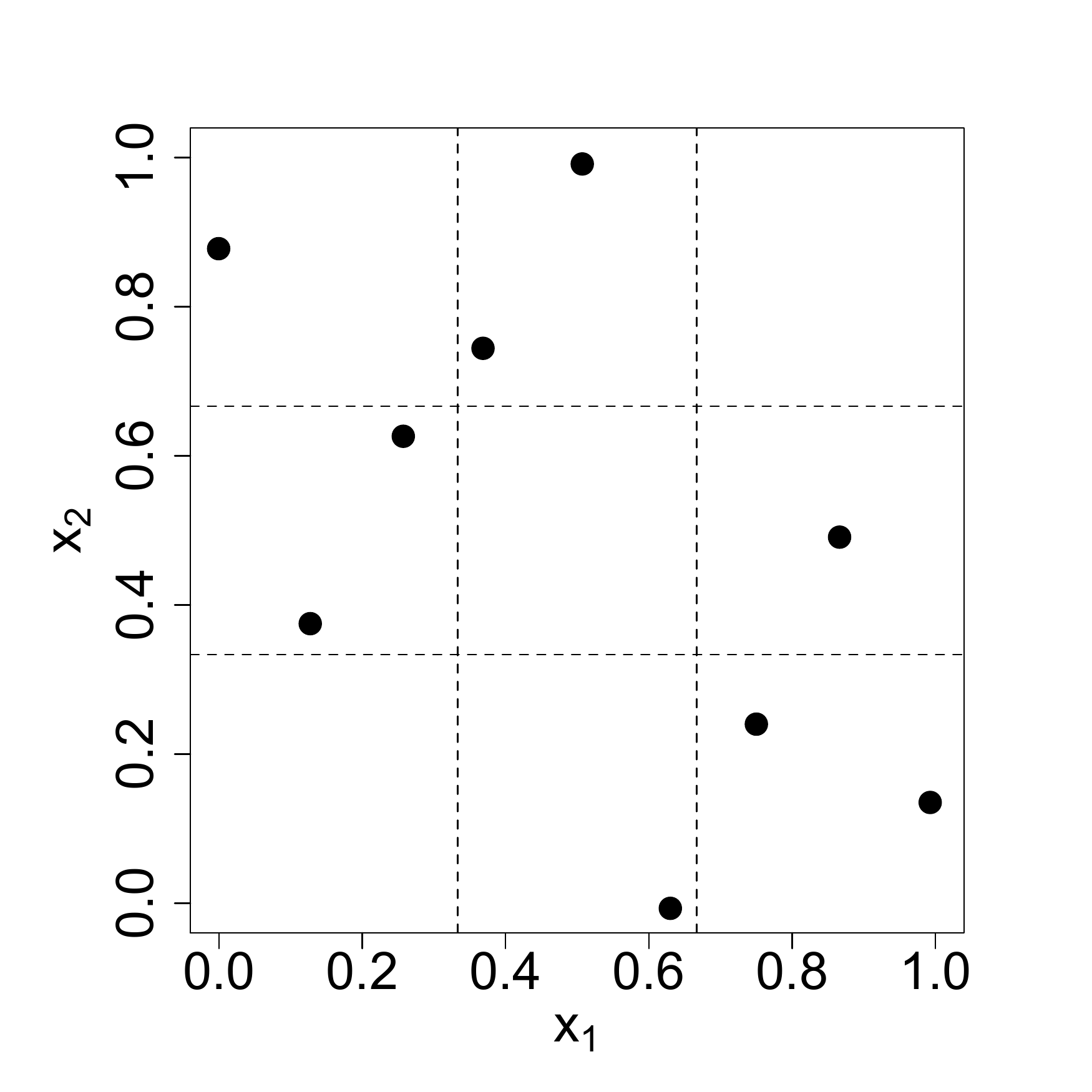} & \includegraphics[scale=0.25]{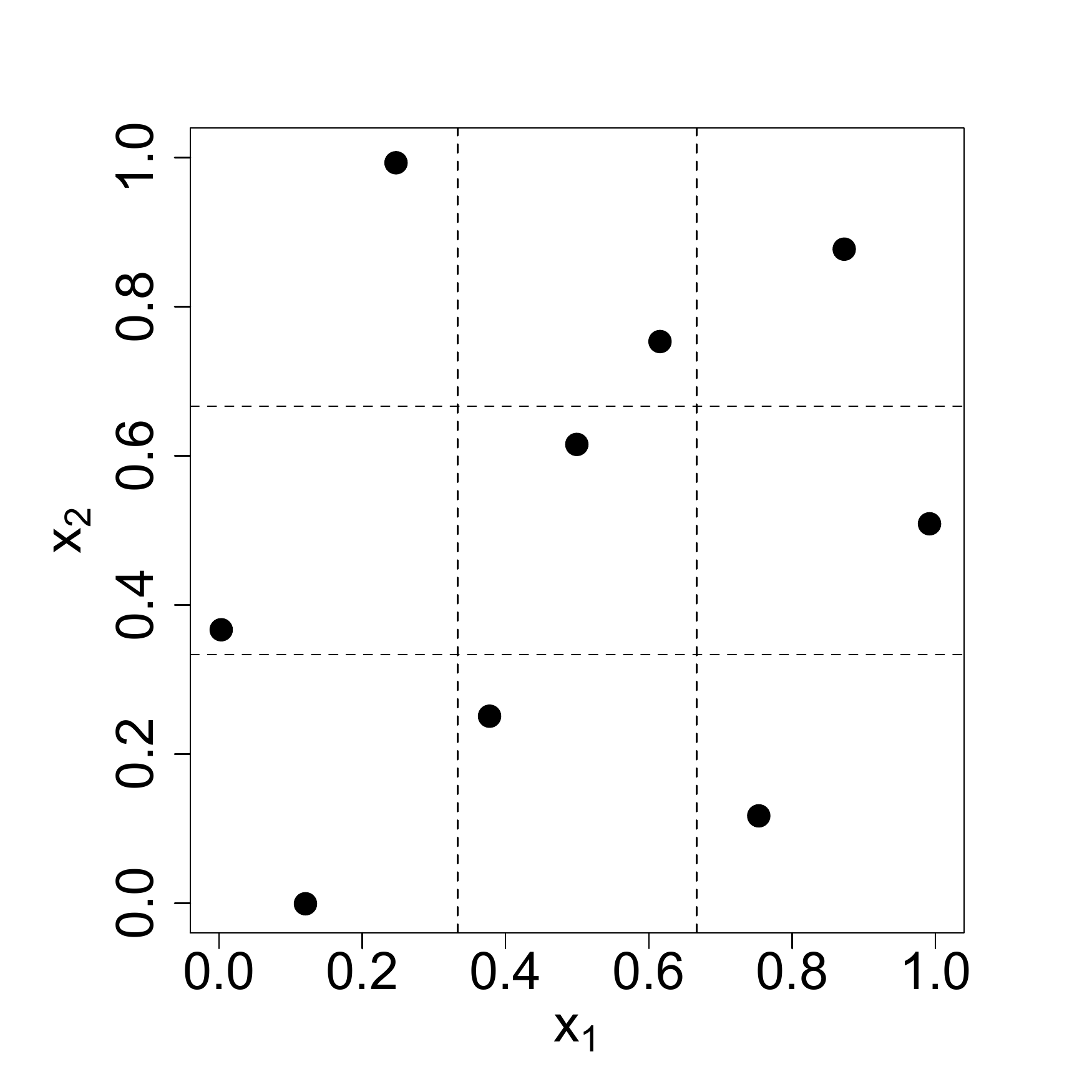} & \includegraphics[scale=0.25]{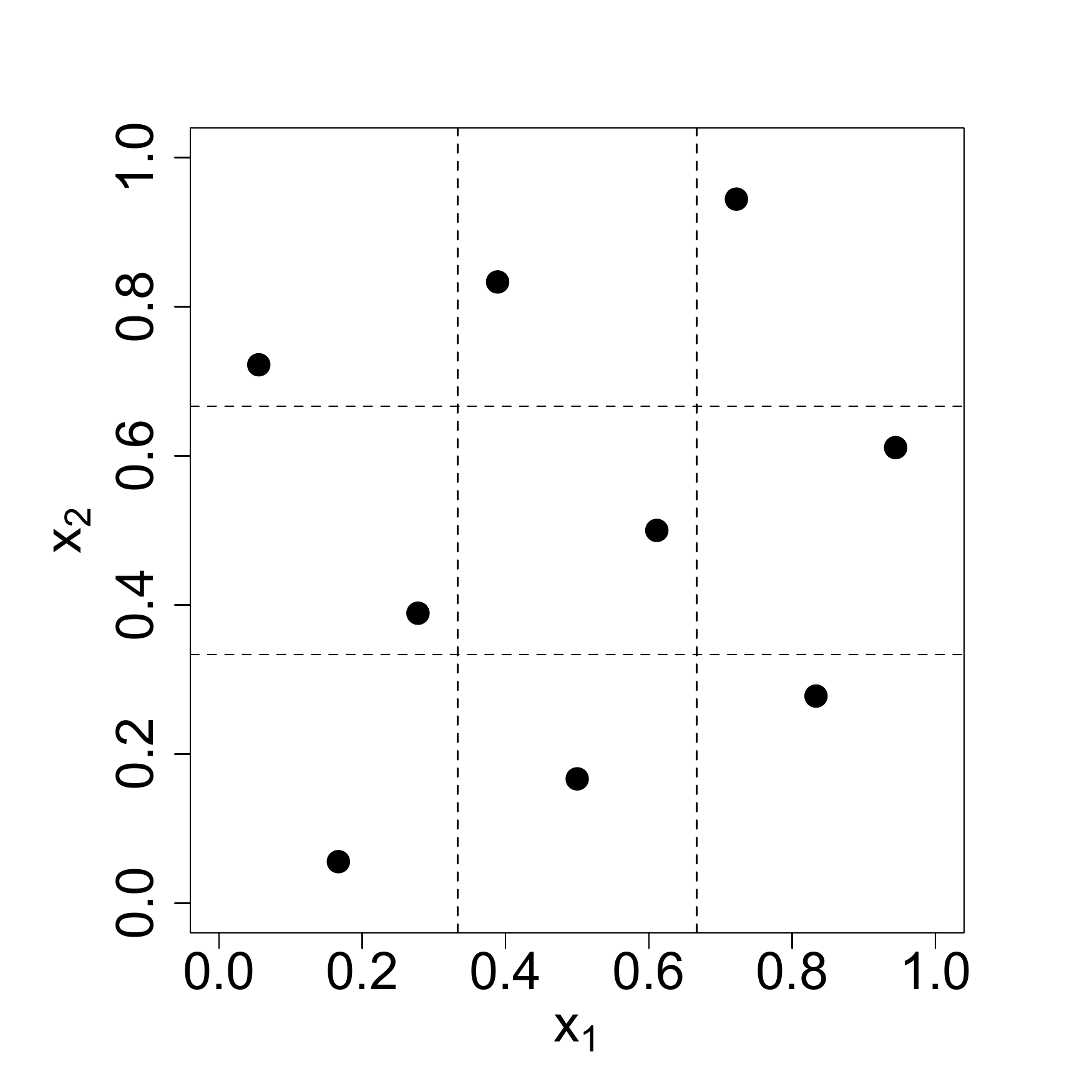}
\end{tabular}
\caption{Latin hypercube samples with $n=9$ and $d=2$: (a) random LHS; (b) random LHS generated from an orthogonal array; (c) maximin LHS.}
\label{fig:lhs}
\end{figure}

A $d$-dimensional LHS is a generalisation of a Latin square, which is a two-dimensional grid with at most one point in each row and each column. Construction of a standard LHS is straightforward: generate $d$ random permutations of the integers $1,\ldots,n$ and arrange them as an $n\times d$ matrix $\mat{D}$ (one permutation forming each column); transform each element of $\mat{D}$ to obtain to a sample 
from a given distribution $F(\cdot)$, that is, define the coordinates of the design points as $x^{(i)}_j = F^{-1}\left\{(d_{j}^{(i)}-1)/(n-1)\right\}$, where $d^{(i)}_j$ is the $ij$th element of $\mat{D}$ ($i=1,\ldots,n;\,j=1,\ldots,d$). Typically, a (small) random perturbation is added to each $d_j^{(i)}$, or some equivalent operation performed, prior to transformation to $x_j^{(i)}$. An LHS design generated by this method is shown in Figure~\ref{fig:lhs}(a). 

There may be great variation in the overall space-filling properties of LHS designs. For example, the LHS design in Figure~\ref{fig:lhs}(a) clearly has poor two-dimensional space-filling properties. Hence, a variety of extensions to Latin hypercube sampling have been proposed. Most prevalent are orthogonal array-based and maximin Latin hypercube sampling. 

To generate an orthogonal array-based LHS \cite{Owen92,Tang93}, the matrix $\mat{D}$ is formed from an orthogonal array. Hence the columns of $\mat{D}$ are no longer independent permutations of $1,\ldots,d$. For simplicity, assume $\mat{O}$ is a symmetric OA($n$, $s^d$, $t$) with symbols $1,\ldots,s$ and $t\ge 2$. The $j$th row of $\mat{D}$ is formed by mapping the $n/s$ occurrences of each symbol in the $j$th column of $\mat{O}$ to a random permutation, $\alpha$, of $n/s$ new symbols, i.e. $1\rightarrow \alpha(1,\ldots,n/s)$, $2\rightarrow \alpha(n/s+1,\ldots,2n/s)$, \ldots, $s\rightarrow \alpha((s-1)n/s+1,\ldots,n)$, where $\alpha(1,\ldots,a)$ is a permutation of the integers $1,\ldots,a$. Figure~\ref{fig:lhs}(b) shows an orthogonal array-based LHS, constructed from an OA($9$, $3^2$, $2$). Notice the improved two-dimensional space-filling compared with the randomly generated LHS. The two-dimensional projection properties of more general space-filling designs have also been considered by other authors \cite{Damblinetal2013}, especially for \textit{uniform} designs minimising specific $L^2$-discrepancies \cite{Fangetal2000}.  

In addition to the orthogonal-array-based LHS, there has been a variety of work on generating space-filling designs that directly minimise the correlation between columns of $\mat{X}^n$ \cite{ImanConover82}, including algorithmic \cite{Tang98} and analytic \cite{Ye98} construction methods. Such designs have good two-dimensional space-filling properties and also provide near-independent estimators of the $\beta_i$ ($i=1,\ldots,d$) in equation~\eqref{eq:firstorder}, a desirable property for screening.

A maximin LHS \cite{MorrisMitchell95} achieves a wide spread of design points across the design region by maximising the minimum distance between pairs of design points within the class of LHS designs with $n$ points and $d$ variables. The Euclidean distance between two points $\vect{x}=(x_1,\ldots,x_d)^\T$ and $\vect{x}^\prime=(x_1^\prime,\ldots,x_d^\prime)^\T$ is given by
\begin{equation}\label{eq:edist}
\mbox{dist}(\vect{x},\vect{x}^\prime) = \left\{\sum_{j=1}^d(x_j - x^\prime_j)^2\right\}^{\frac{1}{2}}\,.
\end{equation} 
Rather than maximise directly the minimum of~\eqref{eq:edist}, most authors \cite{MorrisMitchell95,Baetal2015} find designs by minimisation of
\begin{equation}\label{eq:maximin}
\phi(\mat{X}^n) = \left\{\sum_{1\le i < j \le n}\left[\mbox{dist}(\vect{x}_i^n,\vect{x}_j^n)\right]^{-q}\right\}^{1/q}\,,
\end{equation}
where for $q\rightarrow\infty$, minimisation of~\eqref{eq:maximin} is equivalent to maximising the minimum of~\eqref{eq:edist} across all pairs of design points; see also \cite{PronzatoMuller2012}. Figure~\ref{fig:lhs}(c) shows a maximin LHS, found by this method with $q=15$. This design displays better two-dimensional space-filling than the random LHS, and a more even distribution of the design points than the orthogonal array-based LHS. 

Maximin LHS can be found using the \texttt{R} packages \texttt{DiceDesign} \cite{Francoetal2014} and \texttt{SLHD} \cite{Ba2015}. More general classes of distance-based space-filling designs, without the projection properties of the Latin hypercubes, can also be found \cite[see][]{JohnsonMooreYlvisaker90,BowmanWoods2013}. Studies of the numerical efficiencies of optimization algorithms for LHS designs are available \cite{Jinetal2005,Damblinetal2013}.

Construction of LHS designs is an active area of research, and many further extensions to the basic methods have been suggested. Recently, maximum projection space-filling designs have been found \cite{Josephetal2015} that minimise
\begin{equation}\label{eq:maxpro}
\psi(\mat{X}^n) = \left\{\frac{1}{{n \choose 2}}\sum_{1\le i < j \le n}\frac{1}{\prod_{l=1}^d(x_{i}^{(l)}-x_{j}^{(l)})^{2}}\right\}\,.
\end{equation}
Such a design promotes good space-filling properties, as measured by~\eqref{eq:maximin}, in \textit{all} projections of the design into subspaces of variables. Objective function~\eqref{eq:maxpro} arises from the use of a weighted Euclidean distance, see also \cite{BowmanWoods2013}.

Another important recent development is \textit{sliced} LHS designs \cite{QianWu2009, Qian2012, Baetal2015}, where the design can be partitioned into sets of runs or slices, each of which is an orthogonal or maximin LHS. The overall design, composed of the runs from all the slices, is also an LHS. Such designs may be used to study multiple numerical models having the same inputs where one slice is used to investigate each model; for example, to compare results from different model implementations. They are also useful for experiments on quantitative and qualitative variables when each slice is combined with one combination of levels of the qualitative variables. This latter application is the most important for screening where the resulting data could be used to estimate a surrogate linear model with dummy variables or a GP model with an appropriate correlation structure \cite{QianWuWu2008}. Supersaturated LHS designs, for $d\ge n$, have also been developed \cite{Butler2005}. 

\subsection{Sampling plans for estimating elementary effects (Morris' method)}\label{morris}

As an alternative to estimation of a surrogate model, Morris \cite{morris91} suggested a model-free approach that uses the elementary effect~\eqref{eq:elem} to measure the sensitivity of the response $Y(\vect{x})$ to a change in the $i$th variable at point $\vect{x}$. Each $\mbox{EE}_i(\vect{x})$ may be exactly or approximately (a) zero for all $\vect{x}$, implying a negligible influence of the $i$th variable on $Y(\vect{x})$; (b) a non-zero constant for all $\vect{x}$, implying a linear, additive effect; (c) a non-constant function of $x_i$, implying nonlinearity; or (d) a non-constant function of $x_j$ for $j\ne i$, implying the presence of at least one interaction involving $x_i$. 

In practice, active variables are usually selected using data from a relatively small experiment and therefore it is not possible to reconstruct $\mbox{EE}_i(\vect{x})$ as a continuous function of $\vect{x}$. The use of $r$ ``trajectory vectors'', $\vect{x}_1,\ldots,\vect{x}_r$, enables the following sensitivity indices to be defined for $i=1,\ldots,d$:
\begin{equation}\label{eq:EEmu}
\mu_i = \frac{1}{r}\sum_{j=1}^r\mbox{EE}_i(\vect{x}_j)
\end{equation}  		 	 
and
\begin{equation}\label{eq:EEsigma}
\sigma_i = \sqrt{\sum_{j=1}^r\frac{\left(\mbox{EE}_i(\vect{x}_j)-\mu_i\right)^2}{r-1}}\,.
\end{equation}  
		 	 
A large value of the mean $\mu_i$ suggests the $i$th input variable is active. Nonlinear and interaction effects are indicated by large values of $\sigma_i$. Plots of the sensitivity indices may be used to decide which variables are active and, among these variables, which have complex (non-additive and non-linear) effects. 

In addition to~\eqref{eq:EEmu} and~\eqref{eq:EEsigma}, an additional measure \cite{Campolongoetal2007} of the individual effect of the $i$th variable has been proposed that overcomes the possible cancellation of elementary effects in~\eqref{eq:EEmu} due to non-monotonic variable effects, namely,
\begin{equation}\label{eq:EEmu*}
\mu_i^\star = \frac{1}{r}\sum_{j=1}^r|\mbox{EE}_i(\vect{x}_j)|\,,
\end{equation}  	
where $|\cdot|$ denotes the absolute value. Large values of both $\mu_i$ and $\mu_i^\star$ suggest that the $i$th variable is active and has a linear effect on $Y(\vect{x})$; large values of $\mu_i^\star$ and small values of $\mu_i$ indicate cancellation in~\eqref{eq:EEmu} and a nonlinear effect for the $i$th variable.

\begin{figure}
\centering
\begin{tabular}{cc}
(a) Linear effect: $\mu_i>0$, $\mu_i^\star>0$; $\sigma_i=0$ & (b) Nonlinear effect: $\mu_i^\star>0$; $\sigma_i>0$ \\[-3ex]
\includegraphics[scale=0.4]{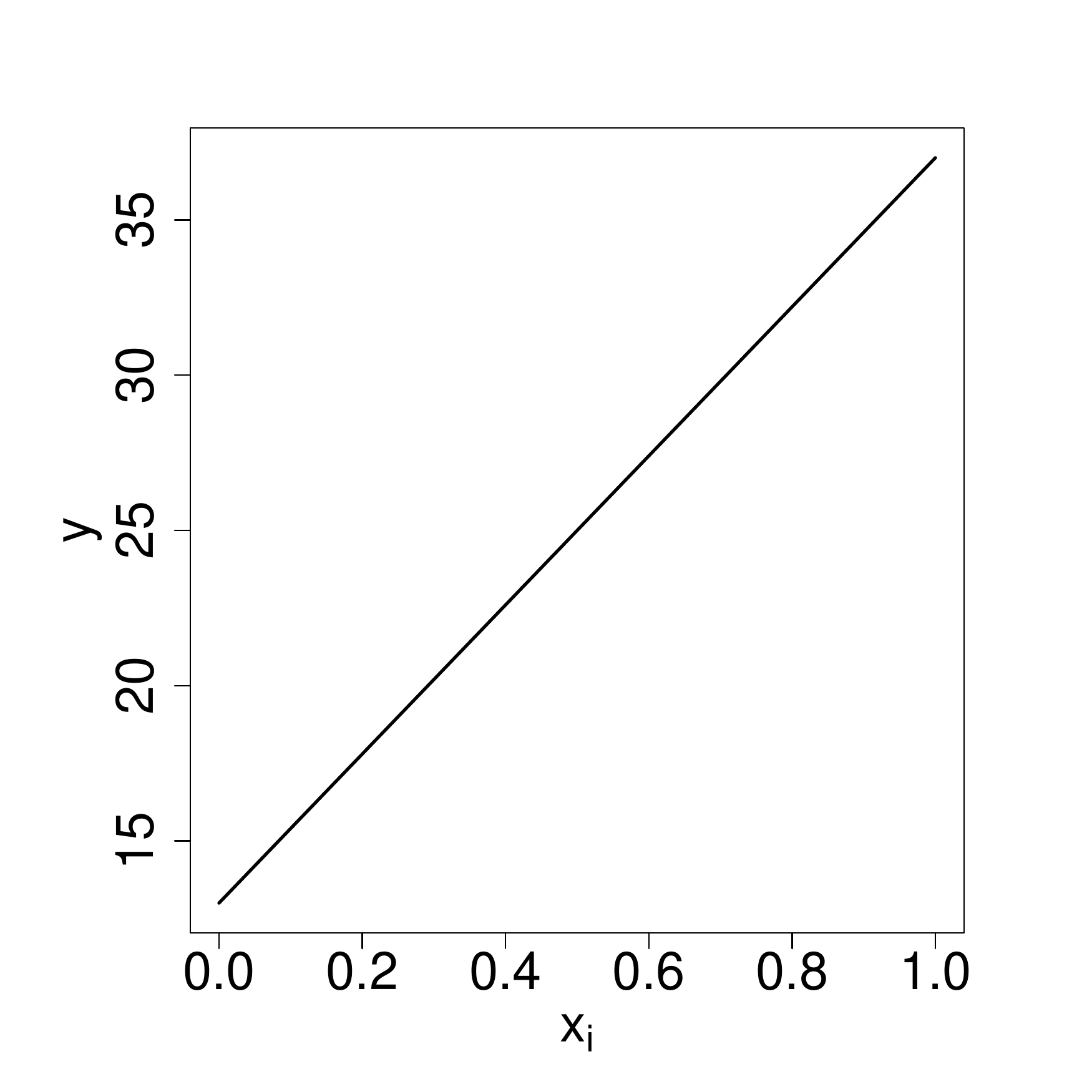} & \includegraphics[scale=0.4]{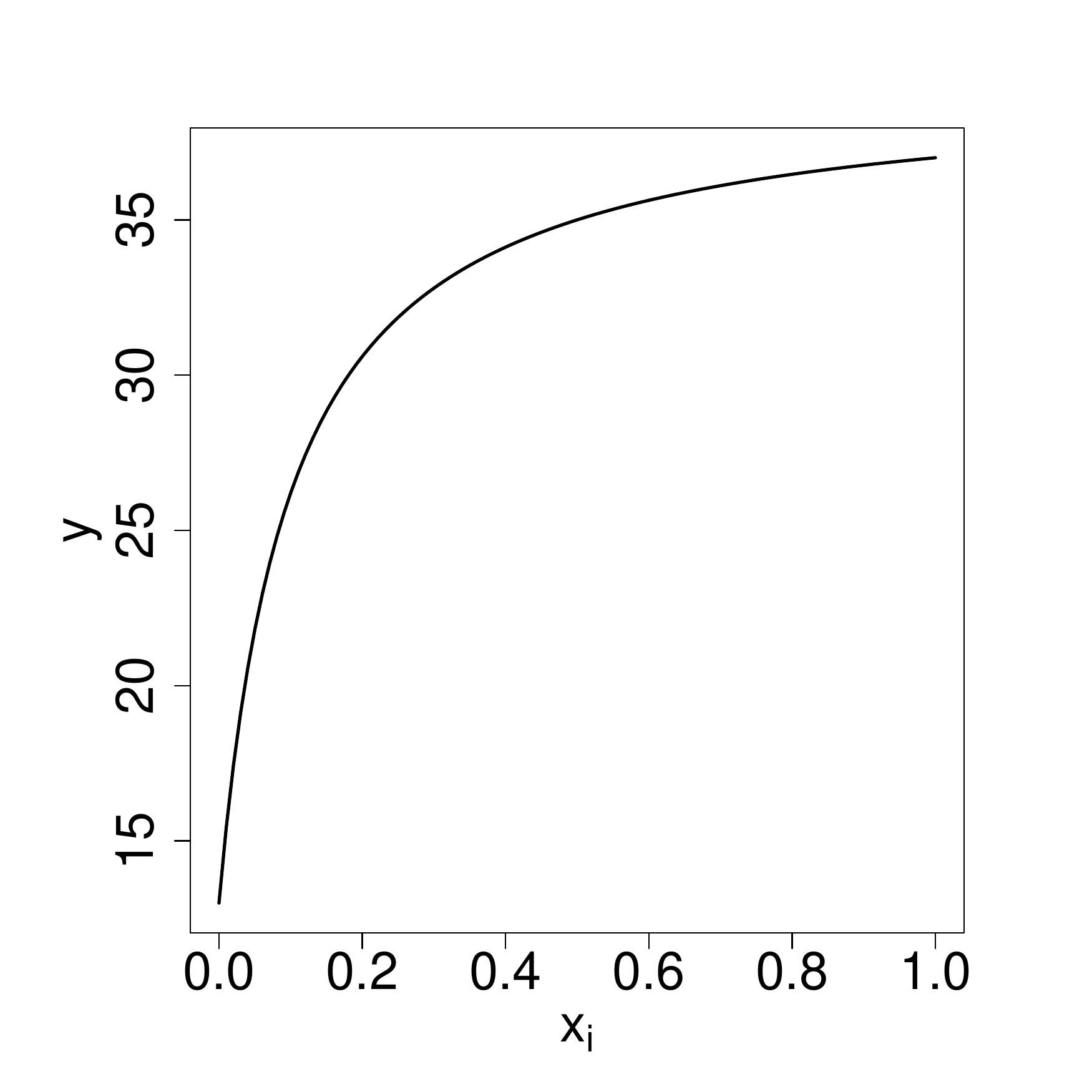} \\
(c) Interaction: $\sigma_i>0$ & (d) Nonlinear effect and interaction: $\mu_i^\star>0$, $\sigma_i>0$ \\[-3ex]
\includegraphics[scale=0.4]{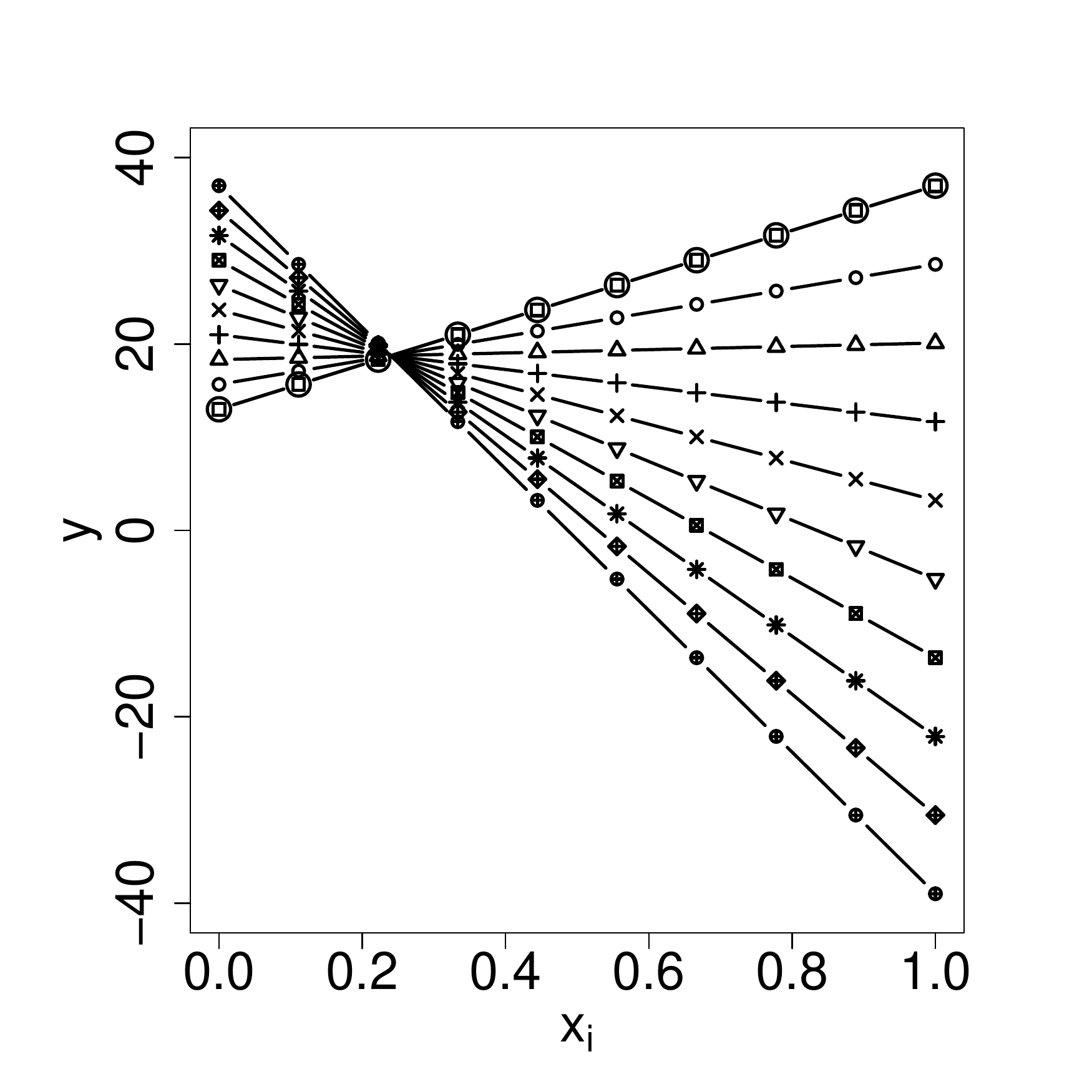} & \includegraphics[scale=0.4]{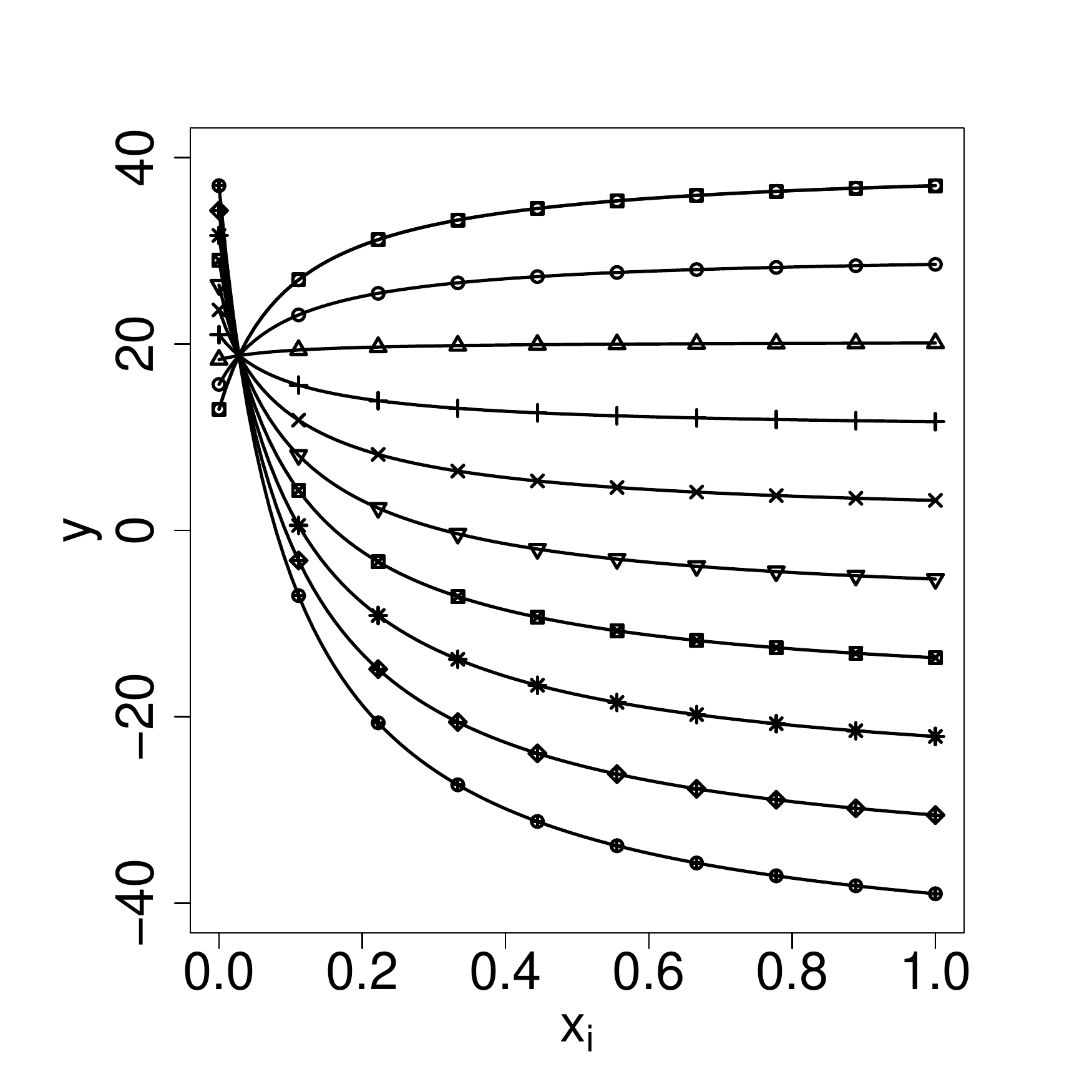} \\
\end{tabular}
\caption{Illustrative examples of effects for an active variable $x_i$ with values of $\mu_i$, $\mu_i^\star$ and $\sigma_i$. In plots (c) and (d), the plotting symbols correspond to the 10 levels of a second variable.}
\label{fig:morrisex}
\end{figure}

Four examples of the types of effects which use of~\eqref{eq:EEmu}--\eqref{eq:EEmu*} seek to identify are shown in Figure~\ref{fig:morrisex}. These are linear, nonlinear and interaction effects corresponding to non-zero values of~\eqref{eq:EEmu} and~\eqref{eq:EEmu*}, and both zero and non-zero values of~\eqref{eq:EEsigma}. It is not possible to use these statistics to distinguish between nonlinear and interaction effects, see Figures~\ref{fig:morrisex}(b) and~\ref{fig:morrisex}(c). 

The future development of a surrogate model can be simplified by separation of the active variables into two vectors, $\vect{x}_{S_1}$ and $\vect{x}_{S_2}$, where the variables in $\vect{x}_{S_1}$ have linear effects and the variables in $\vect{x}_{S_2}$ have nonlinear effects or are involved in interactions \cite{Boukouvalasetal2014}. For example, model~\eqref{eq:lm} might be fitted in which $\vect{h}(\vect{x}_{S_1})$ consists of linear functions and $\varepsilon(\vect{x}_{S_2})$ is modelled via a Gaussian process with correlation structure dependent only on variables in $\vect{x}_{S_2}$. 

In the elementary effects literature, the design region is assumed to be $\mathcal{X}=[0,1]^d$, after any necessary scaling, and is usually approximated by a $d$-dimensional grid, $\mathcal{X}_G$, having $f$ equally-spaced values, $0, 1/(f-1),\ldots,1$, for each input variable. The design of a screening experiment to allow the computation of the sample moments~\eqref{eq:EEmu}--\eqref{eq:EEmu*} has three components: the trajectory vectors $\vect{x}_1,\ldots,\vect{x}_r$, the $m$-run sub-design used to calculate $\mbox{EE}_i(\vect{x}_j)$ ($i=1,\ldots,d$) for each $\vect{x}_j$ ($j=1,\ldots,r$), and stepsize $\Delta$. Choices for these components are now discussed.

\begin{enumerate}
\item Morris \cite{morris91} chose $\vect{x}_1,\ldots,\vect{x}_r$ at random from $\mathcal{X}_G$, subject to the constraint that $\vect{x}_j+\Delta\vect{e}_{id}\in\mathcal{X}_G$ for $i=1,\ldots,d;\,j=1,\ldots,r$. Alternative suggestions include choosing the trajectory vectors as the $r$ points of a space-filling design \cite{Campolongoetal2007} found, for example, by minimising~\eqref{eq:maximin}. Larger values of $r$ result in~\eqref{eq:EEmu}--\eqref{eq:EEmu*} being more precise estimators of the corresponding moments of the elementary effect distribution. Morris used $r=3$ or $4$; more recently, other authors \cite{Campolongoetal2007} have discussed larger values ($r=10$--$50$).  

\item An OFAAT design with $m=d+1$ runs can be used to calculate $EE_1(\vect{x}_j),\ldots,EE_d(\vect{x}_j)$ for $j=1,\ldots,r$. The design matrix is $\mat{X}_j = \vect{1}_{d+1}\vect{x}_j + \Delta\mat{B}$, where $\vect{1}_{m}$ is a column $m$-vector with all entries equal to 1 and $\mat{B}=\sum_{l=2}^{d+1}\sum_{k=1}^{l-1}\vect{e}_{l(d+1)}\vect{e}_{kd}^{\T}$. That is, $\mat{B}$ is the $(d+1)\times d$ matrix
\begin{equation*}
\mat{B} = \left(
\begin{array}{ccccc}
0 & 0 & 0 & \cdots & 0 \\
1 & 0 & 0 & \cdots & 0 \\
1 & 1 & 0 & \cdots & 0 \\
\vdots & \vdots & & & \vdots\\
1 & 1 & 1 & \cdots & 1\\
\end{array}
\right)\,.
\end{equation*}

Designs are generated by swapping 0's and 1's at random within each column of $\mat{B}$ and randomising the column order. The overall design $\mat{X}^n = (\mat{X}_1^\T,\ldots,\mat{X}_r^\T)^\T$ then has $n=(d+1)r$ runs. It can be shown that this choice of $\mat{B}$, combined with the randomisation scheme, minimises the variability in the number of times in $\mat{X}^n$ that each variable takes each of the $f$ possible values \cite{morris91}.

Such OFAAT designs have the disadvantage of poor projectivity onto subsets of (active) variables, compared with, for example, LHS designs. Projection of a $d$-dimensional OFAAT design into $d-1$ dimensions reduces the number of distinct points from $d+1$ to $d$. This is a particular issue when the projection of the screening design onto the active variables is used to estimate a detailed surrogate model. Better projection properties may be obtained by replacing an OFAAT design by a rotated simplex \cite{Pujol2009} at the cost of less precision in the estimators of the elementary effects.

\item The choice of the ``step-size'' $\Delta$ in~\eqref{eq:elem} is determined by the choice of design region $\mathcal{X}_G$. Recommended values are $f=2g$ for some integer $g>0$ and $\Delta=f/2(f-1)$. This choice ensures that all $f^{d}/2$ elementary effects are equally likely to be selected for each variable when the trajectory vectors $\vect{x}$ are selected at random from $\mathcal{X}_G$.
\end{enumerate}	 

Further extensions of the elementary effects methodology include the application to group screening for numerical models with hundreds or thousands of input variables through the study of $\mu^\star$~\eqref{eq:EEmu*} \cite{Campolongoetal2007}, and a sequential experimentation strategy to reduce the number of runs of the numerical model required \cite{Boukouvalasetal2014}. This latter approach performs $r$ OFAAT experiments, one for each trajectory vector, in turn. For the $j$th experiment, elementary effects are calculated only for those variables that have not already been identified as having a nonlinear or interaction effect. That is, if $\sigma_i>\sigma_0$, for some threshold $\sigma_0$, when $\sigma_i$ is calculated from $r_1<r$ trajectory vectors, the $i$th elementary effect is no longer calculated for $\vect{x}_{r_1+1},\ldots,\vect{x}_r$. The threshold $\sigma_0$ can be elicited directly from subject experts or by using prior knowledge about departures from linearity for the effect of each variable \cite{Boukouvalasetal2014}. An obvious generalisation of the elementary effects method is to compute sensitivity indices directly from the (averaged local) derivatives. 

Methodology for the design and analysis of screening experiments using elementary effects is available in the \texttt{R} package \texttt{sensitivity} \cite{Pujoletal2015}.

\section{Model selection methods}\label{links}

The selection and estimation of a surrogate model~\eqref{eq:lm} from an application of a design discussed in this paper generally requires advanced statistical methods. An exception is a regular fractional factorial design for which standard linear modelling methods can be used provided that only one effect from each alias string is included in the model, and it is recognised that $\hat{\vectg{\beta}}$ may be biased. A brief description is now given of variable selection methods for (a) linear models with non-regular and supersaturated designs, and (b) Gaussian process models. 

\subsection{Variable selection for non-regular and supersaturated designs}\label{nonregmodel}

For designs with complex partial aliasing such as supersaturated and non-regular fractional factorial designs, a wide range of model selection methods have been proposed. An early suggestion was forward stepwise selection \cite{Miller2002} but this was shown to have low sensitivity in many situations \cite{Abrahametal99,MarleyWoods2010}. More recently, evidence has been provided for the effectiveness of shrinkage regression for the selection of active effects from these more complex designs \cite{Phoaetal2009,MarleyWoods2010,Draguljicetal2014}, particularly the Dantzig Selector \cite{CandesTao2007}. For this method, estimators $\hat{\vectg{\beta}}$ of the parameters in model~\eqref{eq:lm} are chosen to satisfy 
\begin{equation}\label{eq:ds}
\min_{\hat{\vectg{\beta}}\in\R^p} \sum_{u=1}^p|\hat{\beta}_u| 
\quad\mbox{ subject to } ||\mat{H}^\T (\vect{Y}^n-\mat{H}\hat{\vectg{\beta}})||_\infty\le s\,,
\end{equation}
with $s$ a tuning constant and $||\vect{a}||_\infty=\max |a_i|$, $\vect{a}^\T = (a_1,\ldots,a_p)$. This equation balances the desire for a parsimonious model with the need for the models to adequately describe the data.The value of $s$ can be chosen via an information criterion \cite{BurnhamAnderson2002} (e.g. AIC, AICc, BIC). The solution to~\eqref{eq:ds} may be obtained via linear programming; computationally efficient algorithms exist for calculating coefficient paths for varying $s$ \cite{Jamesetal2009}. 

The Dantzig Selector is applied to choose a subset of potentially active variables, and then standard least squares is used to fit a reduced linear model. The terms in this model whose coefficient estimates exceeded a threshold $t$, elicited from subject experts, are declared active. This procedure is known as the Gauss-Dantzig Selector \cite{CandesTao2007}. 

Other methods for variable selection that have been effective for these designs include Bayesian methods that use mixture prior distributions for the elements of $\vectg{\beta}$ \cite{GeorgeMcCulloch93,Chipmanetal97} and the application of stochastic optimisation algorithms \cite{WoltersBingham2011}.

\subsection{Variable selection for Gaussian process models}\label{gpmodel}

Screening for a Gaussian process model~\eqref{eq:lm}, with constant mean ($\vect{h}(\vect{x})=1$) and $\varepsilon(\vect{x}), \varepsilon(\vect{x})^\prime$ having correlation~\eqref{eq:corrfun}, may be performed by deciding which $\theta_i$ in~\eqref{eq:corrfun} are ``large'' using data obtained from an LHS or other space-filling design. Two approaches to this problem are outlined below: Stepwise Gaussian Process Variable Selection (SGPVS) \cite{Welchetal1992} and Reference Distribution Variable Selection (RDVS) \cite{Linkletteretal2006}.

In the first method, screening is via stepwise selection of $\theta_i$, $\alpha_i$ in~\eqref{eq:corrfun}, analogous to forward stepwise selection in linear regression models. The SGPVS algorithm identifies those variables that differ from the majority in their impact on the response in the following steps: 
\begin{enumerate}[(i)]
\item Find the maximised log-likelihood for model~\eqref{eq:lm} subject to $\theta_i=\theta$ and $\alpha_i=\alpha$ for all $i=1,\ldots,d$; denote this by $l_0$.
\item Set $\mathcal{E} = \{1,\ldots,d\}$.
\item\label{iter} For each $j\in\mathcal{E}$, find the maximised log-likelihood subject to $\theta_k=\theta$ and $\alpha_k=\alpha$ for all $k\in\mathcal{E}\setminus\{j\}$; denote the maximised log-likelihood by $l_j$. 
\item\label{stop} Let $j^\star =\mbox{arg}\max\limits_{j\in\mathcal{E}} l_j$. If $l_{j^\star}-l_0>c$, set $\mathcal{E}=\mathcal{E}\setminus\{j^\star\}$, $l_0=l_{j^\star}$ and go to step~(\ref{iter}). Otherwise stop.
\end{enumerate}
In step~\eqref{stop}, the values $c\approx 6$ (the 5\% critical value for a $\mathcal{X}_2$ distribution) have been suggested \cite{Welchetal1992}. 

The algorithm starts by assuming an isotropic correlation function and, at each iteration, at most one variable is allocated individual correlation parameters. The initial model contains only four parameters and the largest model considered has $4d+2$ parameters. However, factor sparsity suggests that the algorithm usually terminates before models of this size are considered. Hence, smaller experiments can be used than are normally employed for GP regression in $d$ variables (e.g. $d=20$ and $n=40$ or $50$ \cite{Welchetal1992}). This approach essentially adds one variable at a time to the model. Hence it has, potentially, similar issues as stepwise regression for linear models; in particular, the space of possible GP models is not very thoroughly explored. 

The second, RDVS, method is a fully Bayesian approach to the GP variable selection problem. A Bayesian treatment of model~\eqref{eq:lm} with correlation function~\eqref{eq:corrfun} requires numerical methods, such as Markov Chain Monte Carlo (MCMC), to obtain an approximate joint posterior distribution of $\theta_1,\ldots,\theta_d$ (in RDVS, $\alpha_i=2$ for all $i$). Conjugate prior distributions can be used for $\beta_0$ and $\sigma^2$ to reduce the computational complexity of this approximation. In RVDS, a prior distribution, formed as a mixture of a standard uniform distribution on $[0,1]$ and a point mass at 1, is assigned to $\rho_i = \exp\left(-0.25\theta_i\right)$. This reparameterisation of $\theta_i$ aids the implementation of MCMC algorithms and provides an intuitive interpretation: small $0<\rho_i\le 1$ corresponds to an active variable with the response changing rapidly with respect to the $i$th variable. 

To screen variables using RVDS, the design matrix $\mat{X}^n$ for the experiment is augmented by a $(d+1)$th column corresponding to an inert variable (having no substantive effect on the response) whose values are set at random. The posterior median for the correlation parameter, $\theta_{d+1}$, of the inert variable is computed for $b$ different randomly generated design matrices, formed by sampling values for the inert variable, to obtain an empirical \textit{reference distribution} for the median $\theta_{d+1}$. The percentiles of this reference distribution can be used to assess the importance of the ``real'' variables via the size of the corresponding correlation parameters.  

For methods that also incorporate variable selection into the Gaussian process mean function, i.e. incorporating the choice of functions in $\vect{h}(\vect{x})$, see \cite{Marreletal2008,OverstallWoods2015}.

\section{Examples and comparisons}\label{examples}

In this section, six combinations of the design and modelling strategies discussed in this paper are demonstrated and compared for variable screening using two test functions from the literature having $d=20$ variables and $\mathcal{X}=[-1,1]^d$. The functions differ in the number of active variables and the strength of influence of these variables on the output.

\noindent \textbf{Example~1}: A function used to demonstrate stepwise Gaussian process variable selection \cite{Welchetal1992}:
\begin{align}
Y(\vect{x}) = & \frac{5w_{12}}{1+w_1} + 5(w_4-w_{20})^2 + w_5 +40w_{19}^3 -5w_{19} \nonumber\\
& + 0.05w_2 + 0.08w_3 -0.03w_6 + 0.03w_7 - 0.09w_9 -0.01w_{10} \nonumber\\
& -0.07w_{11} +0.25w^2_{13} - 0.04w_{14} + 0.06w_{15} - 0.01w_{17} - 0.03w_{18}\,, \label{eq:welchfun}
\end{align}
where $w_i=0.5x_i$ ($i=1,\ldots,20$). There are six active variables, $x_1$, $x_4$, $x_5$, $x_{12}$, $x_{19}$, $x_{20}$.

\noindent \textbf{Example~2}: A function used to demonstrate the elementary effects method \cite{morris91}:
\begin{align}
Y(\vect{x}) = & \beta_0 + \sum_{j=1}^{20}\beta_jv_j + \sum_{1\le j<k}^{20}\beta_{jk}v_jv_k + \sum_{1\le j<k<l}^{20}\beta_{jkl}v_jv_kv_l \nonumber\\
& + \sum_{1\le j<k<l<u}\beta_{jklu}v_jv_kv_lv_u\,,\label{eq:morrisfun}
\end{align}
where $v_i = x_i$ for $i\ne 3,5,7$ and $v_i = 11(x_i+1)/(5x_i+6)-1$ otherwise; $\beta_j = 20$ ($j=1,\ldots,10$), $\beta_{jk}=-15$ ($j,k=1,\ldots,6$), $\beta_{jkl} = -10$ ($j,k,l=1,\ldots,5$) and $\beta_{jklu} = 5$ ($j,k,l,u=1,\ldots,4$). The remaining $\beta_j$ and $\beta_{jk}$ values are independently generated from a N(0,1) distribution, and these are used in all the analyses; all other coefficients are set to 0. There are 10 active variables, $x_1,\ldots,x_{10}$.

There are some important differences between these two examples: function~\eqref{eq:welchfun} has a proportion of active variables (0.3) in line with factor sparsity but the influence of many of these active variables on the response is only small; function~\eqref{eq:morrisfun} is not effect sparse (with 50\% of the variables being active) but the active variables have much stronger influence on the response. This second function also contains many more interactions. Thus, the examples present different screening challenges. For these two deterministic functions, a single data set was generated for each design employed.

The screening strategies employ experiment sizes chosen to allow comparison of the different methods. They fall into three classes.
\begin{enumerate}
\item Methods using Gaussian processes and space-filling designs:
\begin{enumerate}
\item Stepwise Gaussian Process Variable Selection (SGPVS). 
\item Reference Distribution Variable Selection (RDVS).  
\end{enumerate}
\suspend{enumerate}
These two variable selection methods use $n=16, 41, 84, 200$ runs, where $n=200$ follows the standard guidelines of $n=10d$ runs for estimating a Gaussian process model \cite{Loeppkyetal2009}. For each value of $n$, two designs are found: a maximin Latin hypercube sampling design and a maximum projection space-filling design. These designs were generated from the \texttt{R} packages \texttt{SLHD} \cite{Ba2015} and \texttt{MaxPro} \cite{BaJoseph2015} using simulated annealing and quasi-Newton algorithms. For both methods, $\alpha_i=2$ in~\eqref{eq:corrfun} for all $i=1,\ldots,d$; that is, for SGPVS, stepwise selection is performed for $\theta_i$ only.
\resume{enumerate}
\item One-factor-at-a-time methods:
\begin{enumerate}
\item Elementary Effects method (EE).
\item Systematic Fractional Replicate Designs (SFRD).
\end{enumerate}
\suspend{enumerate}
The elementary effects were calculated using the \text{R} package \texttt{sensitivity} \cite{Pujoletal2015}, with each variable taking $f=4$ levels, $\Delta = 2/3$ and $r=2,4,10$ randomly generated trajectory vectors $\vect{x}_1,\ldots,\vect{x}_r$, giving design sizes of $n=42, 84, 210$, respectively. An $n=42$ run SFRD was used to calculate the sensitivity indices $S(i)$, equation~\eqref{SI}, for $i=1,\ldots,20$.
\resume{enumerate}
\item Linear model methods:
\begin{enumerate}
\item Supersaturated Design (SSD).
\item Definitive Screening Design (DSD).
\end{enumerate}
\end{enumerate}
The designs used are an SSD with $n=16$ runs and a DSD with $n=41$ runs. For each design, variable selection is performed using the Dantzig selector as implemented in the \texttt{R} package \texttt{flare} \cite{Lietal2014} with shrinkage parameter $s$ chosen using AICc. Note that these designs are tailored to screening in linear models and hence may not perform well when the output is inadequately described by a linear model. 

\begin{table}
\begin{center}
\caption{Sensitivity ($\phi_{\mbox{s}}$), type I error rate ($\phi_{\mbox{I}}$) and false discovery rate ($\phi_{\mbox{fdr}}$) for Example~1}
\label{tab:welchex}       
\begin{tabular}{l@{\hskip 10pt}c@{\hskip 10pt}c@{\hskip 10pt}c@{\hskip 30pt}c@{\hskip 10pt}c@{\hskip 10pt}c@{\hskip 30pt}c@{\hskip 10pt}c@{\hskip 10pt}c@{\hskip 30pt}c@{\hskip 10pt}c@{\hskip 10pt}c}
\hline\noalign{\smallskip}
 & $\phi_{\mbox{s}}$ & $\phi_{\mbox{I}}$ & $\phi_{\mbox{fdr}}$ & $\phi_{\mbox{s}}$ & $\phi_{\mbox{I}}$ & $\phi_{\mbox{fdr}}$ & $\phi_{\mbox{s}}$ & $\phi_{\mbox{I}}$ & $\phi_{\mbox{fdr}}$ & $\phi_{\mbox{s}}$ & $\phi_{\mbox{I}}$ & $\phi_{\mbox{fdr}}$ \\
\noalign{\smallskip}\hline\noalign{\smallskip}
\multicolumn{13}{l}{Gaussian processes and space-filling designs} \\
 & \multicolumn{3}{c}{$n=16$} & \multicolumn{3}{c}{$n=41$} & \multicolumn{3}{c}{$n=84$} & \multicolumn{3}{c}{$n=200$} \\
 SGPVS & 0.33 & 0.07 & 0.33 & 1 & 0 & 0 & 1 & 0 & 0 & 0.67 (1)$^\dagger$ & 0 & 0 \\
 RVDS & 0.17 & 0 & 0 &  0.33 & 0 & 0 & 1 & 0 & 0 & 1 & 0 & 0 \\
\multicolumn{4}{l}{One-factor-at-a-time designs} & \multicolumn{3}{c}{$n=42$} & \multicolumn{3}{c}{$n=84$} & \multicolumn{3}{c}{$n=210$} \\ 
EE & -- & -- & -- & 0.5 & 0 & 0 & 0.83 & 0 & 0 & 0.83 & 0 & 0 \\ 
SFRD & -- & -- & -- & 0.83\footnotemark[1] (1)\footnotemark[2] & 0 & 0 & -- & -- & -- & -- & -- & -- \\
\multicolumn{13}{l}{Non-regular fractional factorial designs and linear models} \\
 & \multicolumn{3}{c}{$n=16$} & \multicolumn{3}{c}{$n=41$} \\
 SSD & 0.50 & 0.14 & 0.40 & -- & -- & -- & -- & -- & -- & -- & -- & -- \\
 DSD & -- & -- & -- & 0.17 (0.33)\footnotemark[3] & 0 & 0 & -- & -- & -- & -- & -- & -- \\
\noalign{\smallskip}\hline
\end{tabular}
\end{center}
$^\dagger$ Using~\eqref{eq:corrfun} with $\alpha_i=1$\\
\footnotemark[1] Using a threshold of $5\%$ on the sensitivity indices. \\
\footnotemark[2] Using a threshold of $1\%$ on the sensitivity indices. \\
\footnotemark[3] Sensitivity for a main effects only model.\\
\end{table}
\begin{table}
\begin{center}
\caption{Sensitivity ($\phi_{\mbox{s}}$), type I error rate ($\phi_{\mbox{I}}$) and false discovery rate ($\phi_{\mbox{fdr}}$) for Example~2}
\label{tab:morrisex}       
\begin{tabular}{l@{\hskip 10pt}c@{\hskip 10pt}c@{\hskip 10pt}c@{\hskip 30pt}c@{\hskip 10pt}c@{\hskip 10pt}c@{\hskip 30pt}c@{\hskip 10pt}c@{\hskip 10pt}c@{\hskip 30pt}c@{\hskip 10pt}c@{\hskip 10pt}c}
\hline\noalign{\smallskip}
 & $\phi_{\mbox{s}}$ & $\phi_{\mbox{I}}$ & $\phi_{\mbox{fdr}}$ & $\phi_{\mbox{s}}$ & $\phi_{\mbox{I}}$ & $\phi_{\mbox{fdr}}$ & $\phi_{\mbox{s}}$ & $\phi_{\mbox{I}}$ & $\phi_{\mbox{fdr}}$ & $\phi_{\mbox{s}}$ & $\phi_{\mbox{I}}$ & $\phi_{\mbox{fdr}}$ \\
\noalign{\smallskip}\hline\noalign{\smallskip}
\multicolumn{13}{l}{Gaussian processes and space-filling designs} \\
 & \multicolumn{3}{c}{$n=16$} & \multicolumn{3}{c}{$n=41$} & \multicolumn{3}{c}{$n=84$} & \multicolumn{3}{c}{$n=200$} \\
 SGPVS & 0 & 0 & 0 & 0.40 & 0 & 0 & 1 & 0 & 0 & 1 & 0 & 0 \\
 RVDS & 0 & 0 & 0 &  0.30 & 0 & 0 & 0.80 & 0 & 0 & 1 & 0 & 0 \\
 \multicolumn{4}{l}{One-factor-at-a-time designs} & \multicolumn{3}{c}{$n=42$} & \multicolumn{3}{c}{$n=84$} & \multicolumn{3}{c}{$n=210$} \\ 
 EE & -- & -- & -- & 0.80 & 0 & 0 & 1 & 0 & 0 & 1 & 0 & 0 \\ 
 SFRD & -- & -- & -- & 0.60\footnotemark[1] (1)\footnotemark[2] & 0 & 0 & -- & -- & -- & -- & -- & -- \\
  \multicolumn{13}{l}{Non-regular fractional factorial designs and linear models} \\
  & \multicolumn{3}{c}{$n=16$} & \multicolumn{3}{c}{$n=41$} \\
 SSD & 0.60 & 0.90 & 0.60 & -- & -- & -- & -- & -- & -- & -- & -- & -- \\
 DSD & -- & -- & -- & 0.10 (0)\footnotemark[3] & 0 & 0 & -- & -- & -- & -- & -- & -- \\ 
\noalign{\smallskip}\hline
\end{tabular}
\end{center}
\footnotemark[1] Using a threshold of $5\%$ on the sensitivity indices. \\
\footnotemark[2] Using a threshold of $1\%$ on the sensitivity indices. \\
\footnotemark[3] Sensitivity for a main effects only model.\\
\end{table}

Tables~\ref{tab:welchex} and~\ref{tab:morrisex} summarise the results for examples~1 and~2, respectively, and present the sensitivity ($\phi_{\mbox{s}}$), type I error rate ($\phi_{\mbox{I}}$) and false discovery rate ($\phi_{\mbox{fdr}}$) for the five methods. The summaries reported in these tables use automatic selection of tuning parameters in each method, see below. More nuanced screening, for example using graphical methods, may produce different trade-offs between sensitivity, type I error rate and false discovery rate. In general, with the exception of the EE method, results are better for Example~1, which obeys factor sparsity, than for Example~2, which does not. 

For the Gaussian process methods (SGPVS and RDVS), the assessments presented in Tables~\ref{tab:welchex} and~\ref{tab:morrisex} are the better of the results obtained from the maximin Latin hypercube sampling design and the maximum projection space-filling design. In Example~1, for $n=16,41,84$ the maximin LHS design gave the better performance, while for $n=200$ the maximum projection design was preferred. In Example~2, the maximum projection design was preferred for $n=16,200$ and the maximin LHS design for $n=41,84$.  For Example~1, note that, rather counter-intuitively, for correlation function~\eqref{eq:corrfun} with $\alpha_i=2$, SGPVS has lower sensitivity for $n=200$ than for $n=41$ or $n=84$. A working hypothesis to explain this result is that larger designs, with points closer together in the design space, can present challenges in estimating the parameters $\theta_1,\ldots,\theta_d$ when the correlation function is very smooth. Setting $\alpha_i = 1$, so that the correlation function is less smooth, resulted in better screening for $n=200$. The choice of design for Gaussian process screening is an area for further research, as is the application to screening of extensions of the Gaussian process model to high-dimensional inputs \cite{GramacyLee2008,Durrandeetal2012}

\renewcommand{\RDVSscale}{0.27} 
\begin{figure}
\centering
\begin{tabular}{cc}
(a) Example~1: $n=16$ & (b) Example~2: $n=16$ \\[-3ex]
\includegraphics[scale=\RDVSscale]{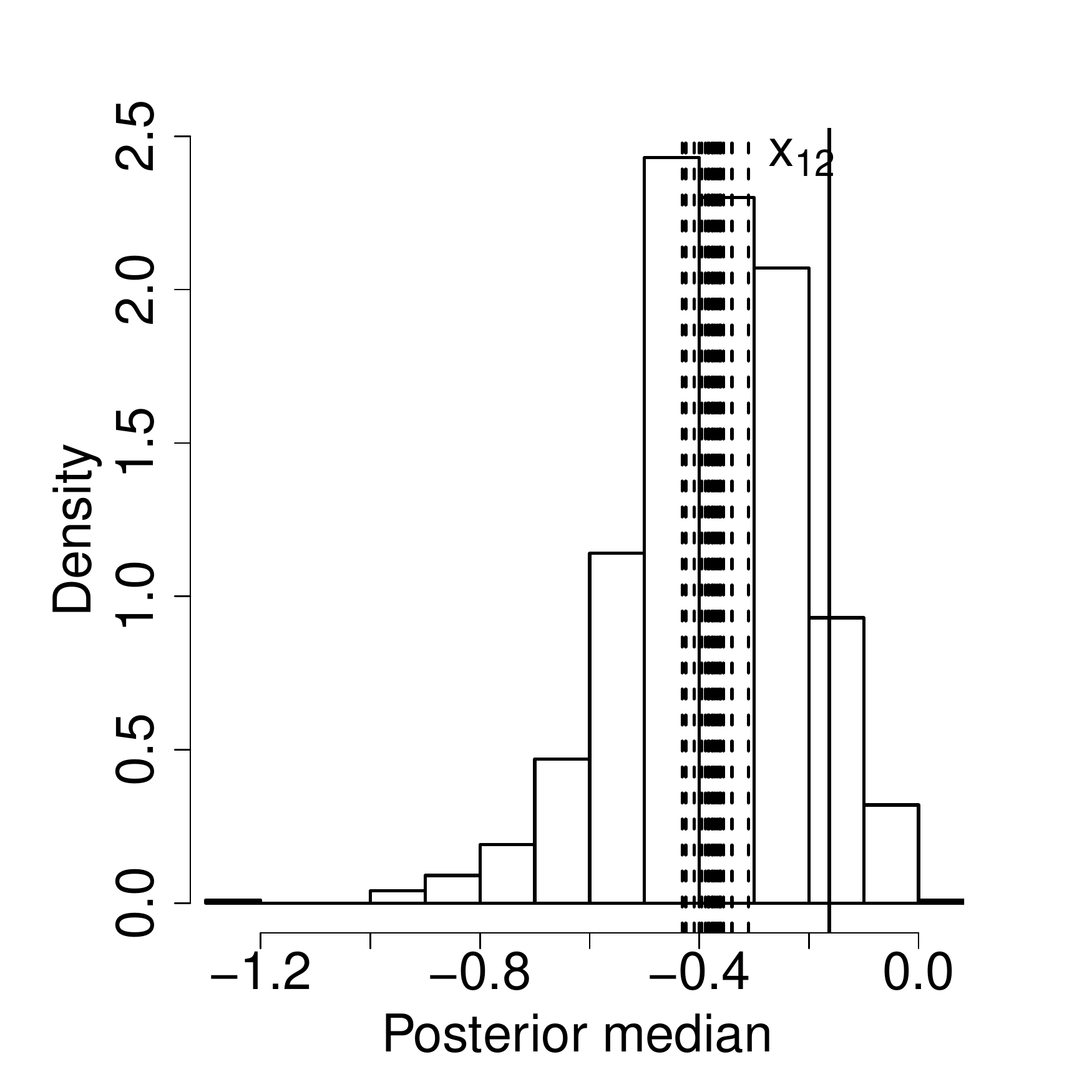} & \includegraphics[scale=\RDVSscale]{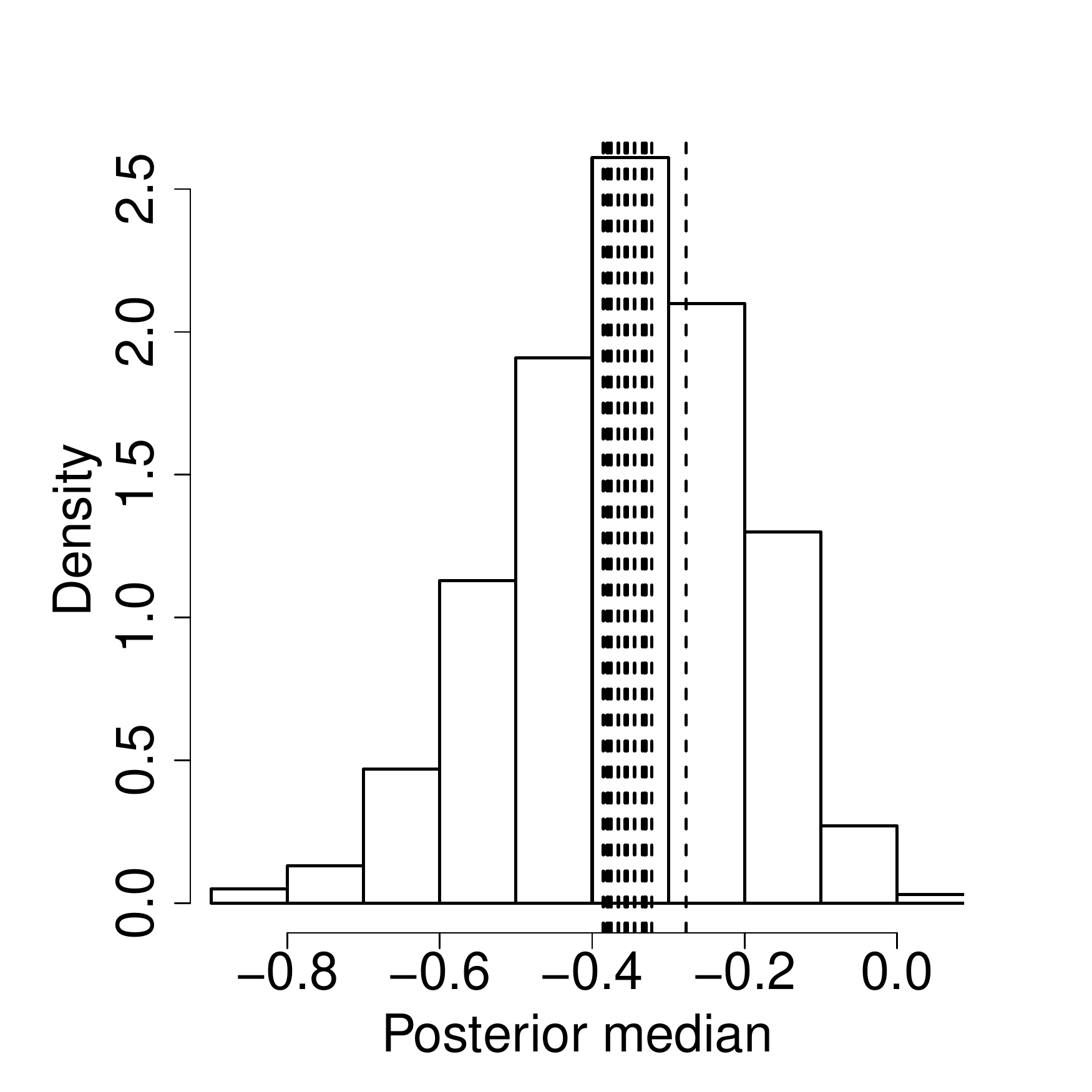}\\
(c) Example~1: $n=41$ & (d) Example~2: $n=41$ \\[-3ex]
\includegraphics[scale=\RDVSscale]{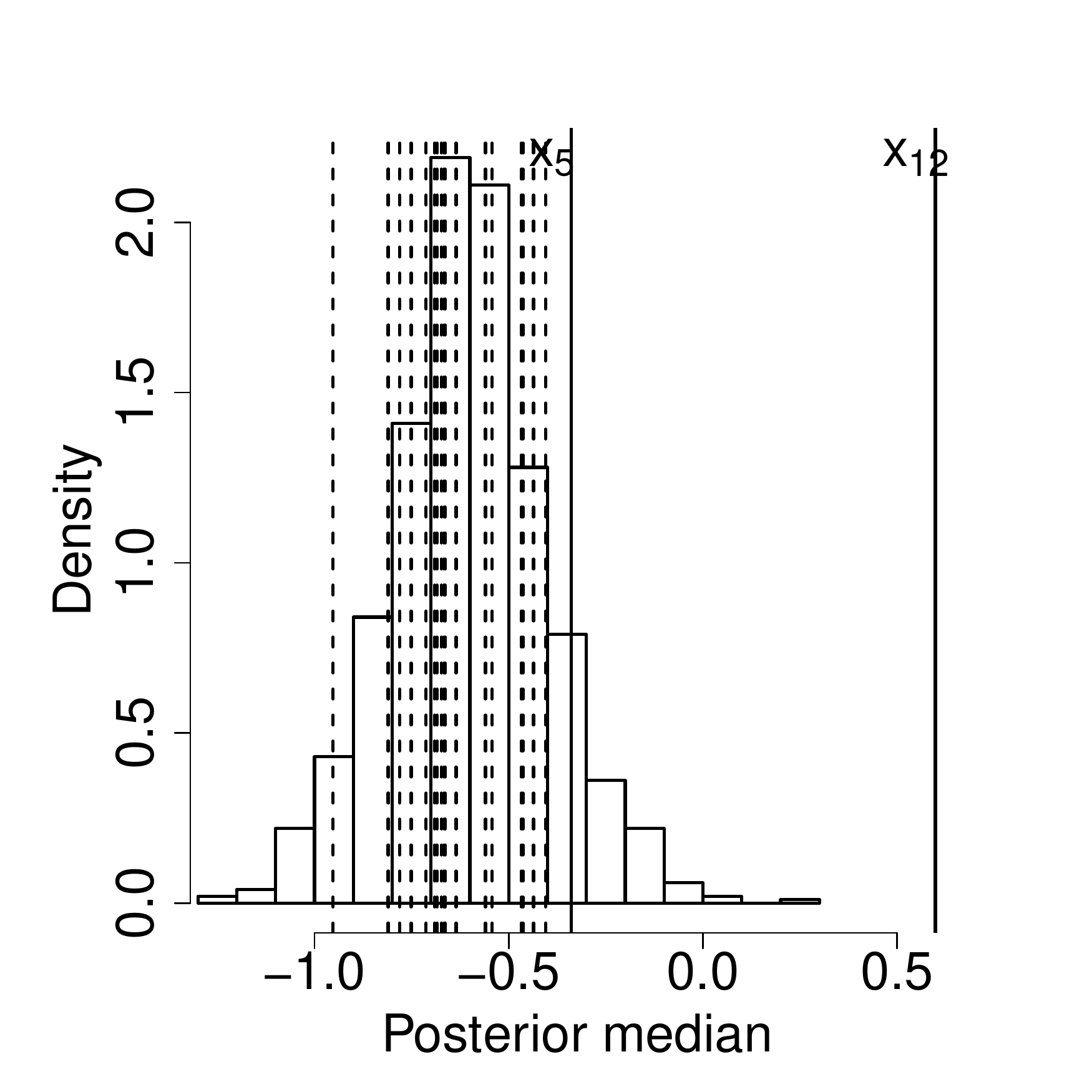} & \includegraphics[scale=\RDVSscale]{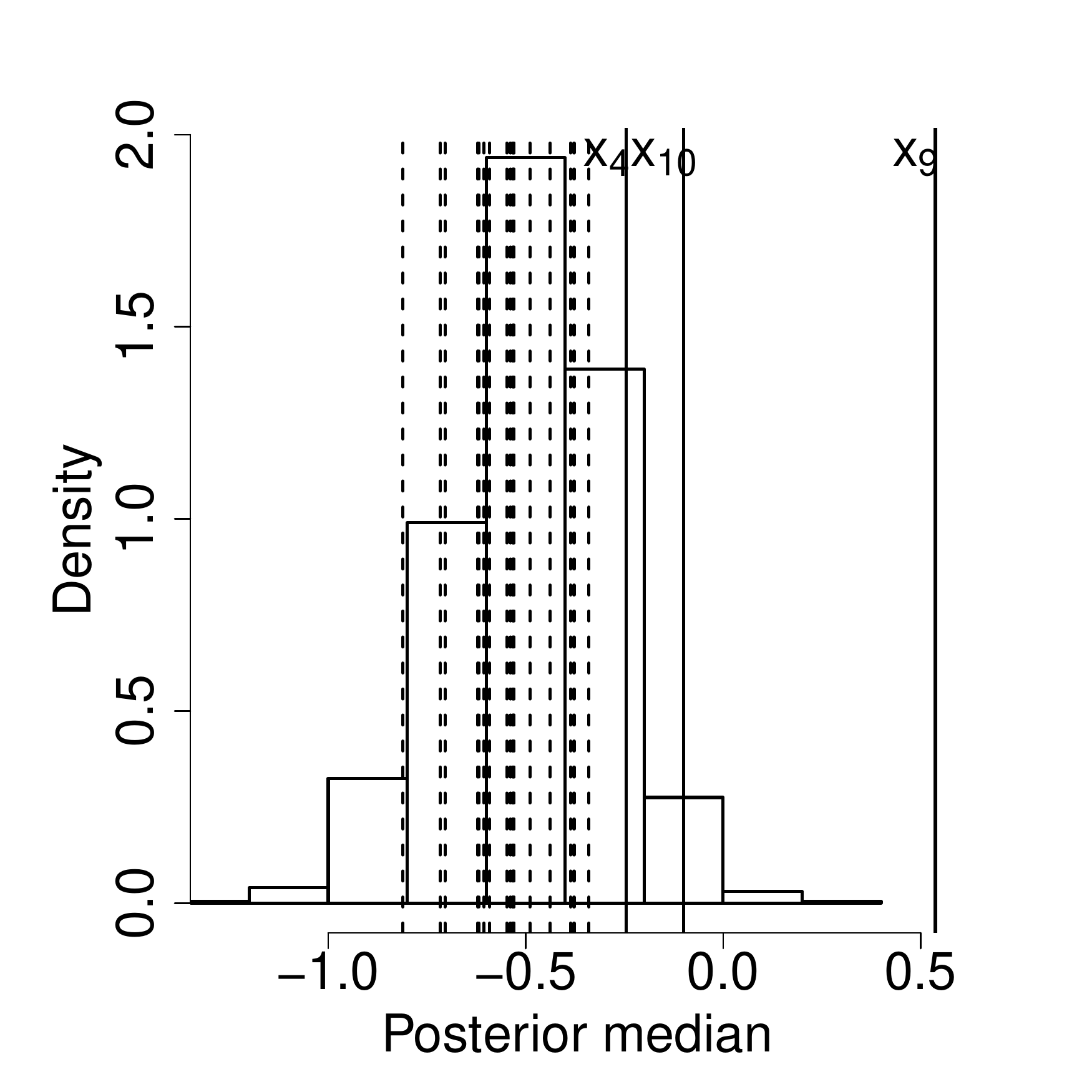}\\
(e) Example~1: $n=84$ & (f) Example~2: $n=84$ \\[-3ex]
\includegraphics[scale=\RDVSscale]{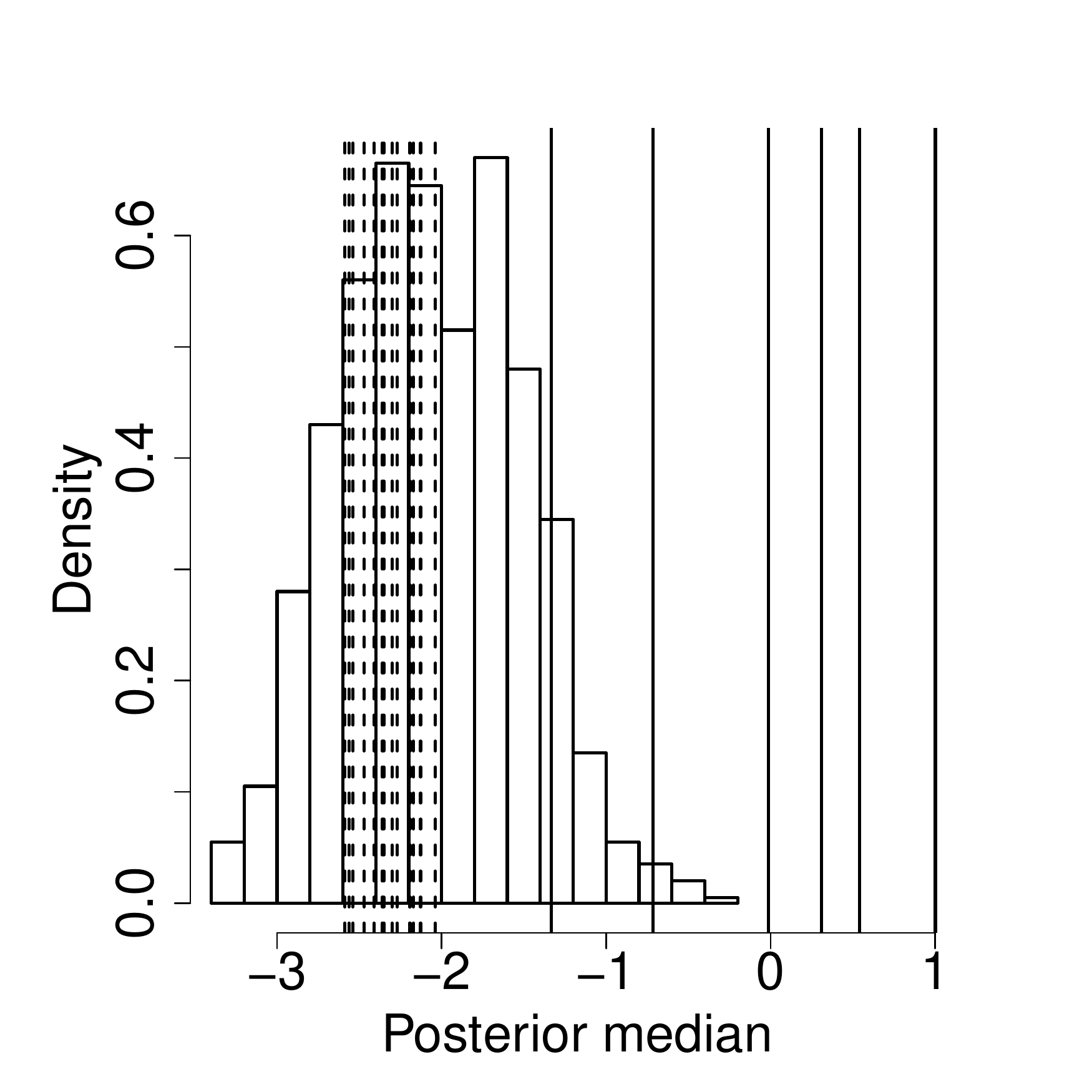} & \includegraphics[scale=\RDVSscale]{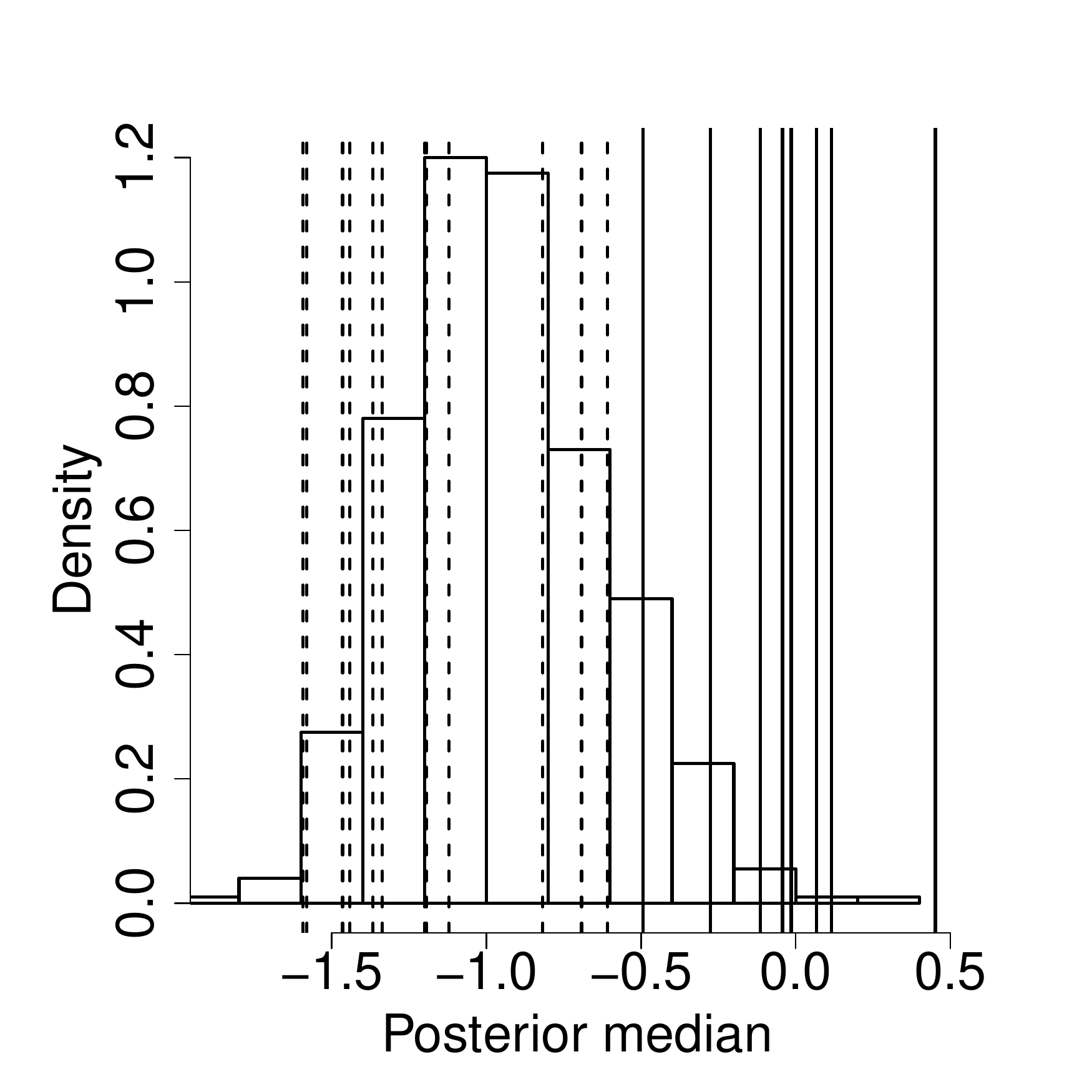}\\
(g) Example~1: $n=200$ & (h) Example~2: $n=200$ \\[-3ex]
\includegraphics[scale=\RDVSscale]{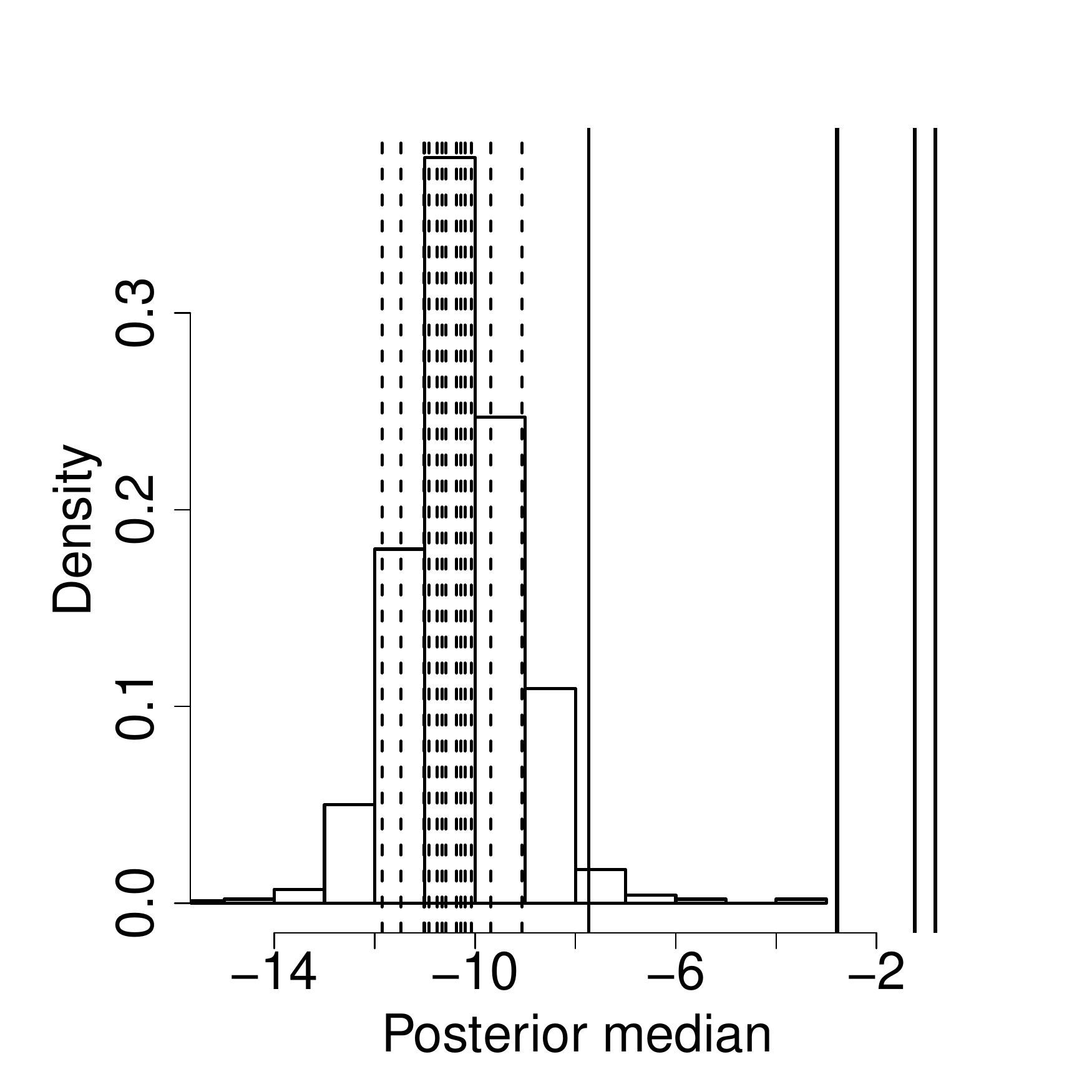} & \includegraphics[scale=\RDVSscale]{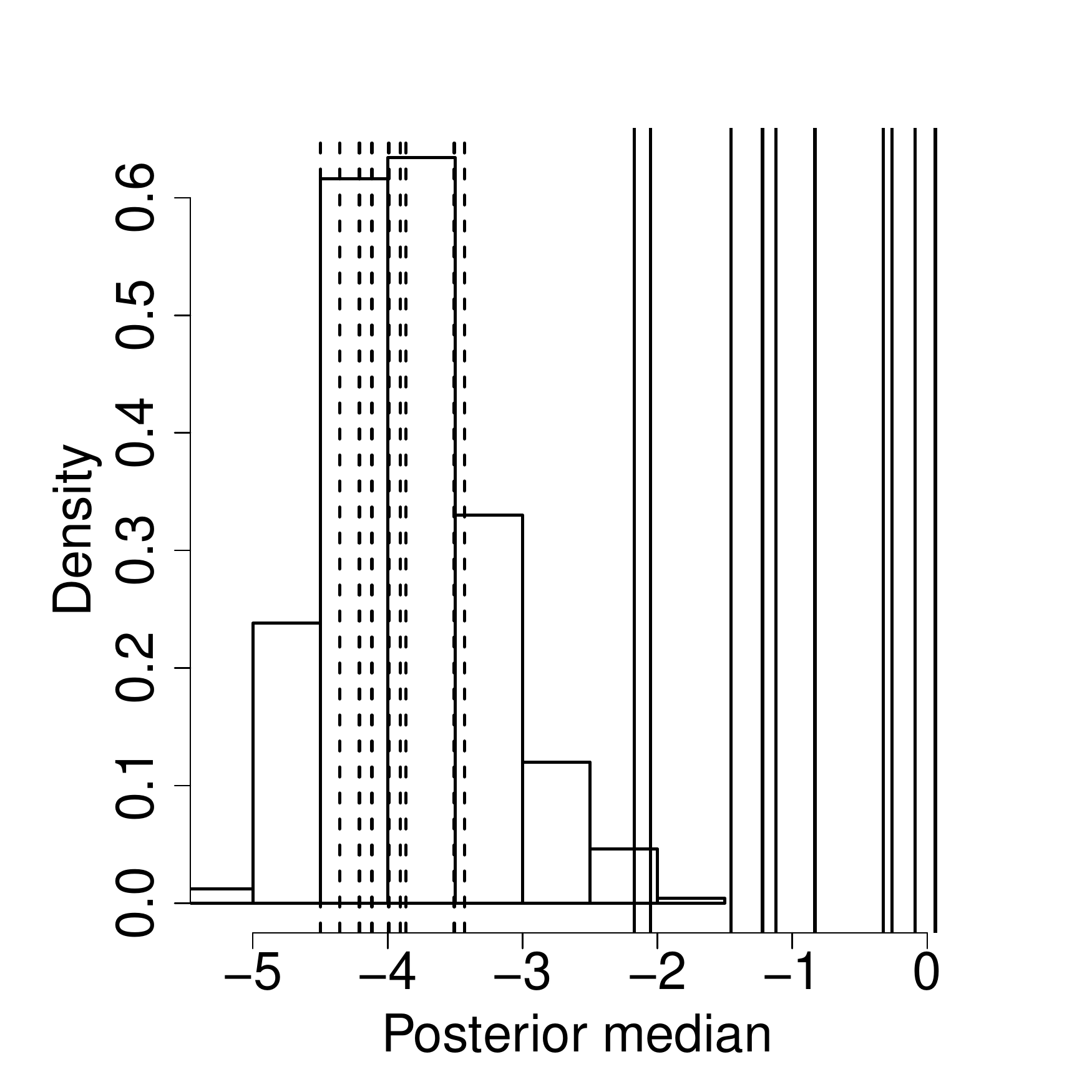}\\
\end{tabular}
\caption{RDVS results: histograms of the empirical distribution of the posterior median of the correlation parameter for 1000 randomly generated inert variables. The posterior medians of the correlation parameters for the 20 variables are marked as vertical lines (unbroken -- declared active; broken -- declared inactive). If there are fewer than 5 variables declared active, the variable names are also given.}
\label{fig:RDVS}
\end{figure}

SGPVS and RDVS performed well when used with larger designs ($n=84,200$) for both examples. For Example~1, the effectiveness of SGPVS has already been demonstrated \cite{Welchetal1992} for a Latin hypercube design with $n=50$ runs. In the current study, the method was also effective when $n=41$. Neither RDVS or SGPVS provided reliable screening when $n=16$. Both methods found Example~2, with a greater proportion of active variables, more challenging; effective screening was only achieved when $n=84,200$. 
 
For RVDS, recall that the empirical distribution of the posterior median of the correlation parameter for the inert variable is used as a reference distribution to assess the size of the correlation parameters for the actual variables. For the assessments in Tables~\ref{tab:welchex} and~\ref{tab:morrisex}, a variable was declared active if the posterior median of its correlation parameter exceeded the 90th percentile of the reference distribution. A graphical analysis can provide a more detailed assessment of the method. Figure~\ref{fig:RDVS} shows the reference distribution and the posterior median of the correlation parameter for each of the 20 variables. Variables declared active have their posterior medians in the right-hand tail of the reference distribution. For both examples, the greater effectiveness of RDVS for larger $n$ is clear. It is interesting to note that choosing a small percentile as the threshold to declare a variable active, for example 80\%, would have resulted in considerably higher type I error and false discovery rates for $n=16,41$. 

\renewcommand{\EEscale}{0.4} 
\begin{figure}
\centering
\begin{tabular}{cc}
(a) Example~1: $n=42$ & (b) Example~2: $n=42$ \\[-3ex]
\includegraphics[scale=\EEscale]{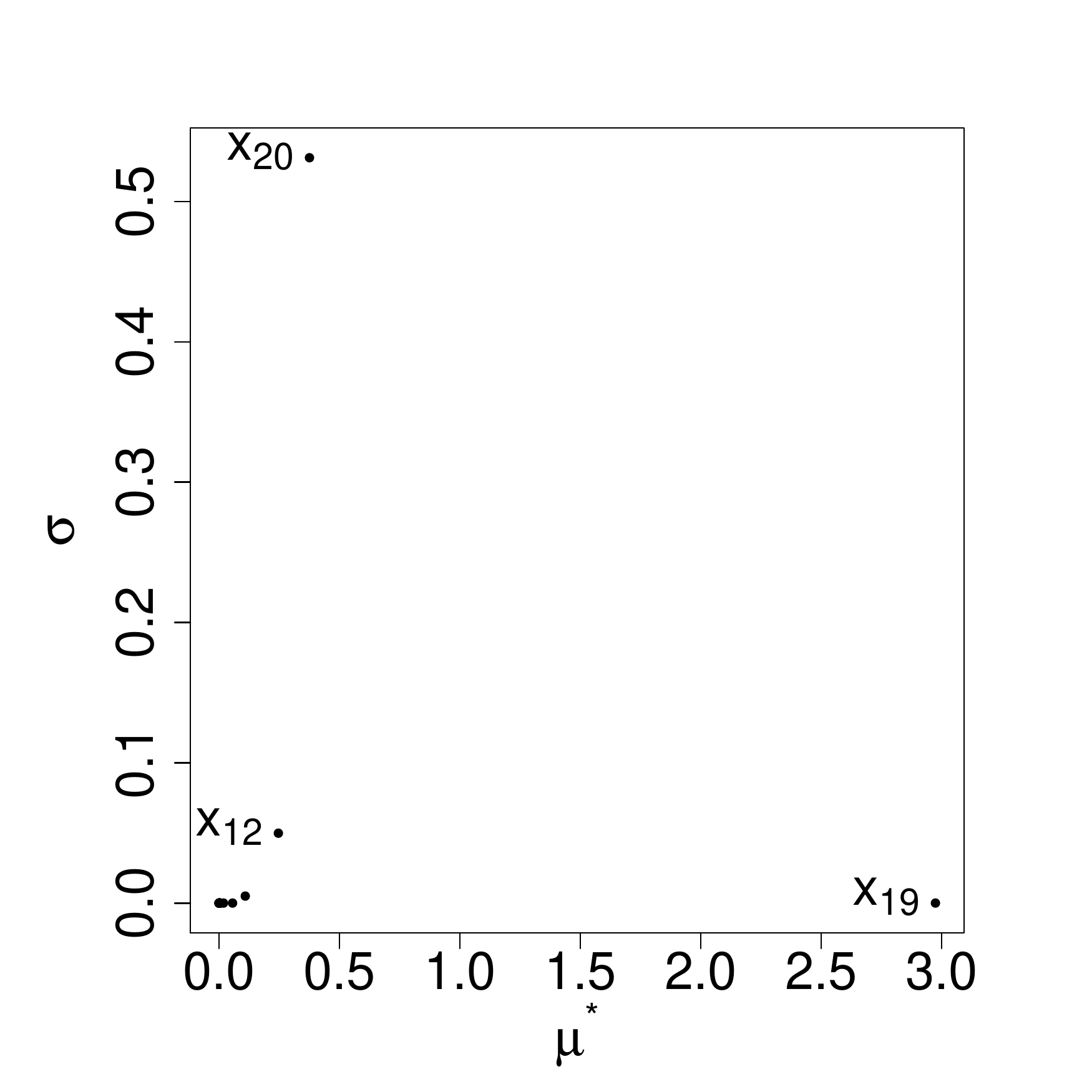} & \includegraphics[scale=\EEscale]{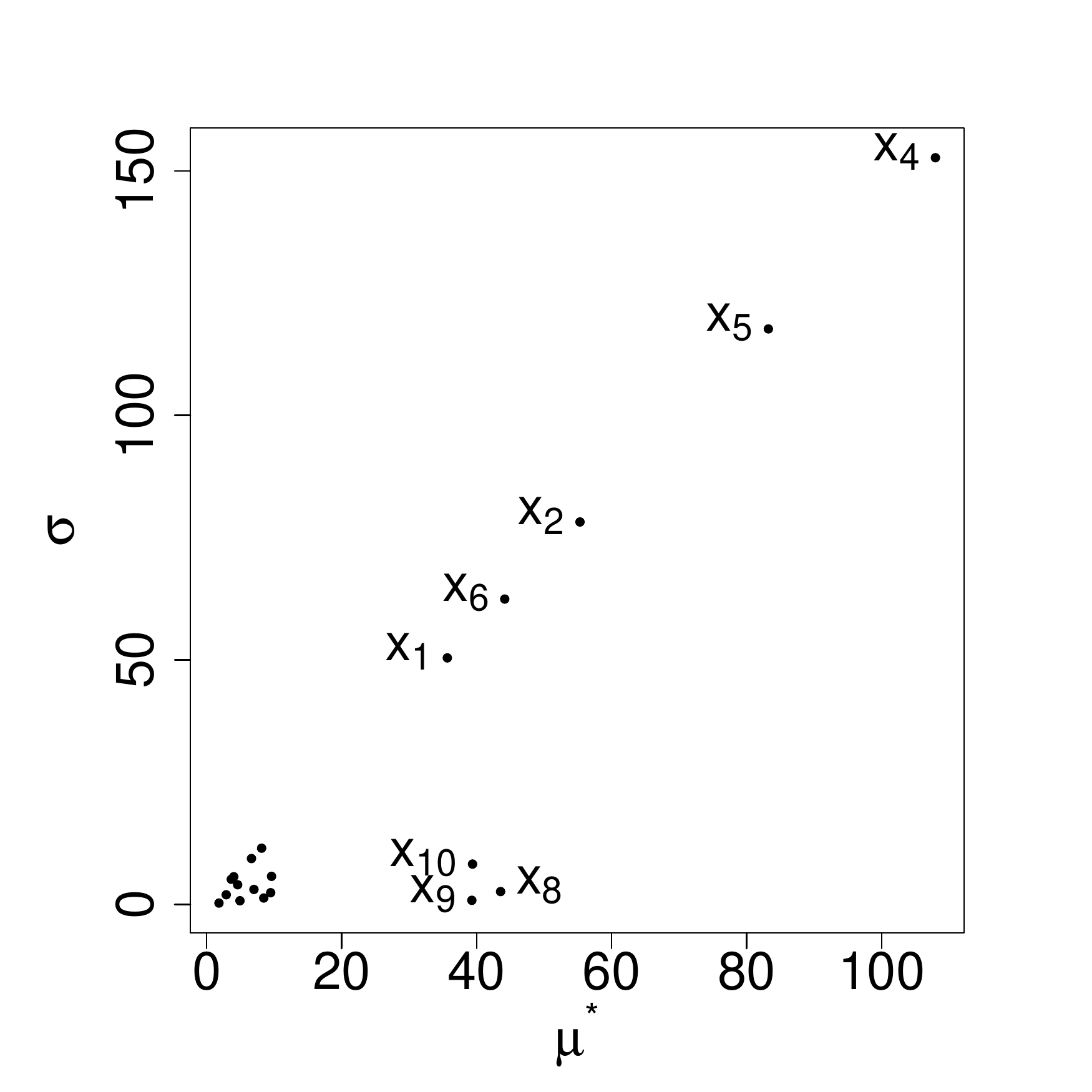}\\
(c) Example~1: $n=84$ & (b) Example~2: $n=84$ \\[-3ex]
\includegraphics[scale=\EEscale]{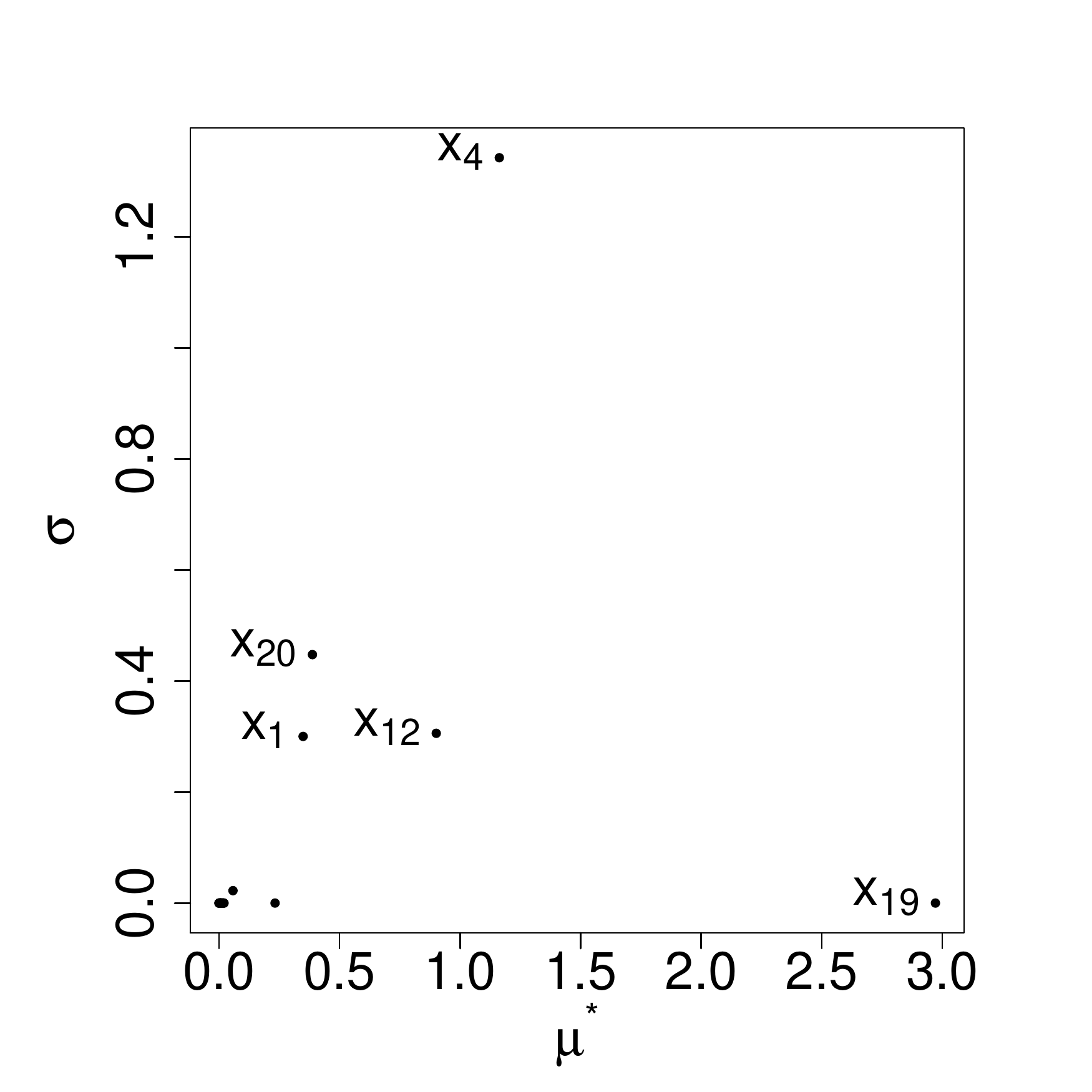} & \includegraphics[scale=\EEscale]{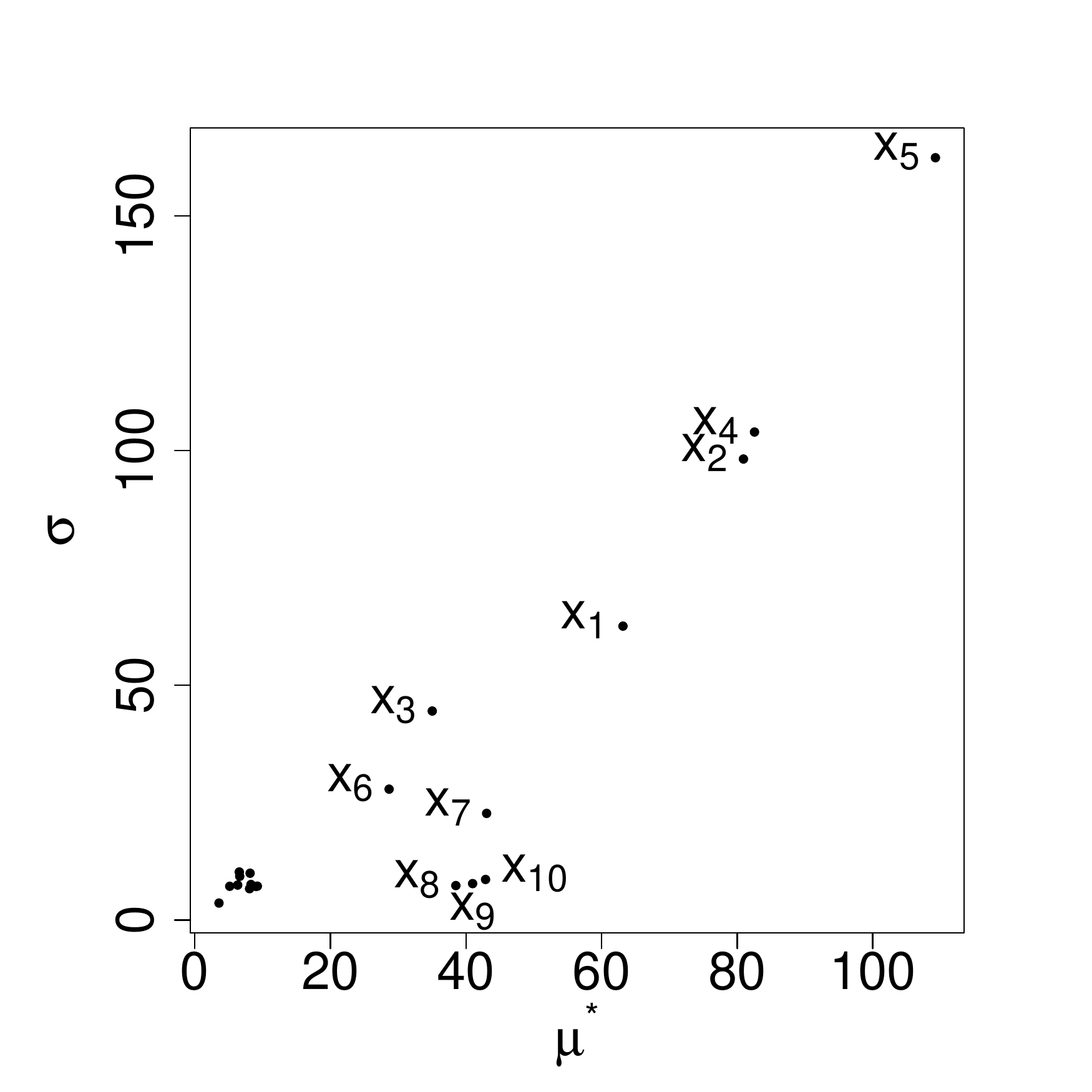}\\
(e) Example~1: $n=210$ & (f) Example~2: $n=210$ \\[-3ex]
\includegraphics[scale=\EEscale]{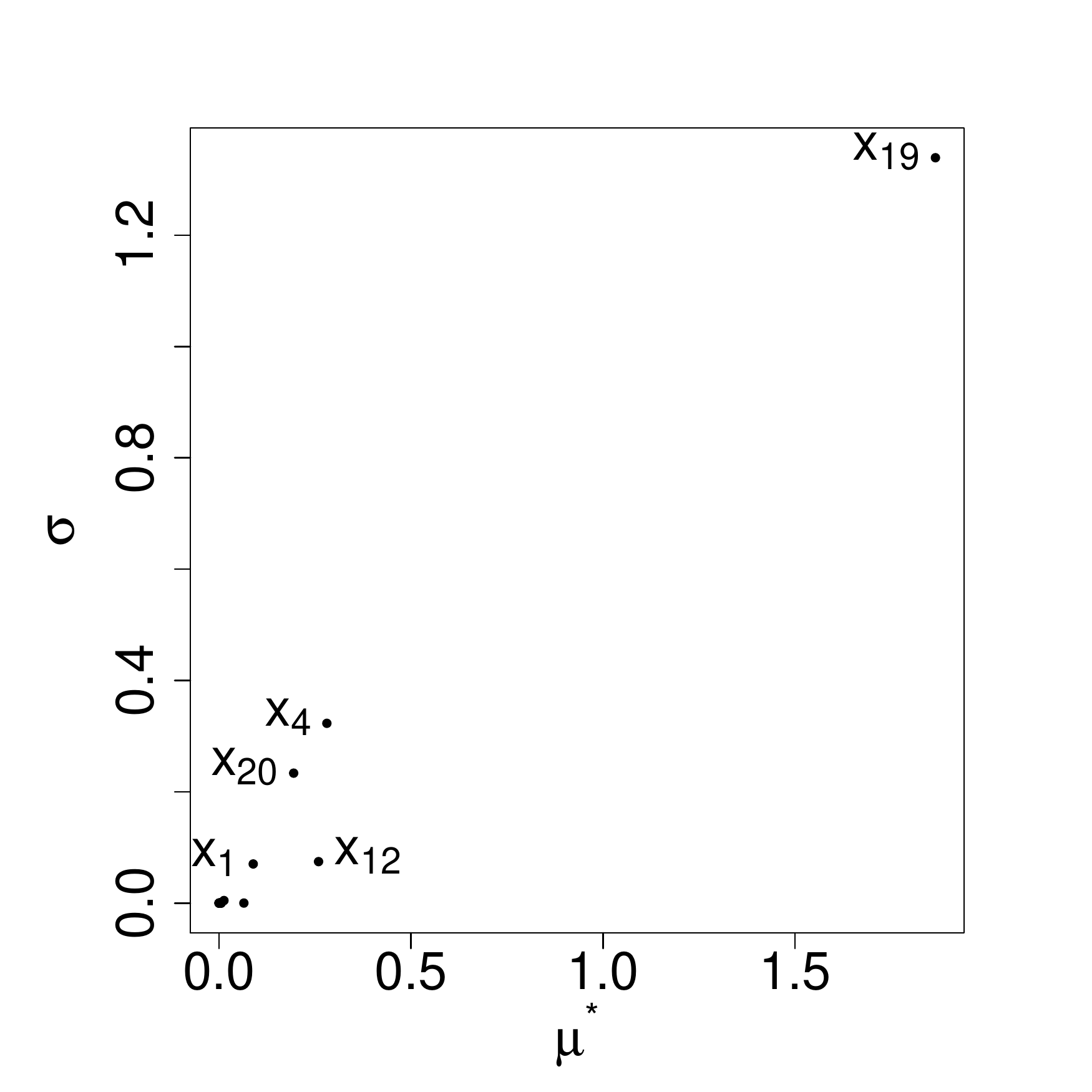} & \includegraphics[scale=\EEscale]{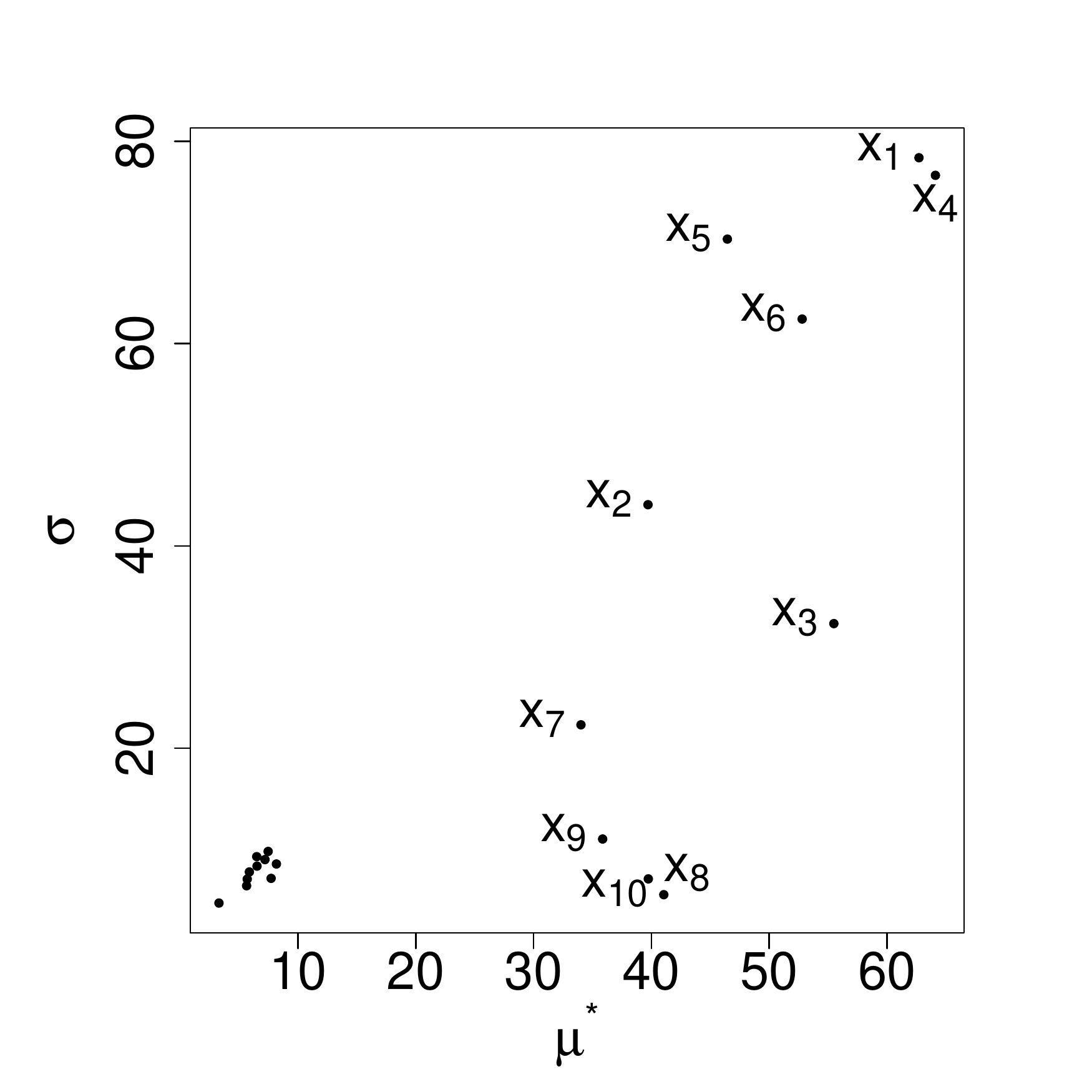}\\
\end{tabular}
\caption{EE results: plots of $\mu^\star_i$~\eqref{eq:EEmu*} against $\sigma_i$~\eqref{eq:EEsigma} for Examples~1 and~2. Labels indicate variables declared active by visual inspection.}
\label{fig:EE}
\end{figure}

For the EE method, performance was assessed by visual inspection of plots of $\mu^\star_i$ against $\sigma_i$ ($i=1,\ldots,20$), see Figure~\ref{fig:EE}. A number of different samples of trajectory vectors $\vect{x}_1,\ldots,\vect{x}_r$ were used and similar results obtained for each.
For Example~1, where active variables have a smaller influence on the response, the EE method struggled to identify all the active variables. Variable $x_{19}$ was consistently declared active, having a nonlinear effect. For larger $n$, variables $x_{1}$, $x_{4}$, $x_{12}$ and $x_{20}$ were also identified. For Example~2, with larger active effects, the performance was better. All active variables are identified when $n=84$ (as also demonstrated by Morris \cite{morris91}) and when $n=200$. Performance was also strong for $n=42$, with only $x_{3}$ and $x_{7}$ not identified.

For both examples, relatively effective screening was achieved through use of a SFRD to estimate sensitivity indices~\eqref{SI}. These estimated indices are displayed in Figure~\ref{fig:SFRD}. Using a threshold of $S(i)>0.05$ leads to a small number of active variables being missed (one in Example~1 and four in Example~2); the choice of the lower threshold of $S(i)>0.01$ results in all active variables being identified in both examples and no type I errors being made. 

A study of the two true functions provides some understanding of the strong performance of the SFRD. Both functions can be reasonably well approximated (via Taylor series expansions) by linear models involving main effect and interaction terms, with no cancellation of main effects with three variable interactions or of two variable with four variable interactions. It is not difficult to modify the two functions to achieve a substantial reduction in the effectiveness of the SFRD. In Example~1, replacement of the term $5(w_4-w_{20})^2$ by $5w_4^2-5w_{20}^2$ produces a function which is highly nonlinear in $w_4$ and $w_{20}$. Screening for this function resulted in these two variables being no longer declared active when the SFRD was used. For Example~2, if $\beta_j=20$ for $j=1,\ldots,10$, $\beta_{jkl} = 0$ for $j,k,l=1,\ldots,5$ and $\beta_{jkl} = -5$ for $jkl = 6, \ldots, 10$, then use of an SFRD failed to detect variables $x_7 - x_{10}$, even when the threshold $S(i)>0.01$ was applied, due to cancellation of main effects and three variable interactions. The performance of the EE method for both these modified functions was the same as that achieved for the original functions.

\renewcommand{\SFRDscale}{0.6} 
\begin{figure}
\centering
\includegraphics[scale=\SFRDscale]{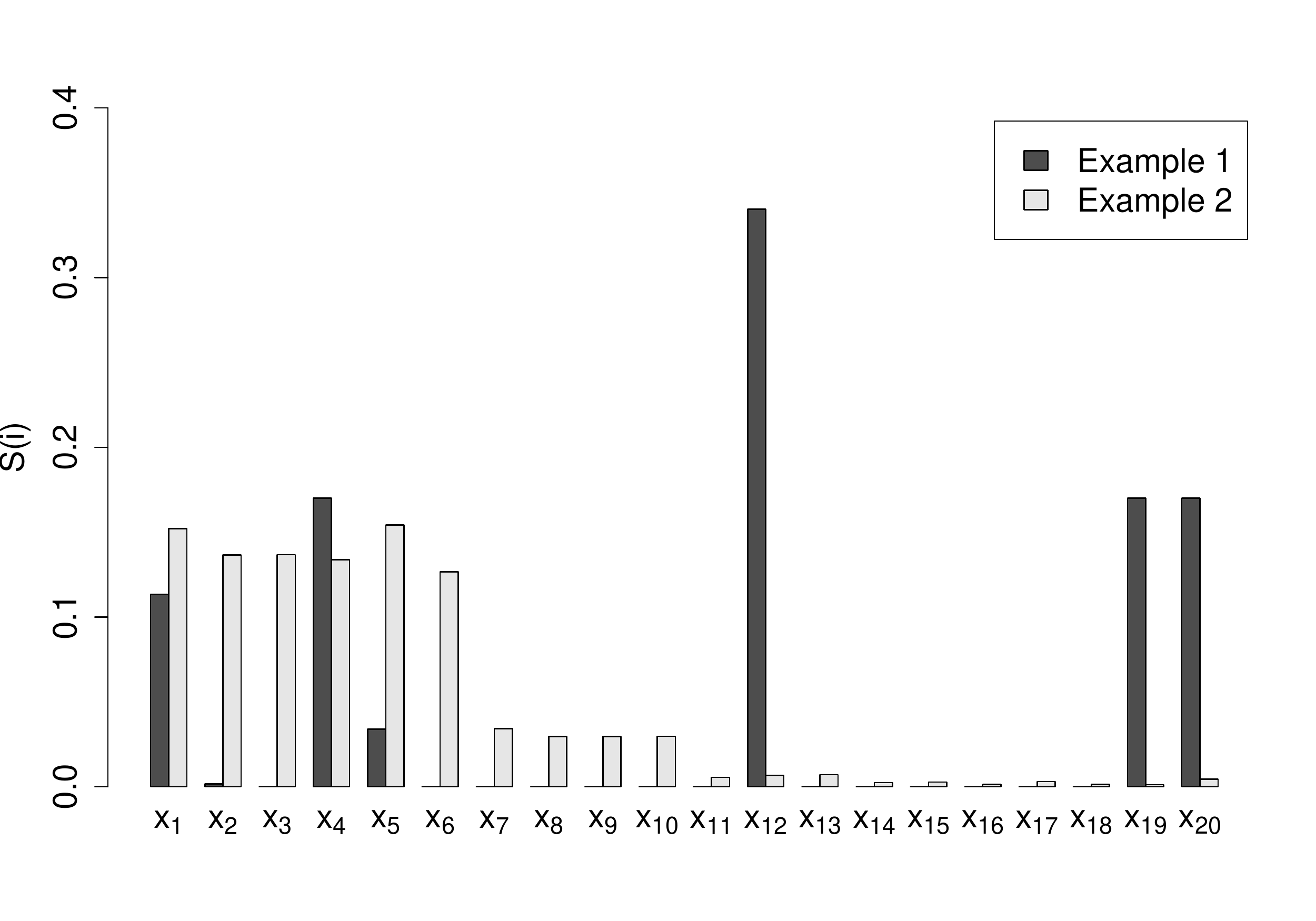}\\
\caption{Sensitivity indices~\eqref{SI} from the SFRD for Examples~1 and~2 for variables $x_1-x_{20}$.}
\label{fig:SFRD}
\end{figure}

\renewcommand{\DSscale}{0.45} 
\begin{figure}[!t]
\centering
\begin{tabular}{cc}
(a) Example~1: SSD $n=16$ & (b) Example~2: SSD $n=16$ \\[-3ex]
\includegraphics[scale=\DSscale]{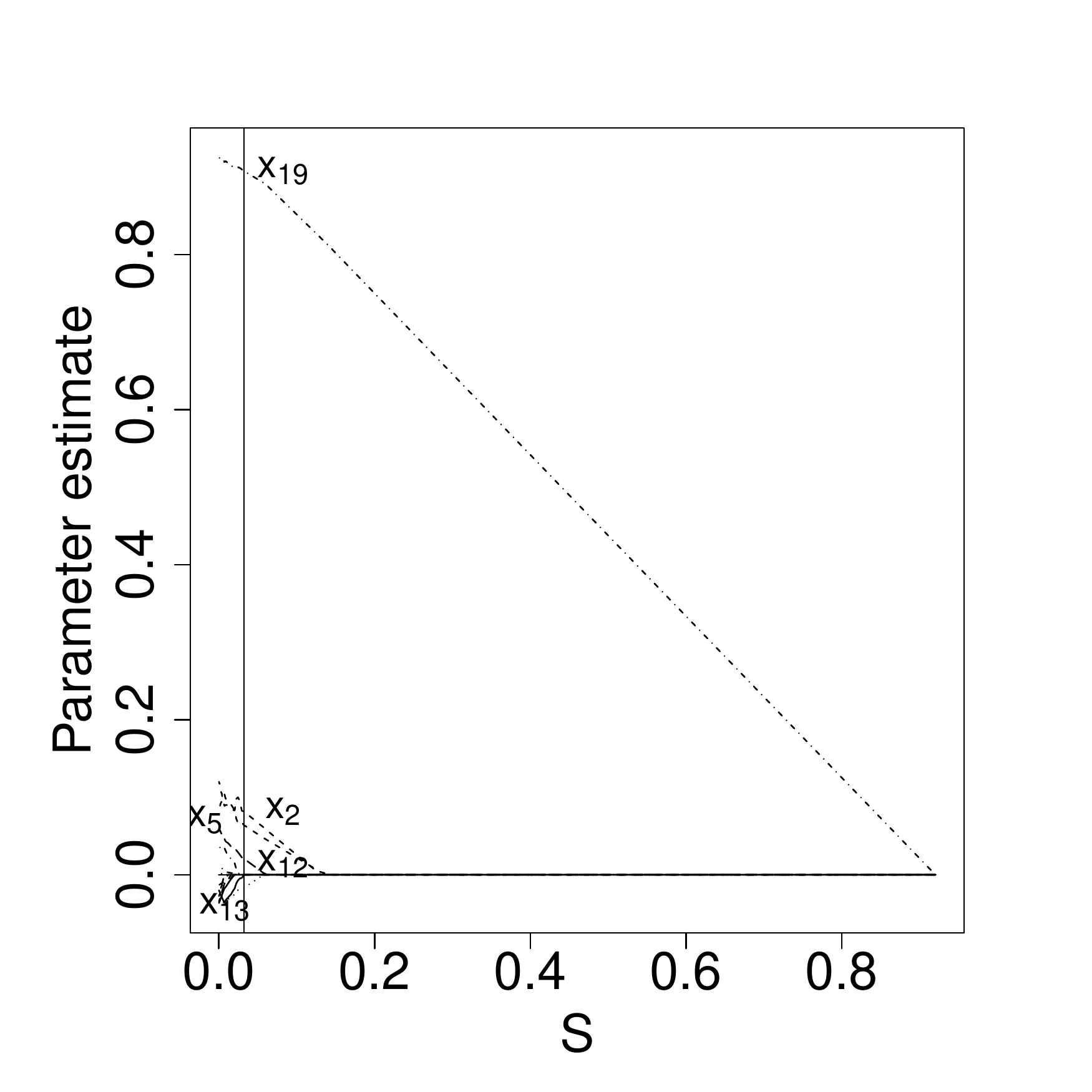} & \includegraphics[scale=\DSscale]{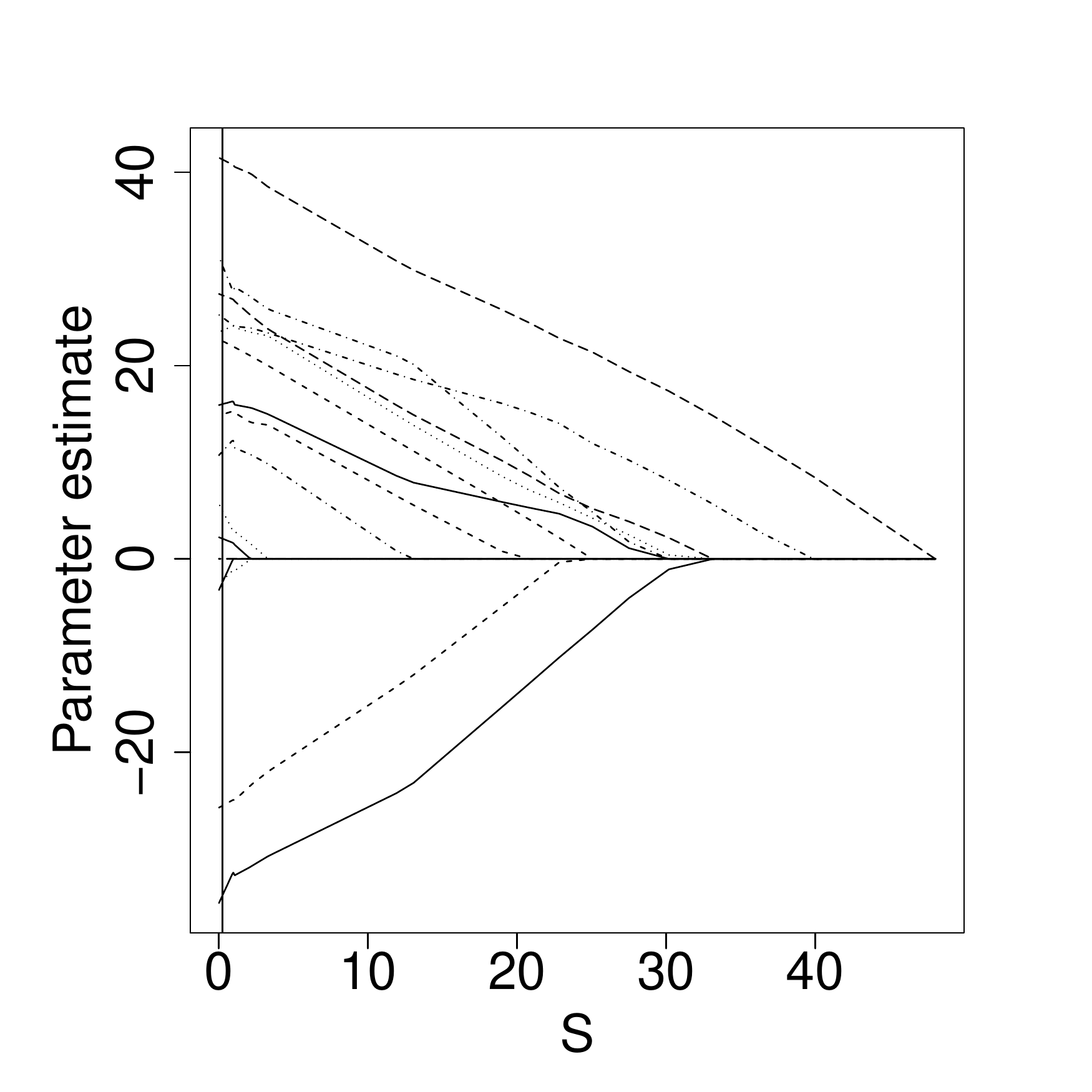}\\
(c) Example~1: DSD $n=41$ & (d) Example~2: DSD $n=41$ \\[-6ex]
\includegraphics[scale=\DSscale]{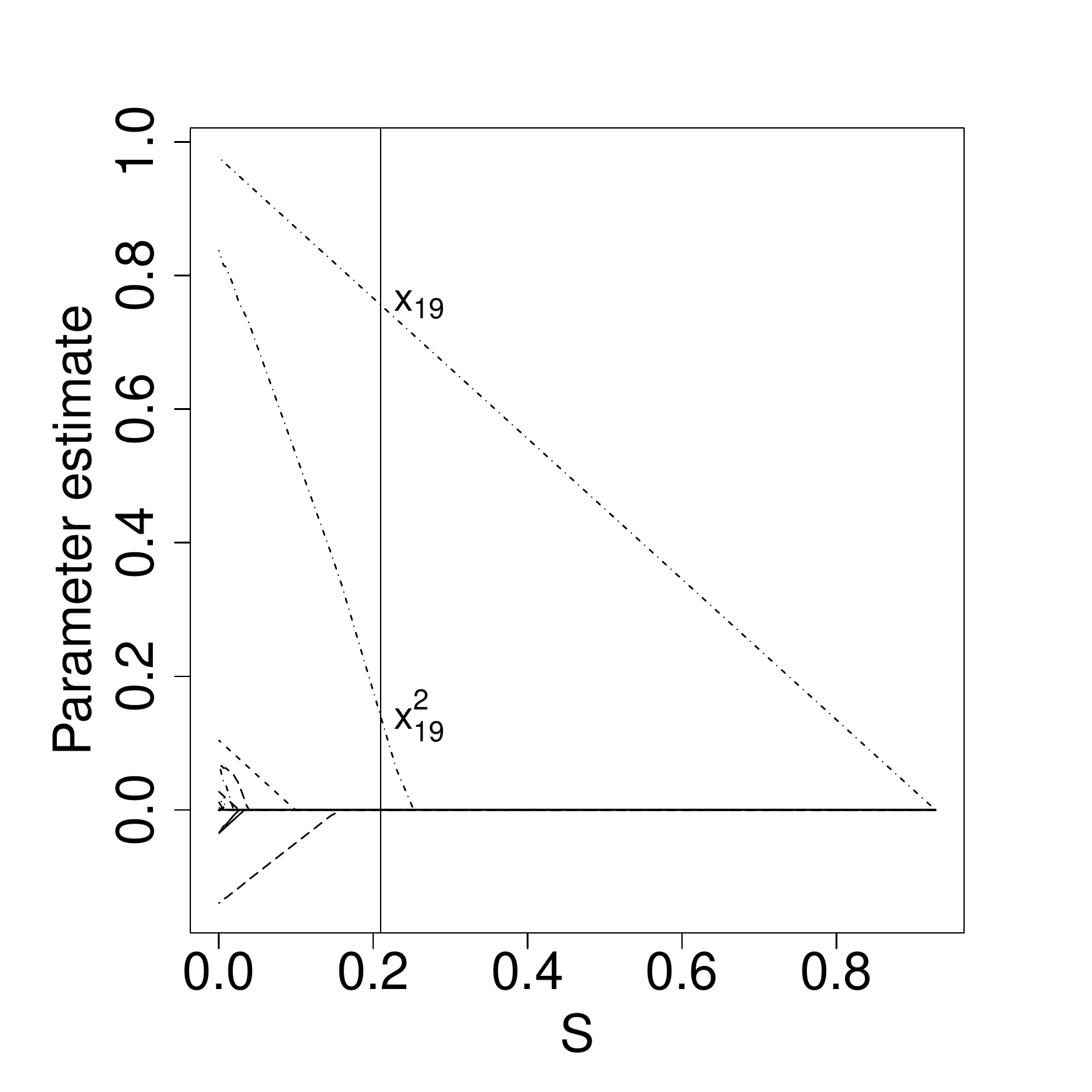} & \includegraphics[scale=\DSscale]{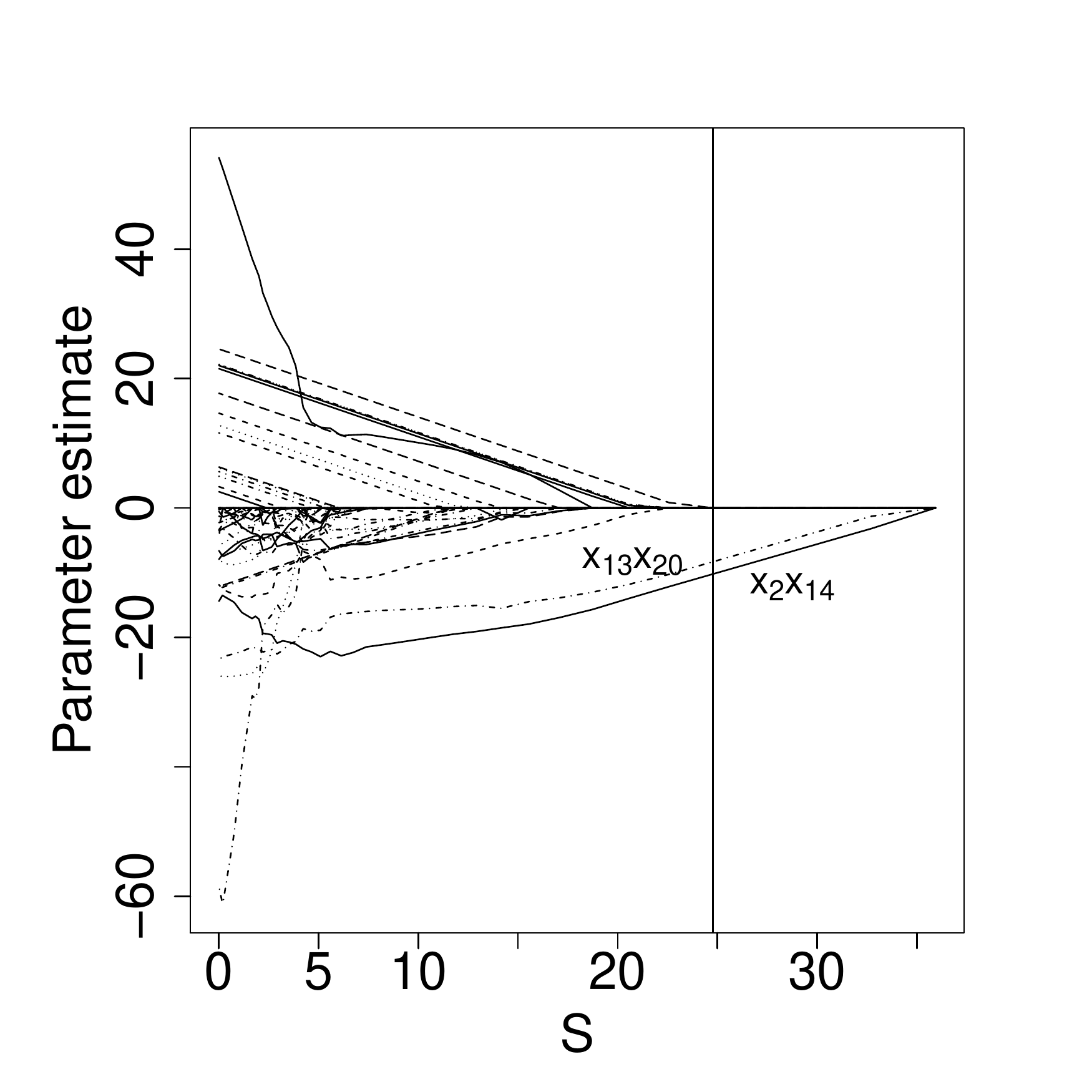}\\
\end{tabular}
\caption{Supersaturated (SSD) and Definitive Screening Design (DSD) results: plots of penalty parameter $s$ against parameter estimates. In each plot, the variables and effects declared active are labelled, with the exception of plot (b) where 15 variables are declared active. }
\label{fig:SSD_DSD}
\end{figure}

The nonregular designs (SSD and DSD) provide an interesting contrast to the other methods, all of which are tailored to, or have been suggested for, variable screening for nonlinear numerical models. For the SSD, only main effects are considered; for the DSD, the surrogate model can include main effects, two-variable interactions and quadratic terms. Figure~\ref{fig:SSD_DSD} gives shrinkage plots for each of the SSD and DSD which show the estimated model parameters against the shrinkage parameter $s$ in~\eqref{eq:ds}. As $s\rightarrow 0$, the shrinkage of the estimated parameters is reduced; at $s=0$, there would be no shrinkage which is not possible for designs with $n<p$. These plots may be used to choose the active variables, namely, those involved in at least one effect whose corresponding estimated model parameter is non-zero for larger values of $s$. In each plot, the value of $s$ chosen by AICc is marked by a vertical line. Figures~\ref{fig:SSD_DSD} (a) and (c) show shrinkage plots for Example~1 from which the dominant active variables are easily identified, although a number of smaller active variables are missed. For Example~2, Figures~\ref{fig:SSD_DSD} (b) and (d) are much harder to interpret, because they have a number of moderately large estimated parameters. This reflects the larger number of active variables in Example~2. Clearly, the effectiveness of both methods for this second example is limited. 

\begin{figure}
\centering
\begin{tabular}{cc}
(a) Example~1: DSD $n=41$ (main effects) & (b) Example~2: DSD $n=41$ (main effects) \\[-6ex]
\includegraphics[scale=\DSscale]{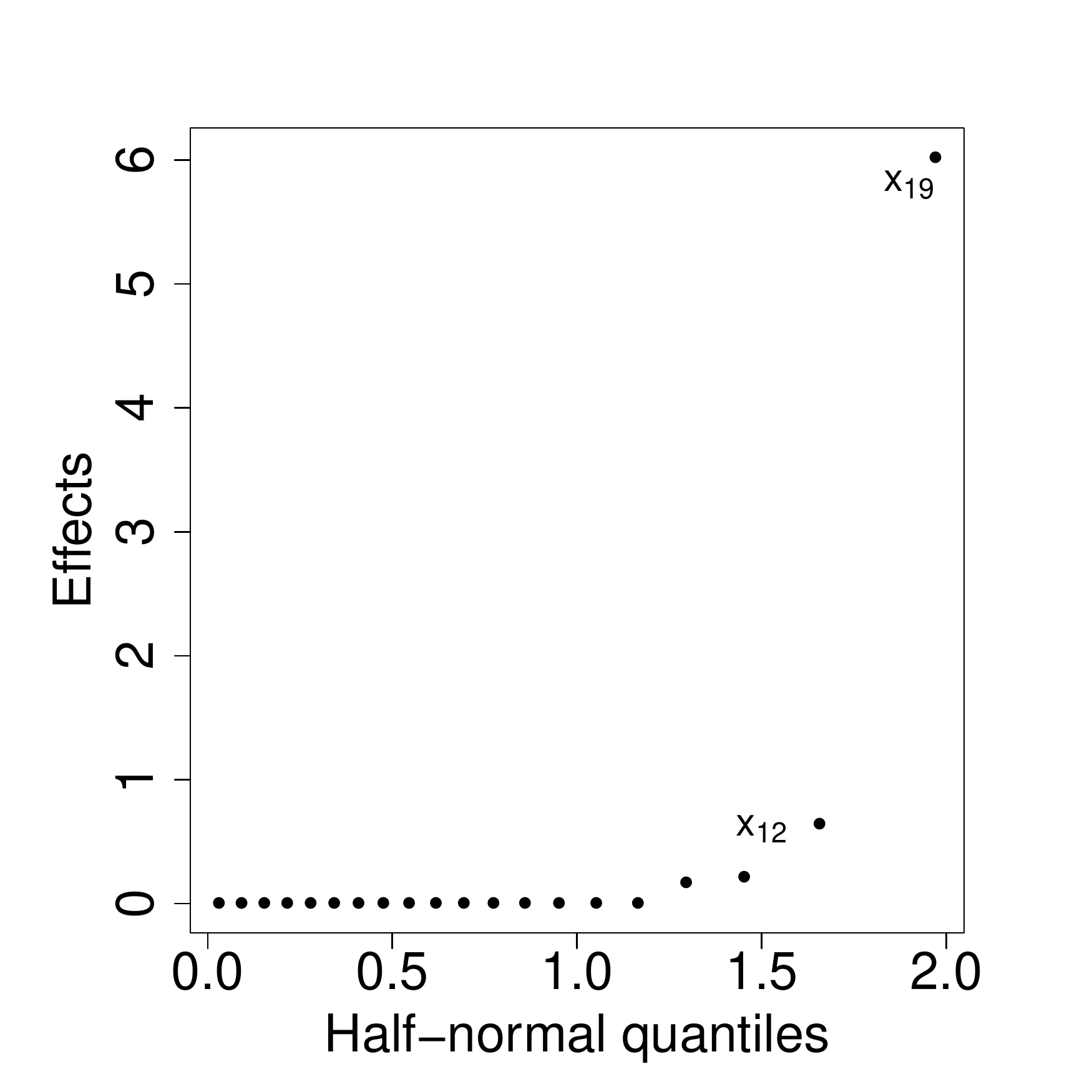} & \includegraphics[scale=\DSscale]{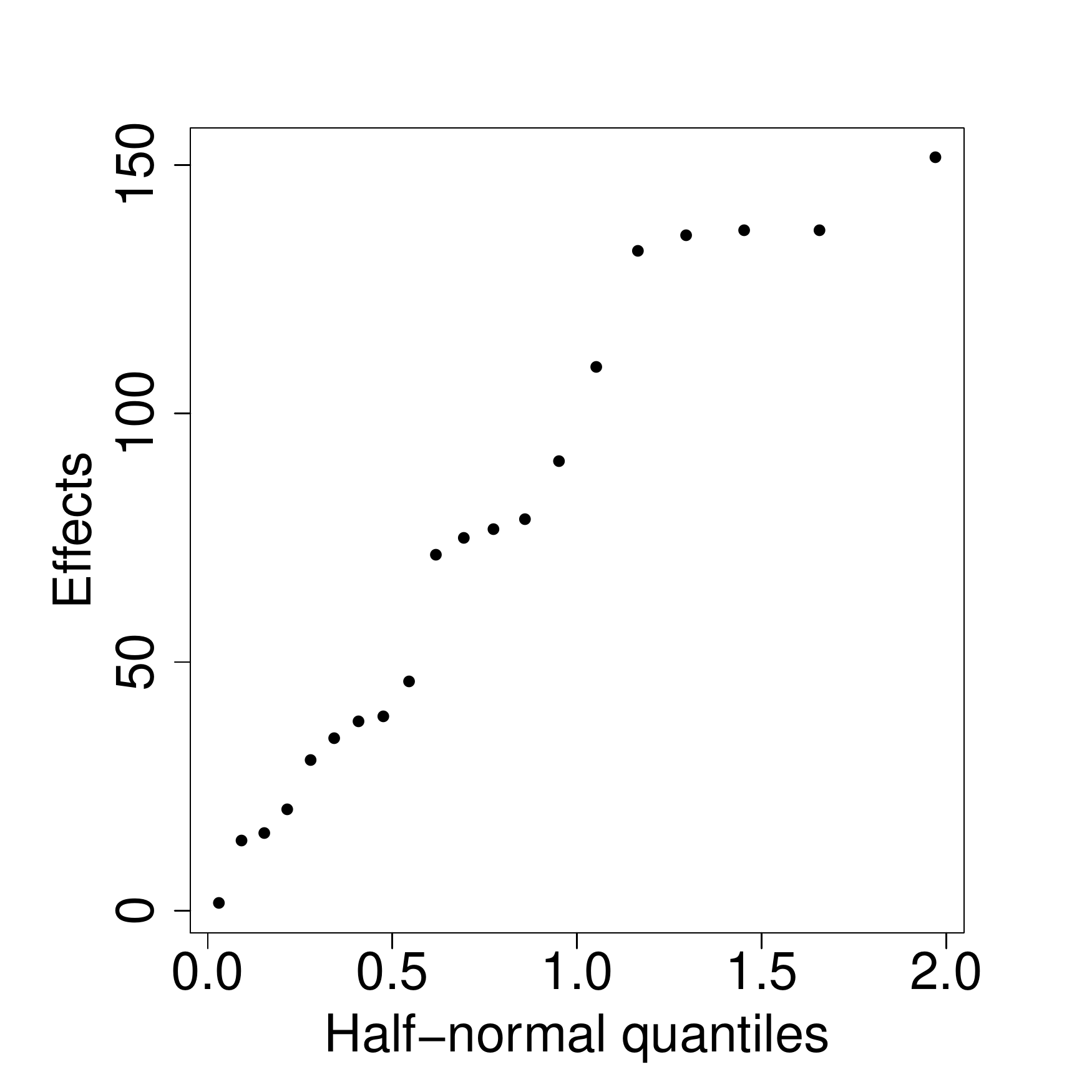}\\
\end{tabular}
\caption{Definitive Screening Design (DSD) results for main effects only: half-normal plots.}
\label{fig:DSDME}
\end{figure}

To provide a further comparison between the SSD and DSD, data from the latter design were also analysed using a main effects only surrogate model~\eqref{eq:firstorder}. For these models, the DSD is an orthogonal design and hence standard linear model analyses are appropriate. To summarise the results, Figure~\ref{fig:DSDME} gives half-normal plots \cite{Daniel59} for each example. Here, the ordered absolute values of the estimated main effects are plotted against theoretical half-normal quantiles. Variables whose estimated main effects stand away from a straight line are declared active, such as $x_{12}$ and $x_{19}$ in Example~1, see Figure~\ref{fig:DSDME}(a). No variables are identified as active for Example~2. These results agree with t-tests on the estimated model parameters. 

\section{Conclusions}\label{conclusions}

Screening with numerical models is a challenging problem, due to the large number of input variables that are typically under investigation, and the complex nonlinear relationships between these variables and the model outputs. 

The results from the study in the previous section highlight these challenges, and the dangers of attempting screening using experiments that are too small or are predicated on linear model methodology. For this study of $d=20$ variables and six screening methods, a sample size of at least $n=40$ was required for effective screening, with more runs needed when factor sparsity did not hold. The EE method was the most effective and robust method for screening, with the highly resource-efficient SSD and DSD being the least effective here. Of course, these two nonregular designs were not developed for the purpose of screening variables in nonlinear functions; in contrast to the SFRD, neither explicitly incorporates higher-order interactions, and the SSD suffers from partial aliasing between main effects. The two Gaussian process methods, RDVS and SSD, required a greater number of runs to provide sensitive screening. 

Methods that use Gaussian process models have the advantage of also providing predictive models for the response. Building such models with the EE or SFRD methods is likely to require additional experimentation. In common with screening via physical experiments, a sequential screening strategy, where possible, is likely to be more effective. Here, a small initial experiment could be run, for example, using the EE method, with more targeted follow-up experimentation and model building focussed on a subset of variables using a Gaussian process modelling approach.

%
\section{Acknowledgements}

D.C. Woods was supported by a Fellowship from the UK Engineering and Physical Sciences Research Council (EP/J018317/1). The authors thank Dr Antony Overstall (University of Glasgow, UK) and Dr Maria Adamou (University of Southampton, UK) for providing code for the RDVS and SGPVS methods, respectively.

\bibliographystyle{spmpsci}
\bibliography{UQscreening}
\end{document}